\documentclass[a4paper,fleqn,usenatbib]{mnras}

\usepackage{newtxtext,newtxmath}

\usepackage[T1]{fontenc}
\usepackage{ae,aecompl}

\usepackage{graphicx}	
\usepackage{amsmath}	
\usepackage{amssymb}	
\usepackage{hyperref}	
\usepackage{pdflscape}	

\title[The outer globular cluster system and stellar halo of M31]{The outer halo globular cluster system of M31 -- III. Relationship to the stellar halo}

\author[Mackey et al.]{A. D. Mackey$^{1}$, A. M. N. Ferguson$^{2}$, A. P. Huxor$^{3}$, J. Veljanoski$^{4}$, G. F. Lewis$^{5}$,
\newauthor A. W. McConnachie$^{6}$, N. F. Martin$^{7,8}$, R. A. Ibata$^{7}$, M. J. Irwin$^{9}$, P. C\^{o}t\'{e}$^{6}$,
\newauthor M. L. M. Collins$^{10}$, N. R. Tanvir$^{11}$, N. F. Bate$^{9}$
\\
$^{1}$Research School of Astronomy and Astrophysics, Australian National University, Canberra, ACT 2611, Australia\\
$^{2}$Institute for Astronomy, University of Edinburgh, Royal Observatory, Blackford Hill, Edinburgh, EH9 3HJ, UK\\
$^{3}$HH Wills Physics Laboratory, University of Bristol, Tyndall Avenue, Bristol, BS8 1TL, UK\\
$^{4}$Kapteyn Astronomical Institute, University of Groningen, PO Box 800, NL-9700 AV Groningen, The Netherlands\\
$^{5}$Sydney Institute for Astronomy, School of Physics, A28, The University of Sydney, Sydney, NSW 2006, Australia\\
$^{6}$NRC Herzberg Astronomy and Astrophysics, 5071 West Saanich Road, Victoria, B.C., V9E 2E7, Canada\\
$^{7}$Universit\'{e} de Strasbourg, CNRS, Observatoire astronomique de Strasbourg, UMR 7550, F-67000 Strasbourg, France\\
$^{8}$Max-Planck-Institut f\"{u}r Astronomie, K\"{o}nigstuhl 17, D-69117 Heidelberg, Germany\\
$^{9}$Institute of Astronomy, University of Cambridge, Madingley Road, Cambridge, CB3 0HA, UK\\
$^{10}$Department of Physics, University of Surrey, Guildford, GU2 7XH, UK\\
$^{11}$Department of Physics and Astronomy, University of Leicester, University Road, Leicester, LE1 7RH, UK\\
}

\date{Draft version \today.}

\pubyear{2019}

\begin{document}
\label{firstpage}
\pagerange{\pageref{firstpage}--\pageref{lastpage}}
\maketitle

\begin{abstract}
We utilise the final catalogue from the Pan-Andromeda Archaeological Survey to investigate the links between the globular cluster system and field halo in M31 at projected radii $R_{\rm proj}=25-150$\ kpc. In this region the cluster radial density profile exhibits a power-law decline with index $\Gamma=-2.37\pm0.17$, matching that for the stellar halo component with $[$Fe$/$H$]<-1.1$. Spatial density maps reveal a striking correspondence between the most luminous substructures in the metal-poor field halo and the positions of many globular clusters. By comparing the density of metal-poor halo stars local to each cluster with the azimuthal distribution at commensurate radius, we reject the possibility of no correlation between clusters and field overdensities at $99.95\%$ significance. We use our stellar density measurements and previous kinematic data to demonstrate that $\approx35-60\%$ of clusters exhibit properties consistent with having been accreted into the outskirts of M31 at late times with their parent dwarfs. Conversely, at least $\sim40\%$ of remote clusters show no evidence for a link with halo substructure. The radial density profile for this subgroup is featureless and closely mirrors that observed for the apparently {\it smooth} component of the metal-poor stellar halo. We speculate that these clusters are associated with the smooth halo; if so, their properties appear consistent with a scenario where the smooth halo was built up at early times via the destruction of primitive satellites. In this picture the {\it entire} M31 globular cluster system outside $R_{\rm proj}=25$\ kpc comprises objects accumulated from external galaxies over a Hubble time of growth.
\end{abstract}

\begin{keywords}
galaxies: individual (M31) -- galaxies: halos -- globular clusters: general --  galaxies: formation -- Local Group
\end{keywords}



\section{Introduction}
Globular clusters are widely used as key tracers of the main astrophysical processes driving the formation and evolution of galaxies \citep[e.g.,][]{brodie:06,harris:10}. Their utility stems in part from a variety of convenient characteristic properties: {\it ubiquity}, being found in essentially all galaxies with stellar masses greater than $\sim 10^9 M_\odot$ as well as many below this limit; {\it observability}, usually being both compact and luminous (with a typical size $r_h \sim 3$\ pc, and brightness $M_V \sim -7.5$); and {\it longevity}, commonly surviving in excess of a Hubble time unless subjected to a disruptive tidal environment. However, their usefulness as tracers of galaxy assembly is mainly a consequence of the apparently close, although not necessarily straightforward, couplings found between the features of a given globular cluster system and the overall properties of the host and its constituent stellar populations. These connections can give rise to surprisingly simple scaling relations, such as the nearly one-to-one linear correlation observed between the halo mass of a galaxy and the total mass in its globular cluster population spanning more than five orders of magnitude \citep[see e.g.,][]{hudson:14,harris:15,forbes:18}. 

While it was once thought that globular clusters formed as a result of special conditions found only in the high-redshift universe \citep[e.g.,][]{peebles:68,peebles:84,fall:85}, more recent work has shown that the simple assumption that globular clusters form wherever high gas densities, high turbulent velocities, and high gas pressures are found -- i.e., in intense star forming episodes -- leads self-consistently to many of the observed properties of globular cluster systems at the present day \citep[e.g.,][]{kravtsov:05,muratov:10,elmegreen:10,griffen:10,tonini:13,li:14,katz:14,kruijssen:14,kruijssen:15,pfeffer:18}. This provides a natural explanation for the tight links observed between cluster systems and their host galaxies, and motivates the empirically successful use of globular clusters as tracers of galaxy development across all morphological types \citep[e.g.,][]{strader:04,strader:06,peng:08,georgiev:09,georgiev:10,forbes:11,romanowsky:12,harris:13,brodie:14}.

Globular clusters played a central role in helping develop our understanding of the formation of the Milky Way, providing some of the first experimental evidence that the hierarchical accretion of small satellites might represent an important assembly channel \citep{searle:78,zinn:93}. This picture was spectacularly verified with the discovery of the disrupting Sagittarius dwarf galaxy \citep{ibata:94,ibata:95}, presently being assimilated into the Milky Way's halo along with its retinue of globular clusters \citep[e.g.,][]{dacosta:95,md:02,bellazzini:03,carraro:07}. It is now known that the extended low surface brightness stellar halos that surround large galaxies are a generic product of the mass assembly process in $\Lambda$CDM cosmology \citep[e.g.,][]{bullock:05,cooper:10}; this accreted material typically includes a substantial portion of the associated globular cluster system \citep[cf.][]{beasley:18}.

Additional evidence in favour of the idea that a significant fraction of globular clusters in the Milky Way is accreted comes from precision stellar photometry with {\it HST}, which revealed that the Galactic globular clusters follow a bifurcated age-metallicity distribution \citep{marinfranch:09,dotter:11,leaman:13}. The properties of the cluster age-metallicity relationship have been used to infer that the Milky Way must have accreted at least three significant satellites including Sagittarius \citep{kruijssen:18}; overall, around $40\%$ or more of the Galactic globular cluster system is likely to have an {\it ex situ} origin \citep[see also][]{mackey:04,mackey:05,forbes:10}. The second data release from the Gaia mission \citep{gaia:18} has recently facilitated the derivation of full 6D phase-space information for many of the Milky Way's globular clusters \citep[e.g.,][]{gaiagc:18,vasiliev:18}, adding further support for the idea that many are accreted objects \citep{myeong:18,helmi:18}.

Despite this vast array of indirect evidence, and despite the discovery of abundant substructure and numerous stellar streams criss-crossing the Milky Way's inner halo \citep[e.g.,][]{belokurov:06,bell:08,bernard:16,grillmair:16,shipp:18,malhan:18}, surveys targeting the outskirts of globular clusters in search of the expected debris from their now-defunct parent systems have proven largely fruitless \citep[e.g.,][]{carballobello:14,carballobello:18,kuzma:16,kuzma:18,myeong:17,sollima:18}. Indeed, apart from several Sagittarius members, there is no unambiguous example of a Milky Way globular cluster that is embedded in a coherent tidal stream from a disrupted dwarf galaxy. While this observation might find a natural explanation if the majority of significant accretion events occurred very early in the Galaxy's history \citep[cf.][]{myeong:18,helmi:18,kruijssen:18}, the lack of any obvious association between the supposedly-accreted subset of Milky Way clusters and substructures in the stellar halo inevitably places some doubt on the fidelity with which the properties of the globular cluster system reflect the accretion and merger history inferred directly from the field.

The Andromeda galaxy (M31) provides the next nearest example of a large stellar halo beyond the Milky Way, and constitutes an ideal location to explore in detail the links between the field halo populations and the globular cluster system in an L$^*$ galaxy. Indeed, in many ways M31 offers clear advantages for such study relative to our own Milky Way \citep[as outlined in e.g.,][]{ferguson:16}, and we arguably possess a significantly more complete understanding of both its periphery (i.e., at projected radii $R_{\rm proj} \ga 40$\ kpc) and its low-latitude regions. Considering the stellar halo as a whole, it is well established that the system belonging to M31 contains a higher fraction of the overall galaxy luminosity, is significantly more metal-rich, and is apparently more heavily substructured than that of the Milky Way \citep[e.g.,][]{mould:86,pritchet:88,ibata:01,ibata:07,ibata:14,ferguson:02,irwin:05,mcconnachie:09,gilbert:09,gilbert:12,gilbert:14}, while the M31 globular cluster population is more numerous than that of the Milky Way by at least a factor of three \citep[e.g.,][]{galleti:06,galleti:07,huxor:08,huxor:14,caldwell:16}. 

These observations all suggest that the accretion history of M31 is quite different from that of our own Galaxy, in that M31 has likely experiened more accretions and/or a more prolonged history of accretion events. Beyond this, it is clear that many globular clusters in the outer halo of M31 (at $R_{\rm proj} \ga 25$\ kpc) exhibit distinct spatial and/or kinematic associations with stellar streams or overdensities in the field \citep[e.g.,][]{mackey:10b,mackey:14,veljanoski:14}, and a subset of these objects possess properties \citep[including red horizontal branches possibly indicating younger ages,][]{mackey:13a} similar to those displayed by the apparently accreted subsystem in the Milky Way \citep[e.g.,][]{searle:78,zinn:93,mackey:04}. 

In this paper we utilise the final catalogue from the Pan-Andromeda Archaeological Survey (PAndAS), in combination with an essentially complete census of the globular cluster system \citep[e.g.,][the first paper in this series]{huxor:14}, to conduct the first global, quantitative investigation of the links between the globular clusters and the field halo in M31 at projected radii $R_{\rm proj} = 25-150$\ kpc. This updates and extends our previous work on this topic that either considered only a fraction of the outer halo \citep[e.g.,][]{mackey:10b,huxor:11}, or only a spectroscopically-observed subsample of the cluster population \citep[][the second paper in this series]{veljanoski:14}. PAndAS \citep{mcconnachie:09,mcconnachie:18}, was a Large Program awarded $226$ hours on the Canada-France-Hawaii Telescope (CFHT) during 2008-2010 to survey the outskirts of M31 and M33 with the \mbox{$1$\ deg$^2$} MegaCam imager. This project built upon a set of earlier CFHT/MegaCam imaging by our group over the period 2003-2007 which covered the southern quadrant of the M31 halo \citep[see][]{ibata:07,ibata:14} and was itself built on an earlier survey using the Wide-Field Camera on the Isaac Newton Telescope \citep[e.g.,][]{ibata:01,ferguson:02}. 

The paper is structured as follows. In Section \ref{s:data} we describe the stellar halo and globular cluster catalogues used in this work; in Section \ref{s:analysis} we investigate the spatial distribution of the clusters relative to the halo field populations both qualitatively and statistically; and in Section \ref{s:properties} we use the results of this investigation to identify and measure the properties of cluster subsets exhibiting robust association, and no evident association, with stellar substructures in the halo. We finish with a discussion of our results in the context of the properties of the M31 stellar halo and its inferred accretion history (Section \ref{s:discussion}), and an overall summary (Section \ref{s:summary}).

Throughout this work we assume an M31 distance modulus $(m-M)_0 = 24.46$ \citep{conn:12}, corresponding to a physical distance of $780$\ kpc and an angular scale of $3.78$\ pc per arcsecond.

\section{Data}
\label{s:data}
\subsection{The Pan-Andromeda Archaeological Survey}
\label{ss:pandas}
The basis of the present work is the publically-released PAndAS source catalogue described by \citet{mcconnachie:18}. The final PAndAS data set comprises $406$ individual MegaCam pointings that almost completely cover the area around M31 to a projected galactocentric radius $R_{\rm proj} \sim 150$\ kpc, as well as a conjoined region reaching out to $R_{\rm proj} \sim 50$\ kpc around M33. The total mapped area is roughly $400$\ deg$^{2}$ on the sky. 

All imaging was conducted in the MegaCam $g$- and $i$-band filters, with $3\times 450$s exposures taken in each filter at a given pointing. The PAndAS image quality is typically excellent with a $g$-band mean of $0\farcs 66$ full-width half-maximum (FWHM) and an $i$-band mean of $0\farcs 59$ FWHM (with an rms scatter of $0\farcs 10$ in both filters).  As described in detail by \citet{ibata:14} and \citet{mcconnachie:18}, initial data reduction occurred at CFHT, followed by additional processing, source detection and photometry using pipelines developed at the Cambridge Astronomical Survey Unit (CASU). Some $96$ million objects are listed in the full PAndAS photometric catalogue, of which roughly one-third are classified as ``stellar'' (i.e., point sources). The astrometric solution is based on cross-matching with Gaia DR1 \citep{gaia:16} and has typical residuals smaller than $0.02\arcsec$ rms, while the overall photometric calibration is derived using overlapping Pan-STARRS DR1 fields \citep{flewelling:16} and is good to $\sim 0.01$ mag. The median PAndAS $5\sigma$ point source depth is $g = 26.0$ and $i = 24.8$. 

\subsection{Globular cluster catalogue ($R_{\rm proj} > 25$\ kpc)}
\label{ss:outercat}
In this paper we are primarily interested in the {\it outer halo} globular cluster system of M31, which we define as objects lying at $R_{\rm proj} > 25$ kpc. Our catalogue of such clusters comes predominantly from a survey conducted by our group, that utilised the PAndAS imaging \citep{huxor:14}. From our search of these data we located $52$ previously-unknown clusters with $25 \le R_{\rm proj} \le 150$\ kpc\footnote{Note that in \citet{huxor:14} we actually catalogued $53$ previously-unknown clusters; however, we show in Appendix \ref{a:gemini} of the present work that the borderline object PA-55 is in fact a background galaxy. The two low-confidence candidate clusters identified in \citet{huxor:14} are also background galaxies.}, augmenting another $32$ already known from our various pre-PAndAS surveys \citep{huxor:05,huxor:08,martin:06,mackey:06,mackey:07} plus $6$ already known from a number of earlier works as compiled in Version 5 of the Revised Bologna Catalogue \citep[RBC,][]{galleti:04}\footnote{See \href{http://www.bo.astro.it/M31/}{http://www.bo.astro.it/M31/}}.  We were also able to rule out, as either background galaxies or foreground stars, almost all of the candidate clusters with $R_{\rm proj} > 25$ kpc listed in the RBC V5.

The uniform spatial coverage and excellent quality of the PAndAS imaging mean that our catalogue of remote clusters is largely complete.  In \citet{huxor:14} we used an extensive series of artificial cluster tests to show that the detection efficiency only begins to degrade at luminosities below $M_V = -6.0$, with the $50\%$ completeness limit at $M_V \approx -4.1$. Furthermore, the PAndAS filling factor is high: $\ga 96\%$ out to $R_{\rm proj} = 105$\ kpc, falling to $80\%$ at $130$\ kpc and $\sim 20\%$ at $150$\ kpc. Combining this with the observed radial distribution of clusters suggests that we plausibly missed $\la 5$ objects over the range $25 \le R_{\rm proj} \le 150$\ kpc due to gaps in the coverage \citep[see][and Section \ref{ss:profile} in the present work]{huxor:14}. This is independent of luminosity but subject to the same detection function outlined above.

\begin{figure}
\begin{center}
\includegraphics[width=75mm,clip]{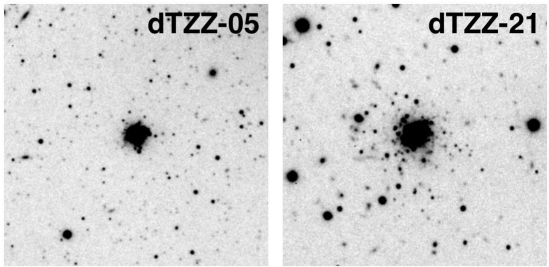}
\caption{Gemini/GMOS images of the two confirmed outer halo globular clusters from the sample of \citet{dtz:15} that do not appear in \citet{huxor:14}. Each image is a $1\arcmin \times 1\arcmin$ cut-out from a full GMOS $i^\prime$-band frame. North is to the top and east to the left.}
\label{f:gmos}
\end{center}
\end{figure}

Simultaneously with our PAndAS work, \citet{dtz:13,dtz:14,dtz:15} utilised the Sloan Digital Sky Survey (SDSS) to search for new M31 globular clusters. They ultimately produced a sample of $22$ high-confidence objects, of which $12$ lie at $R_{\rm proj} > 25$ kpc. Ten of these remote clusters appear independently in our catalogue from \citet{huxor:14}; as detailed in Appendix \ref{a:gemini}, we have used imaging with the GMOS instrument on Gemini North to independently verify that the remaining two are also {\it bona fide} globular clusters.  These objects are dTZZ-05 (also known as SDSS-D), which is a small compact cluster at $R_{\rm proj} = 32.0$\ kpc falling partially in a PAndAS chip-gap; and dTZZ-21 (also known as SDSS-G), which is a more luminous and extremely remote cluster at $R_{\rm proj} = 137.8$\ kpc on the extreme north-eastern edge of the PAndAS footprint. GMOS image cutouts are shown in Figure \ref{f:gmos}.

In summary, the catalogue of M31 outer halo globular clusters that we use in the present work consists of $92$ objects spanning $25 \le R_{\rm proj} \la 150$\ kpc. The full list, along with ancillary photometric and kinematic data, is presented in Appendix \ref{a:data}.  Since the vast majority of the sample was identified in PAndAS imaging we assume the completeness limits described above.

\subsection{Globular cluster catalogue ($R_{\rm proj} \le 25$\ kpc)}
\label{ss:innercat}
In the following analysis we will sometimes, largely for illustrative purposes, supplement our outer halo catalogue with a list of globular clusters belonging to the inner parts of the M31 system. For this we first select all objects listed in the RBC V5 as ``confirmed'' globular clusters (i.e., with a classification flag `f' of either $1$ or $8$) lying at $R_{\rm proj} \le 25$\ kpc. We then exclude from this list the subset possessing indicators of a young age $\la 2$\ Gyr, which are predominantly clusters set against the stellar disk.  This is achieved by considering the following three RBC flags: `yy', which is an age classification from \citet{fusipecci:05} based on the integrated $(B-V)$ colour and/or the strength of the H$\beta$ spectral index; `ac', which relies on spectroscopy by \citet{caldwell:09}; and `pe', which comes from the broadband photometry of \citet{peacock:10}. For those objects with multiple classifications we take the majority view, although in general the agreement between the three studies is quite good.  In the case where an object has no available age data, we retain it in the list. Finally, we edit the list to incorporate the few updated classifications and new clusters detailed in \citet{huxor:14}, as well as the clusters discovered by \citet{dtz:13,dtz:14,dtz:15}.  That is, we remove SK002A, SK004A, and BA11, and add B270D, SK255B, SK213C, PA-28, PA-29, PA-32, PA-34 ($=\,$dTZZ-11), PA-35, PA-59, dTZZ-04 ($=\,$SDSS1), dTZZ-06, dTZZ-07 ($=\,$SDSS-E), dTZZ-08, dTZZ-09, dTZZ-10 ($=\,$SDSS3), dTZZ-12 ($=\,$SDSS6), dTZZ-13, and dTZZ-14.  Overall, this process returns a sample of $425$ M31 globular clusters with $R_{\rm proj} \le 25$\ kpc, consistent with ($\approx 10\%$ larger than) the ensemble compiled by \citet{caldwell:16}. 

\subsection{Globular clusters in M31 satellite galaxies}
\label{ss:sat}
Several of the major satellites of M31 sitting inside the PAndAS footprint possess their own globular cluster systems, and in what follows it will also, at times, be of interest for us to consider the spatial distribution of these objects. Fortunately, searches of the PAndAS data have ensured that the censuses of clusters in NGC 147, NGC 185, and the outskirts of M33 are now essentially complete, building on earlier compilations extending back many years. For NGC 147 we use the catalogue presented by \citet{veljanoski:13b}, which lists $10$ globular clusters including three discovered in PAndAS, three found by \citet{sharina:09}, and four noted by \citet{hodge:76}\footnote{But see also \citet{baade:44}, as well as the Appendix in \citet{veljanoski:13b} which details inconsistencies in the naming of these four clusters throughout the literature over the intervening seventy years.}. For NGC 185 we again employ the \citet{veljanoski:13b} catalogue, which lists $8$ globular clusters including one from PAndAS and seven identified by \citet*{ford:77}\footnote{But again see \citet{baade:44}, as well as \citet{hodge:74}, \citet{dacosta:88}, and \citet{geisler:99}.}. 

M33 presents a more complicated case because it possesses an extensive population of both young and intermediate-age clusters set against its face-on disk, which makes identifying a robust set of {\it ancient} globular clusters in this galaxy a difficult task. The outskirts of M33, at projected radii larger than $\sim 10$ kpc, have been thoroughly searched and for this region we utilise a catalogue consisting of the $6$ clusters identified by \citet{stonkute:08}, \citet{huxor:09}, and \citet{cockcroft:11}. We supplement this with a list of $27$ objects inside $10$ kpc taken from \citet{sarajedini:07} and \citet{beasley:15}, that have age estimages greater than $7$\ Gyr (i.e., a limit that corresponds, approximately, to the youngest globular clusters seen in the Milky Way halo). Since these inner objects are used only for illustrative purposes, we are not concerned about incompleteness in this region, nor errors in the age estimates (which largely come from integrated photometry and spectroscopy). We note that the true number of ancient clusters projected against the inner parts of M33 may be significantly larger than the size of the sample adopted here \citep[e.g.,][]{ma:12,fan:14}. 

The compact elliptical galaxy M32 is not known to possess any globular clusters \citep[although may harbour a few younger objects -- e.g.,][]{rudenko:09}. On the other hand, the nucleated dwarf NGC 205 likely contains $\sim 6-8$ globular clusters \citep[see e.g.,][]{hubble:32,dacosta:88}; however due to the close proximity of this satellite to the centre of M31 ($R_{\rm proj} = 8.3$\ kpc) we do not worry about explicitly separating these objects from the list of $425$ "inner" M31 clusters discussed in Section \ref{ss:innercat} above. It is also likely that extremely faint star clusters may be present in the dwarf spheroidal satellites Andromeda I and XXV \citep{cusano:16,caldwell:17}. Since the exact nature of these objects remains ambiguous we elect to exclude them from our present analysis; this choice is of little consequence given our overall focus on exploring the links between globular clusters and the field star populations in the M31 halo.

\subsection{Stellar halo catalogue \& contamination model}
\label{ss:halocat}
To quantify the spatial distribution of stars in the M31 halo we use the PAndAS point source catalogue described above in Section \ref{ss:pandas}. Photometry for each detection is de-reddened using the \citet{schlegel:98} extinction maps with the corrections derived by \citet{schlafly:11}. As described by \citet{mcconnachie:18} the median colour excess across the PAndAS footprint (but excluding the central $2\degr$ around M31 and $1\degr$ around M33) is $E(B-V) = 0.072$, with minimum and maximum values of $E(B-V) = 0.032$ and $0.220$, respectively

\begin{figure}
\begin{center}
\includegraphics[width=75mm,clip]{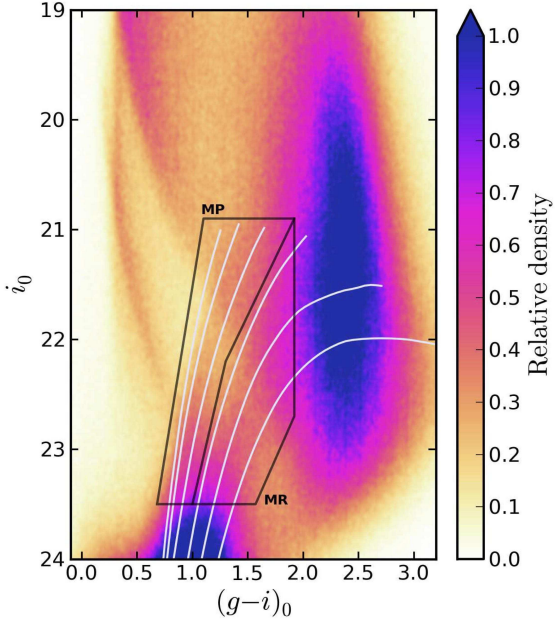}
\caption{Colour-magnitude diagram for all stars in the PAndAS survey area excluding the regions within $30$\ kpc of the centre of M31, $15$\ kpc of the centre of M33, and $10$\ kpc of the centres of NGC 147 and 185. The colour-map represents the stellar density in $0.02\times0.02$ mag pixels, smoothed with a Gaussian kernel of $\sigma = 0.02$ mag, and normalised to $70\%$ of the maximum pixel value. This has the effect of saturating regions of the CMD, but increases the low-density contrast without moving to a non-linear scale. The fiducial sequences are isochrones from the Dartmouth Stellar Evolution Database \citep{dotter:08} for $12.5$ Gyr old stars and a range of metallicities, shifted by $(m-M)_0 = 24.46$ to indicate the region occupied by M31 halo stars. From left to right, $[$Fe$/$H$] = -2.5$, $-2.0$, $-1.5$, $-1.0$, $-0.75$, and $-0.5$. For $[$Fe$/$H$] \le -1.5$ we assume $[\alpha/$Fe$] = +0.4$, decreasing to $[\alpha/$Fe$] = +0.2$ at higher metallicities. Most of the features on the CMD are due to non-M31 populations. Stars in the thin disk of the Milky Way form the dominant vertical sequence near ($g-i)_0 \sim 2.3$, while the diagonal sequence starting near $i_0 = 19.0$ and $(g-i)_0 \approx 0.3$ is due to the thick disk of the Milky Way. The narrow sequence below that of the thick disk is from stars in the Galactic halo, while unresolved background galaxies occupy the region near $i_0 = 24.0$ and $(g-i)_0 \sim 1.0$. The black boxes delineate our selection criteria for M31 stars with $-2.5 \la [$Fe$/$H$] \la -1.1$ (``metal-poor'' $\equiv$\ MP) and $-1.1 \la [$Fe$/$H$] \la 0.0$ (``metal-rich'' $\equiv$\ MR).}
\label{f:cmd}
\end{center}
\end{figure}

In Figure \ref{f:cmd} we plot the colour-magnitude diagram (CMD) for all stars in the PAndAS survey area barring those that lie in the central regions of M31 and M33, and the dwarf elliptical satellites NGC 147 and 185. More specifically, we have excised all stars within $30$\ kpc of the centre of M31, $15$\ kpc of the centre of M33, and $10$\ kpc of the centres of NGC 147 and 185, leaving the outer halo populations that the present work is mainly focused on. These populations are of low spatial density -- the majority of stars visible in Figure \ref{f:cmd} do not, in fact, belong to the Andromeda system; rather they are members of the thin disk, thick disk, and halo of the Milky Way that happen to lie along the PAndAS line of sight \citep[e.g.,][]{martin:14}. Unresolved background galaxies also populate the faint end of the CMD. The overplotted isochrones, which come from the Dartmouth Stellar Evolution Database \citep{dotter:08} and have been shifted to our assumed M31 distance modulus, show where we expect to find red giant branch (RGB) stars of age $12.5$ Gyr and varying $[$Fe$/$H$]$ in the Andromeda halo. 

Because M31 sits at relatively low Galactic latitude ($b \approx -20\degr$) the foreground contamination is quite heavy, and in fact overwhelms the sparse stellar halo of Andromeda in some places -- especially to the north where the star counts increase exponentially in the direction towards the Galactic plane. Moreover, the CMD region occupied by unresolved background galaxies tends to substantially overlap the faint part of the domain populated by M31 halo RGB stars. For these reasons, \citet{martin:13} constructed an empirical model describing the density of non-M31 sources as a function of spatial and colour-magnitude position:
\begin{multline}
\Sigma_{(g_0-i_0,i_0)} (\xi,\eta) =\\ 
\exp \left(\alpha_{(g_0-i_0,i_0)} \xi + \beta_{(g_0-i_0,i_0)} \eta + \gamma_{(g_0-i_0,i_0)} \right) .
\label{e:contam}
\end{multline}
Here, the location on the (de-reddened) CMD is given by $(g_0-i_0,i_0)$, while the spatial location is defined by the coordinates $(\xi,\eta)$ on the tangent-plane projection centred on M31. The model is valid over the full span of the PAndAS footprint, and within the colour-magnitude box bounded by $0.2 \le (g_0-i_0) \le 3.0$ and $20.0 \le i_0 \le 24.0$. At any given point within the survey area we can use the tabulated values of $(\alpha,\beta,\gamma)$ to generate a finely-gridded contamination CMD to be subtracted from the observations, allowing the creation of largely contamination-free M31 halo CMDs and spatial density maps. One important caveat is that the \citet{martin:13} model was necessarily defined using the outermost reaches of the PAndAS survey area at $R_{\rm proj} \ga 120$\ kpc. Despite their remoteness, these regions are not completely free of M31 halo stars \citep[see e.g.,][]{ibata:14} meaning that there is a very small, but non-zero, M31 halo component included in the contamination model. In what follows we will note the effect of this where appropriate, although it does not alter any of our conclusions.

\begin{figure*}
\begin{center}
\includegraphics[width=138mm,clip]{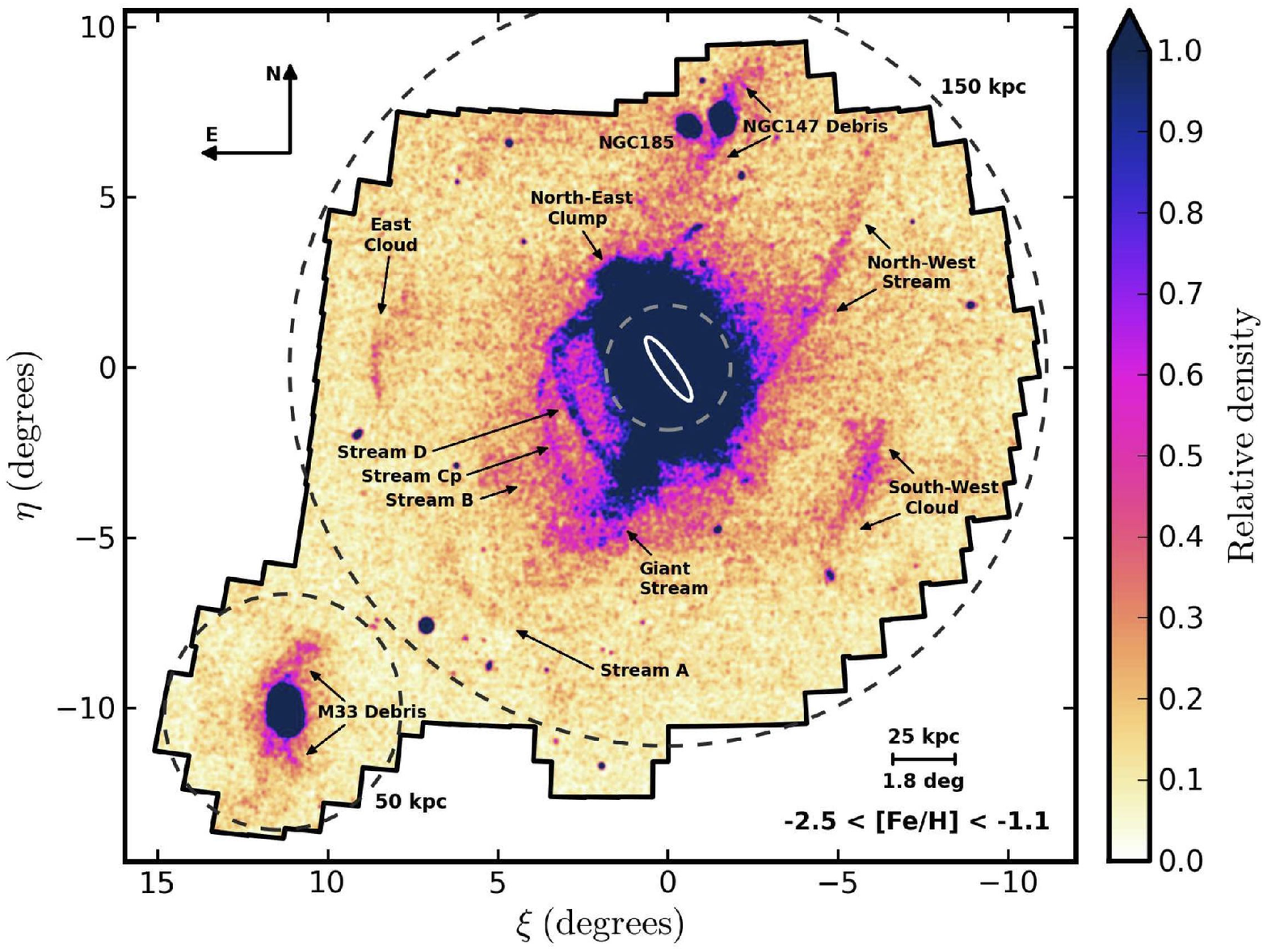}\\
\vspace{-1mm}
\includegraphics[width=138mm,clip]{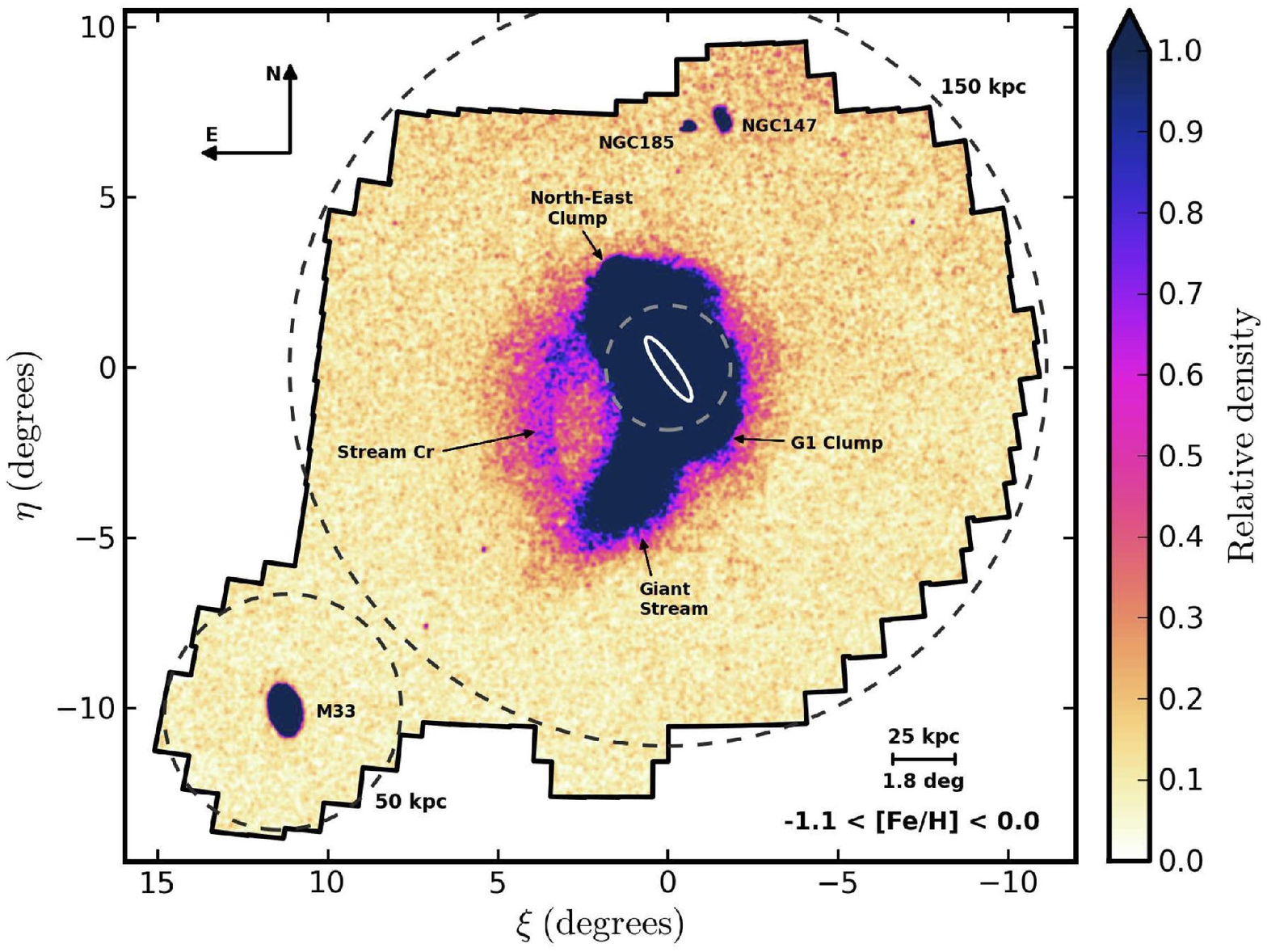}
\caption{Spatial density maps for ``metal-poor'' (upper panel) and ``metal-rich'' (lower panel) RGB stars in the M31 halo. These were selected from the PAndAS point source catalogue using the CMD boxes marked in Figure \ref{f:cmd} to identify red giants at the M31 distance with $-2.5\la[$Fe$/$H$]\la-1.1$ and $-1.1\la[$Fe$/$H$]\la0.0$, respectively. The colour-maps represent the stellar density in $2\arcmin \times 2\arcmin$ bins, after subtraction of the \citet{martin:13} contamination model; smoothing in the spatial dimensions has been applied using a Gaussian kernel of $\sigma = 2.5\arcmin \approx 570$\ pc at our assumed M31 distance.  We set the saturation points in the colour-maps to enhance low-density features in the outer halo; the main stellar streams and overdensities are labelled \citep[see also][]{mcconnachie:18}. The two dashed circles centred on M31 represent $R_{\rm proj} = 25$\ and $150$\ kpc, respectively. The white ellipse indicates a central stellar disk of radial extent $15$\ kpc, inclination angle $77.5\degr$, and position angle $38.1\degr$ east of north, and is provided to help emphasise the overall scale of the map. M33 lies to the south-east of the PAndAS footprint; the dashed circle centred on this galaxy represents $R_{\rm proj} = 50$\ kpc. $(\xi,\eta)$ are coordinates on the tangent-plane projection centred on M31.}
\label{f:maps}
\end{center}
\end{figure*}
   
In Figure \ref{f:maps} we show maps of the spatial density of ``metal-poor'' and ``metal-rich'' M31 halo RGB stars inside the PAndAS footprint. To construct these maps, we first used the isochrones plotted in Figure \ref{f:cmd} to define CMD selection boxes that, allowing for the photometric uncertainties, encompass stars with $-2.5 \la [$Fe$/$H$] \la -1.1$ and $-1.1 \la [$Fe$/$H$] \la 0.0$, respectively. The metal-rich box is truncated towards the red in order to avoid the heaviest regions of foreground contamination on the CMD. Following \citet{ibata:14} the faint limit of both selection boxes is set at $i_0 = 23.5$ as this minimises any pointing-to-pointing variation in star counts due to photometric incompleteness at the faint end. Next, we divided the area inside the PAndAS footprint into small bins, in this case $2\arcmin \times 2\arcmin$ in size, and counted the number of stars falling within each bin and the appropriate CMD selection box. Finally, for each bin we subtracted the number of stars predicted to lie inside the CMD selection box by the \citet{martin:13} contamination model described above.

Numerous authors have previously presented, and discussed in detail, various incarnations of the maps shown in Figure \ref{f:maps} -- most recently \citet{mcconnachie:18}, but see also \citet{ibata:07,ibata:14,mcconnachie:09,mcconnachie:10,richardson:11,lewis:13,martin:13,bate:14,mcmonigal:16a,ferguson:16}. Our main reasons for showing them here are (i) to illustrate our selected metallicity cuts and the use of the \citet{martin:13} contamination model (both of which are integral to the following analysis), and (ii) to provide a labelled set of the main stellar substructures in the outer halo of M31 for ease of reference. The metallicity cut $-2.5 \la [$Fe$/$H$] \la -1.1$ picks out the majority of the stellar substructures visible in the M31 outer halo, although we note that it contains only the minority fraction of halo luminosity over the range $25-150$\ kpc \citep[$\sim15-30\%$, depending on the assumed age of the halo -- see Table $4$ in][]{ibata:14}. For stars with $-1.1 \la [$Fe$/$H$] \la 0.0$ there is one dominant feature -- the Giant Stream -- plus a structure (Stream Cr) that loops to the east, overlapping, in projection, the metal-poor Stream Cp and the upper part of Stream D. 

\section{Spatial distributions of clusters and stars}
\label{s:analysis}
In this section we examine how the spatial distribution of remote globular clusters in M31 compares with that of stars in the halo.  We first consider the radial surface density profiles for clusters and stars, and then use the PAndAS stellar density maps (i.e., Figure \ref{f:maps}) to conduct a detailed exploration of the correlation between clusters and the various components that make up the stellar halo.

\subsection{Radial surface density profiles}
\label{ss:profile}
The fall-off in the radial surface density of remote M31 globular clusters has most recently been considered by \citet{huxor:11}, who used the catalogue presented by \citet{huxor:08} as their starting point. This catalogue consists of clusters discovered using imaging data from the Isaac Newton Telescope (INT) spanning the inner $\approx 30-50$\ kpc of the M31 halo with a contiguous but irregular footprint, plus a few CFHT/MegaCam fields extending the coverage to $\sim 100$\ kpc in one quadrant due south of the galactic centre.  The main features observed by \citet{huxor:11} were: (i) a clear break in the profile, from a relatively steep decline to a much flatter one, at $R_{\rm proj} \approx 25$\ kpc, corresponding to a similar break seen in the metal-poor field population in the same southern quadrant by \citet{ibata:07}; and (ii) a power-law slope of $\Gamma = -0.87 \pm 0.52$ outside this break radius -- i.e., over the range $25 \la R_{\rm proj} \la 100$\ kpc.

\begin{figure}
\begin{center}
\includegraphics[width=77mm,clip]{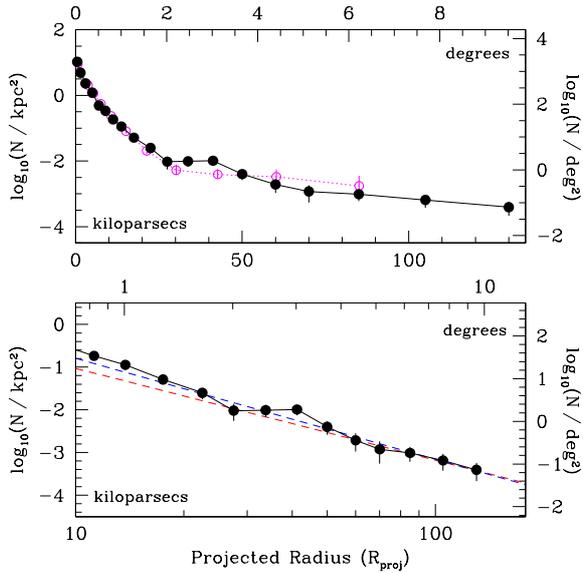}
\caption{Radial surface density profiles for globular clusters in M31. The upper panel shows, with a linear $x$-axis, the profile from \citet{huxor:11} (magenta open points and dotted line) together with our new updated profile (black solid points and unbroken line). The lower panel shows our new profile with a logarithmic $x$-axis, along with the two power-law fits discussed in the text -- with index $\Gamma = -2.15$ (red dashed line) and $\Gamma = -2.37$ (blue dashed line). All points have Poissonian error bars.}
\label{f:profile}
\end{center}
\end{figure}

In Figure \ref{f:profile} we present an updated radial surface density profile for M31 globular clusters. We constructed this using the catalogues described in Sections \ref{ss:outercat} and \ref{ss:innercat} as our starting point, and adopting concentric circular annuli\footnote{While remote M31 clusters are, unfortunately, too sparsely distributed to robustly infer the shape of the system as a function of radius, we believe that the use of circular annuli is appropriate given that the M31 stellar halo outside $R_{\rm proj} = 25$\ kpc appears to be close to spherical \citep[e.g.,][]{gilbert:12,ibata:14}.} with approximately equidistant spacing in $\log (R_{\rm proj})$. We calculated the fraction of each annulus covered by the PAndAS footprint using the spatial compeleteness function we previously derived in \citet{huxor:14} -- see Figure 11 in that work -- and corrected the number of clusters per annulus using this information.  To ensure self-consistency, we first identified and removed from our catalogue any clusters that do not appear in a PAndAS image due to either imperfect tiling of the mosaic, or incompletely dithered inter-chip gaps on the MegaCam focal plane. There are three such objects (B339, B398, and H9), only one of which (H9) sits outside $R_{\rm proj} = 25$\ kpc.

Our new profile traces the globular cluster population to very large radii ($R_{\rm proj} \approx 150$\ kpc). It possesses several interesting features.  Most noticeably, the break from a steep decline to a shallow decline observed by \citet{huxor:11} is still clearly present, occuring at $R_{\rm proj} \approx 27$\ kpc.  Beyond this break, the profile exhibits a prominent bump spanning $R_{\rm proj} \approx 30-50$\ kpc.  Apart from the bump, the radial surface density is close to a power law -- a simple least-squares fit to all points outside $R_{\rm proj} = 25$\ kpc yields an index $\Gamma = -2.37 \pm 0.17$. If the three points comprising the bump are excluded, the power law is a little flatter, with $\Gamma = -2.15 \pm 0.10$ over the range $25-150$\ kpc.

\begin{figure}
\begin{center}
\includegraphics[width=77mm,clip]{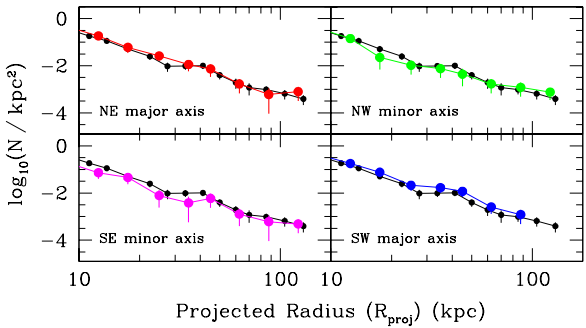}
\caption{Azimuthal variation in the radial surface density profile for M31 globular clusters. In each panel the coloured line and large points represent the profile for the specified quadrant; that for the whole system (i.e., Figure \ref{f:profile}) is shown in black with small points. All error bars are Poissonian.}
\label{f:profile-az}
\end{center}
\end{figure}
 
In Figure \ref{f:profile-az} we show the profile split into four quadrants, to examine the azimuthal variation in globular cluster surface density. We defined the quadrants by using dividing lines due north, south, east and west of the galactic centre, and identify each one according to the galactic axis that lies within it. In general there is good agreement between the profiles calculated in each of the four quadrants -- the observed cluster densities typically match to within $\sim 1\sigma$ of each other and the full profile.  The two locations where this is not the case are the $\approx 30-50$\ kpc bump, and at the very largest radii.  The bump clearly originates predominantly from clusters falling in the NE major axis quadrant and the SW major axis quadrant.  This is perhaps not too surprising, as the two most significant globular cluster overdensities seen in the outer M31 halo fall in these two quadrants at radii corresponding precisely to the observed bump \citep[see e.g.,][]{mackey:10b,veljanoski:14}. Over the radial span of the bump the azimuthal variation in cluster surface density is quite striking -- at $R_{\rm proj} \approx 35$\ kpc, the densities across the four quadrants are discrepant by a factor of up to $\sim 4.5$. 
 
The profiles are also mildly divergent at $R_{\rm proj} \ga 100$\ kpc. The azimuthal variation in cluster surface density is a factor $\sim 2$ between the outermost bins of the NE major axis, SE minor axis, and NW minor axis quadrants, while the SW major axis quadrant has no known clusters beyond $R_{\rm proj} \approx 90$\ kpc. It is difficult to say whether this apparent divergence is simply a result of stochastic variation in the small number of clusters at these large radii; however \citet{ibata:14} observed a similar scatter at the outer edge of the metal-poor stellar halo.

Figures \ref{f:profile} and \ref{f:profile-az} help to explain why our measured power-law slope is substantially steeper than that obtained by \citet{huxor:11}. First, the INT data from which the \citet{huxor:08} catalogue was derived did not reveal the two cluster overdensities responsible for the bump in the PAndAS profile between $30-50$\ kpc, mainly because of its irregular spatial coverage at these radii. Thus the \citet{huxor:11} profile under-estimates the cluster density at these radii.  Second, by chance the CFHT/MegaCam imaging used for the \citet{huxor:08} catalogue covered a region of slightly above-average cluster density predominantly to the south and south-west of the M31 centre between $\sim 50-100$\ kpc. Hence the \citet{huxor:11} profile is a mild over-estimate of the azimuthally-averaged density at these radii. Overall, these two factors lead to the \citet{huxor:11} profile appearing significantly flatter than our final PAndAS profile. As the latter is based on higher quality imaging and uniform spatial coverage in all directions, it should be considered the more robust result.

It is informative to compare the properties of our cluster profile to results for the M31 stellar halo.  The most comprehensive study on this front is by \citet{ibata:14} who used the PAndAS point source catalogue to construct projected star-count profiles for stellar populations spanning different metallicity ranges, and for which the various substructures visible in the M31 halo had either been masked out or not.  Despite considerable variations in density from quadrant to quadrant, \citet{ibata:14} found the azimuthally averaged profiles to be surprisingly featureless and exhibit well-defined power-law behaviour, with the radial fall-off becoming steeper with increasing metallicity. They also observed that masking the substructures suppressed the degree of azimuthal variation in the radial profiles and resulted in somewhat flatter radial declines at given $[$Fe$/$H$]$, leading them to infer the presence of an apparently {\it smooth} (at least to the sensitivity of PAndAS) stellar halo component.

Since we have not masked any of the known cluster substructures in the M31 halo, our density profile is most directly comparable to the unmasked profiles of \citet{ibata:14}. They measured a power-law decline of index $\Gamma = -2.30 \pm 0.02$ for the stellar population with $-2.5 < [$Fe$/$H$] < -1.7$, a decline of index $\Gamma = -2.71 \pm 0.01$ for the population with $-1.7 < [$Fe$/$H$] < -1.1$, and a much steeper fall-off of index $\Gamma = -3.72 \pm 0.01$ for the metal-rich population with $-1.1 < [$Fe$/$H$] < 0.0$. The huge radial extent and comparatively shallow decline of our cluster profile, which has $\Gamma = -2.37 \pm 0.17$, firmly associates the majority of the remote globular cluster population in M31 (i.e., outside $R_{\rm proj} = 25$ kpc) with the metal-poor stellar halo. 

It is interesting that when we exclude the $30-50$\ kpc bump from our fit we obtain a shallower power-law index of $\Gamma = -2.15 \pm 0.10$. This slope is most comparable to those for the {\it masked} metal-poor profiles from \citet{ibata:14}, which have $\Gamma = -2.08 \pm 0.02$ and $\Gamma = -2.13 \pm 0.02$ for the $-2.5 < [$Fe$/$H$] < -1.7$ and $-1.7 < [$Fe$/$H$] < -1.1$ populations, respectively. It is perhaps not too surprising to see such a close match -- we already know that a substantial fraction of the clusters outside $R_{\rm proj} = 25$\ kpc are associated with luminous field substructures \citep[see][and Sections \ref{ss:halomaps} and \ref{ss:correlation} below]{mackey:10b,veljanoski:14}, and removing the $30-50$\ kpc bump from the power-law fit could be considered a crude masking of the two most significant globular cluster overdensities known in the outer M31 halo. The fact that the remote globular cluster population in M31 behaves in such a similar fashion to the field is suggestive of a composite cluster system where some fraction is associated with the smooth halo and some fraction with halo substructures; we will return to this issue in Section \ref{s:properties}.

Finally, we note that there is no evidence of a turn-down in the cluster profile to $R_{\rm proj} \approx 150$\ kpc.  This is consistent with the results of \citet{ibata:14}, who observed no steepening of their projected metal-poor stellar profiles at large radius.  It is thus reasonable to expect a few extremely remote clusters to be lurking beyond the edge of the PAndAS footprint.  Assuming the observed power-law decline holds\footnote{Here we assume the power law of index $\Gamma = -2.15 \pm 0.10$ obtained by masking the $30-50$\ kpc bump, as this provides a marginally better fit to the outer points of the profile than does the power law of index $\Gamma = -2.37 \pm 0.17$.}, we suggest that there may be $11 \pm 3$ clusters in the range $150 \la R_{\rm proj} \la 200$\ kpc waiting to be discovered. At present, two clusters with 3D galactocentric radii in this range are known \citep[MGC1 and PA-48, see][]{mackey:10a,mackey:13b}.

\subsection{Halo maps}
\label{ss:halomaps}
\subsubsection{Outer halo ($R_{\rm proj}\,\ga\,25$\ kpc)}
\begin{figure*}
\begin{center}
\includegraphics[width=145mm,clip]{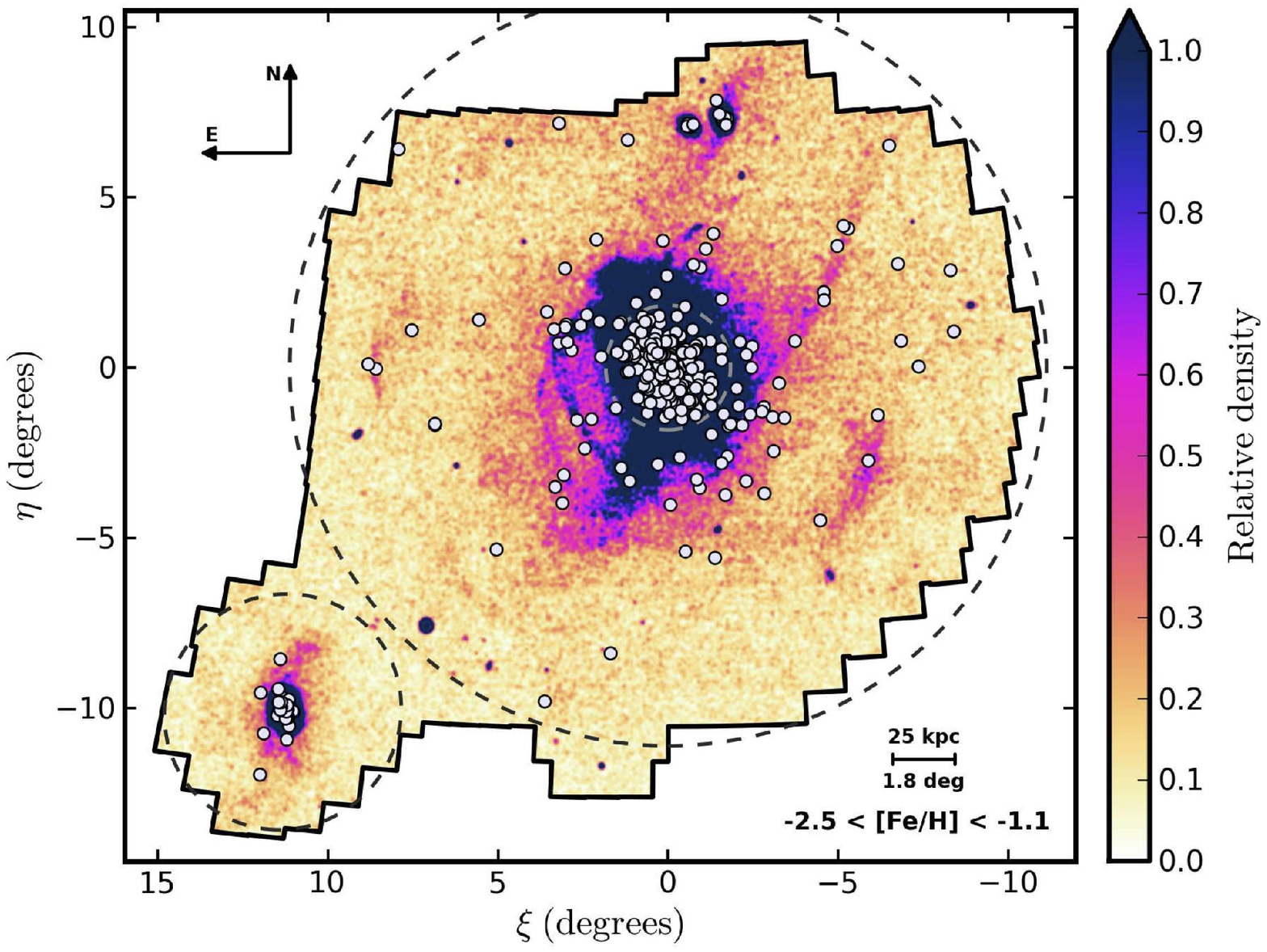}\\
\vspace{-1mm}
\includegraphics[width=145mm,clip]{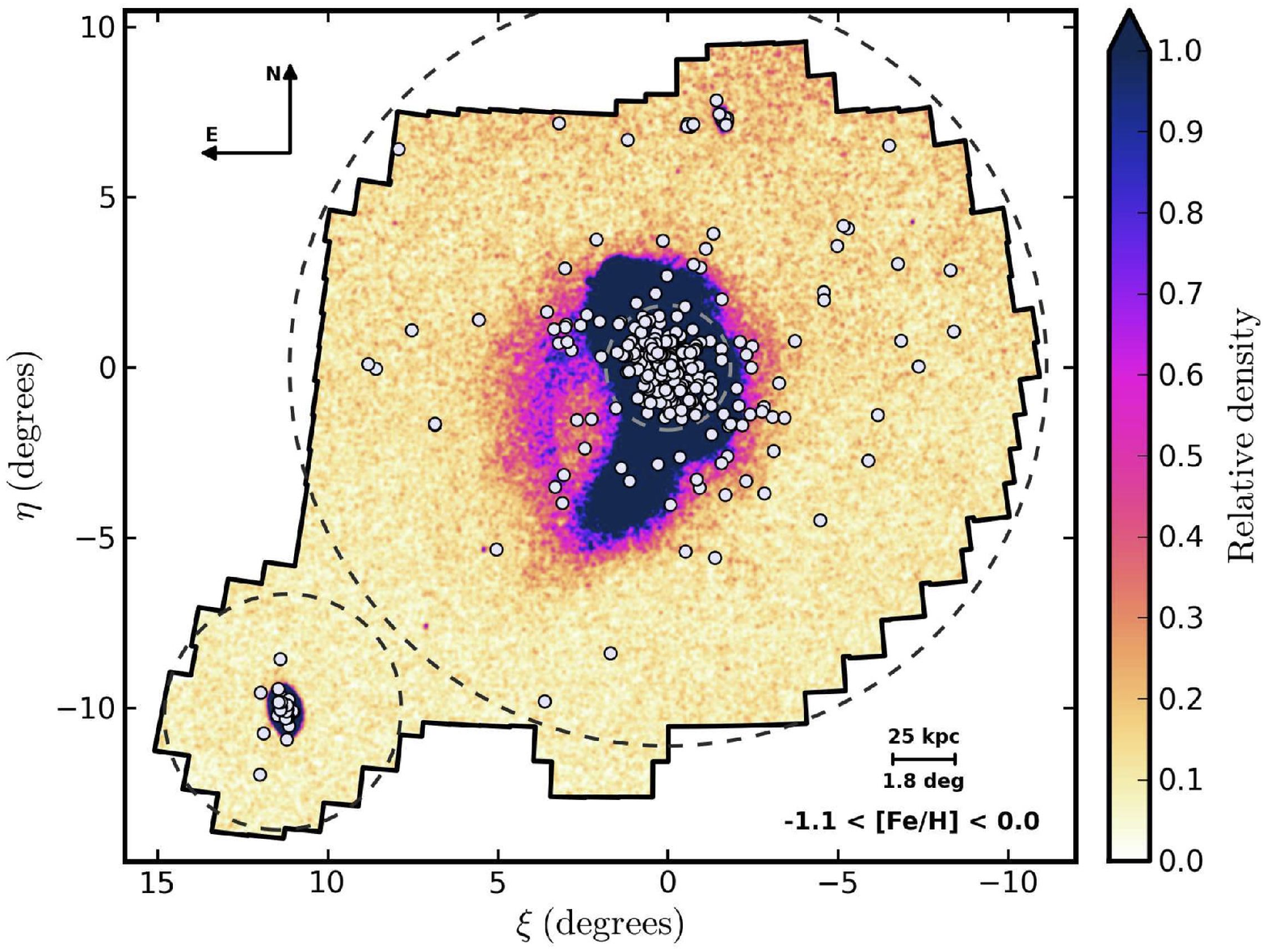}
\caption{PAndAS spatial density maps for metal-poor (upper panel) and metal-rich (lower panel) RGB stars in the M31 halo, with the positions of all globular clusters plotted (light grey points). Apart from the clusters, all details of the maps are the same as in Figure \ref{f:maps}.}
\label{f:gcmaps}
\end{center}
\end{figure*}

In Figure \ref{f:gcmaps} we reproduce the PAndAS spatial density maps for metal-poor and metal-rich RGB stars in the M31 halo, and mark the positions of all globular clusters according to the catalogues described in Section \ref{s:data}. This includes, for illustrative purposes, those near the centre of M31, and those belonging to the large satellite galaxies M33, NGC 147, and NGC 185.

It is evident from this Figure that beyond $R_{\rm proj} \approx 25$\ kpc there is a striking correlation between the most luminous substructures in the M31 stellar halo and the positions of many globular clusters.  This association is clearest in the metal-poor map, which exhibits the majority of the known halo streams and overdensities. The correlation between clusters and substructures was previously discovered and analysed by \citet{mackey:10b} using roughly half of the PAndAS survey area, and then explored in more detail by \citet{veljanoski:14} across a much larger area for a spectroscopically-observed cluster subsample. The present maps extend the coverage to span the entire PAndAS footprint, revealing a number of additional halo streams over those identified in the original \citet{mackey:10b} analysis -- the most noticeable being the East Cloud at $R_{\rm proj} \approx 115$\ kpc \citep[e.g.,][]{mcmonigal:16a}, and the tidal tails of NGC 147 \citep[e.g.,][]{crnojevic:14,mcconnachie:18}. It is also worth emphasising that the present maps incorporate the {\it complete} outer halo globular cluster catalogue (as opposed to the earlier studies by \citealt{mackey:10b} and \citealt{veljanoski:14}).

\begin{figure*}
\begin{center}
\includegraphics[height=95mm,clip]{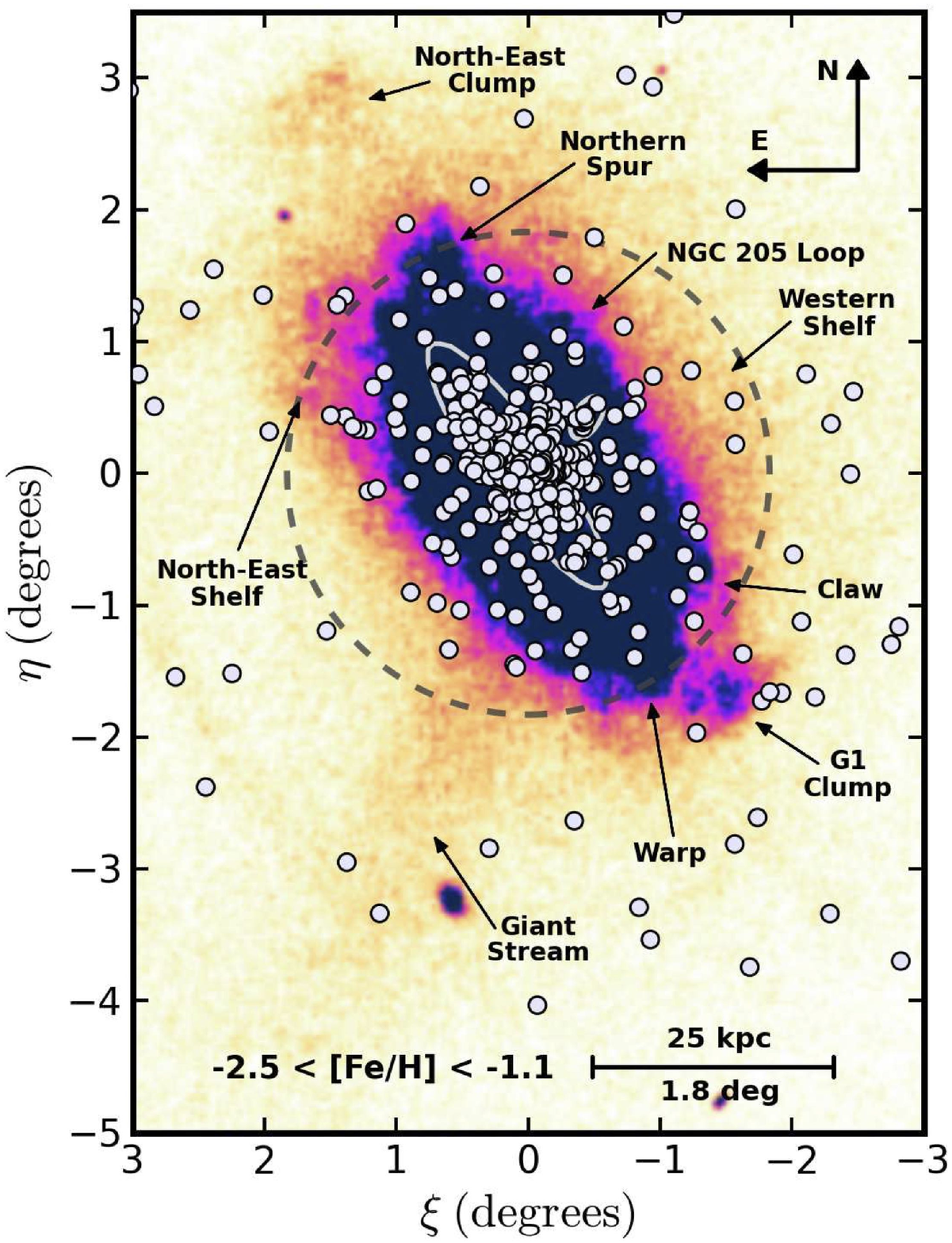}
\hspace{-3mm}
\includegraphics[height=95mm,clip]{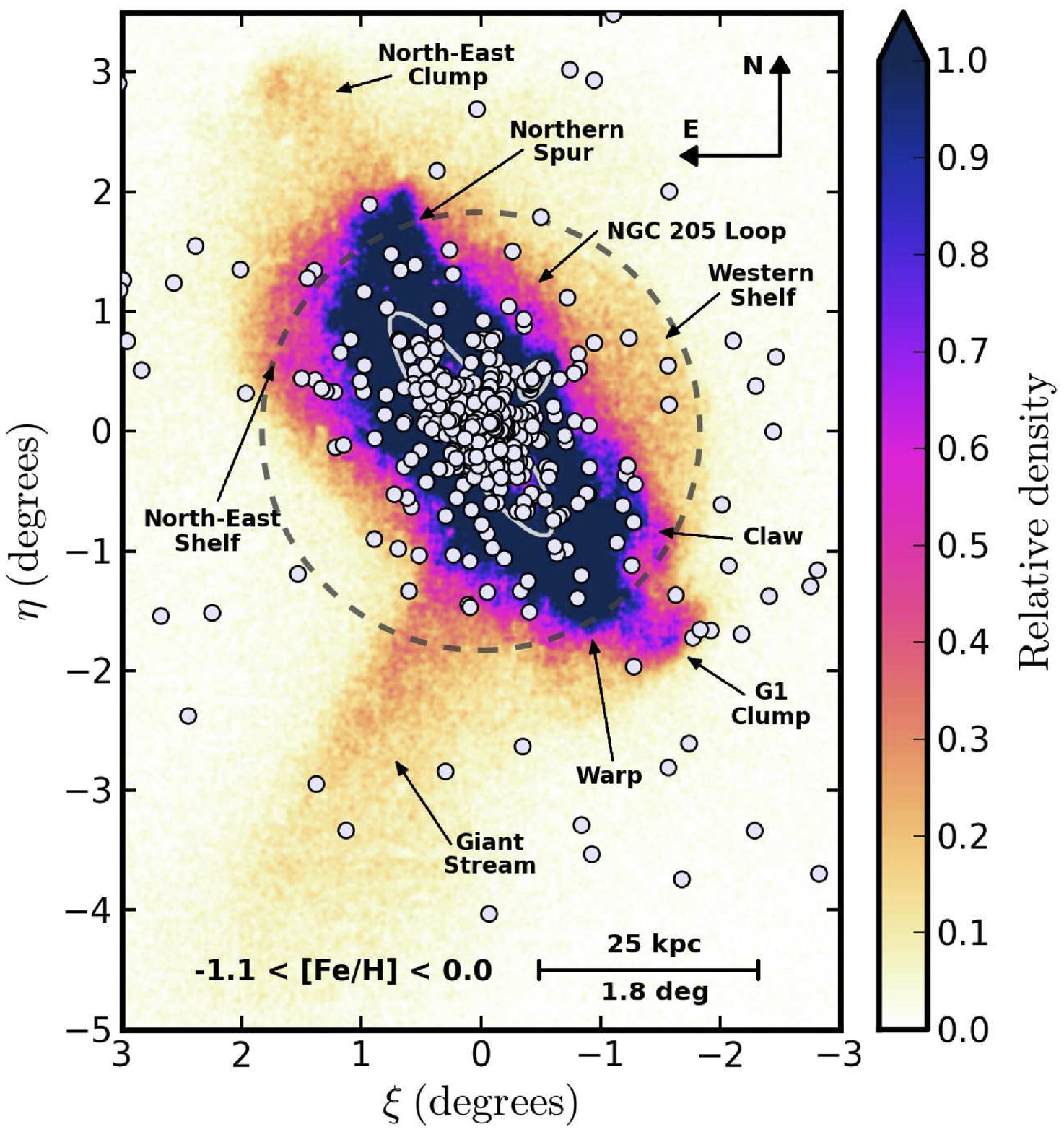}
\caption{PAndAS spatial density maps for metal-poor (left panel) and metal-rich (right panel) RGB stars in the M31 inner halo, with the positions of all globular clusters in our catalogue plotted (light grey points). The main stellar streams and overdensities identified by previous studies of the inner parts of M31 are labelled \citep[see, e.g.,][]{ferguson:05,richardson:08,bernard:15,ferguson:16}. Many of these features are more clearly defined in the metal-rich cut as they are predominantly due to the accreted progenitor of the Giant Stream or the extended M31 disk -- both of which are comparatively metal-rich systems. The dashed circle indicates a projected radius $R_{\rm proj} = 25$\ kpc. The main body of the M31 disk is schematically indicated as in Figure \ref{f:maps}, while the small white ellipse just to the north-west marks the dwarf elliptical satellite NGC 205.}
\label{f:innermaps}
\end{center}
\end{figure*} 

The most prominent potential associations between clusters and streams are straightforward to identify by eye -- there are seven clusters projected onto the North-West Stream; three onto the South-West Cloud; three onto the East Cloud; at least nine onto the region where Streams D, Cp and Cr all overlap; up to three each on the lower portions of Stream D and Stream Cp/Cr; and between three and five on the portion of the Giant Stream outside $R_{\rm proj} = 25$\ kpc.  Many of these apparent associations were considered individually by \citet{mackey:10b} and shown to be statistically significant; subsequent work incorporating radial velocity measurements has typically reinforced those results \citep[see][]{mackey:13a,mackey:14,veljanoski:13a,veljanoski:14,bate:14}. However, these associations account for only around one-third of the known outer halo globular clusters in M31; moreover there are substantial fluctuations in surface density across the stellar halo even when the most luminous streams are masked \citep[see][]{ibata:14}. It is thus important to quantify the significance of the cluster-substructure association across the system as a whole.  We analyse this problem below in Section \ref{ss:correlation}, using an updated and superior methodology to that employed by \citet{mackey:10b}.

The distribution of the outer M33 clusters is elongated in a north-south direction and may possibly trace the low-luminosity tidal features evident in the outskirts of this galaxy \citep{mcconnachie:10}.  This is unlikely to be due to a selection effect, as the region around M33 out to $R_{\rm proj} \approx 50$\ kpc has been uniformly and thoroughly searched for clusters \citep[see][]{huxor:09,cockcroft:11}; however, there are too few remote clusters for statistical tests of the possible association to give meaningful results.  The substructure consists of old and metal-poor stars believed to have been stripped from the M33 disk due to the gravitational influence of M31.  Velocity information for the clusters would help test whether they fit consistently into this picture or, for example, whether they might belong to a true halo-like population\footnote{Although note that \citet{mcmonigal:16b} have placed an upper limit of $\sim 10^6 L_\odot$ on the total luminosity of any stellar halo around M33 (excluding globular clusters).}. 

The clusters belonging to NGC 147 and 185 are centrally concentrated against the main bodies of these dwarf elliptical satellite systems. NGC 147 exhibits striking tidal tails, but there is no evidence that any globular clusters are associated with these features.  The NGC 147 cluster system is mildly elongated from north-east to south-west, in keeping with the position angle of the inner isophotes of the dwarf; the outermost clusters do not obviously follow the isophotal twisting seen in the stellar component at comparable radii \citep{crnojevic:14}.
 
\subsubsection{Inner halo ($R_{\rm proj}\,\la\,25$\ kpc)}
It is also interesting to briefly examine the central portion of M31, which is saturated in Figures \ref{f:maps} and \ref{f:gcmaps}.  We reproduce this region in Figure \ref{f:innermaps}, using the same metal-poor and metal-rich CMD selection boxes as for the previous maps but now adjusting the intensity scaling to reveal the main stellar features inside $R_{\rm proj} \approx 25$ kpc.  It is well known that the inner halo of M31 is very heavily substructured \citep[e.g.,][]{ibata:01,ferguson:02,zucker:04}; however, studies of stellar populations and kinematics in the various different overdensities have revealed that almost all are due to the extended M31 disk and/or the disruption of the satellite galaxy that produced the Giant Stream \citep[e.g.,][]{ferguson:05,ibata:05,guhathakurta:06,gilbert:07,richardson:08,fardal:12,bernard:15,ferguson:16}.

The degree of substructure is so great that it is impossible to associate clusters with any of the main features by eye, or even statistically if using only spatial information -- unambiguous association requires, at a minimum, the inclusion of velocity measurements for the clusters \citep[e.g.,][]{ashman:93,perrett:03}, and preferably kinematic data for the stellar component as well; such an analysis is beyond the scope of the present paper. Nonetheless, it is evident that several of the most luminous overdensities in the inner parts of the halo apparently do not exhibit similar concentrations of globular clusters.  More specifically, it is the features identified as being disturbances in the M31 outer disk: the North-East Clump\footnote{Sometimes called the ``North-East Structure'' \citep{mcconnachie:18}.}, the Northern Spur, the warp to the south, and the G1 Clump \citep[see e.g.,][]{ibata:05,bernard:15} that have relatively few clusters projected on top of them\footnote{Of these four overdensities, the G1 Clump has the most clusters projected near it, and indeed is named after one of these objects. Nevertheless, kinematic measurements have shown that G1, as well as several other nearby clusters, are unlikely to be related to this substructure \citep[e.g.,][]{reitzel:04,faria:07,veljanoski:14}}. This observation is perhaps not too surprising -- after all, in large galaxies globular clusters are typically considered to be a halo population rather than a disk population\footnote{Although note that \citet{caldwell:16} demonstrated that the $\approx 20$ most metal-rich globular clusters outside the bulge in M31 apparently {\it do} possess disk-like kinematics.}; however, it does reinforce the interpretation of these specific overdensities as being part of the extended disk of M31.

Two of the other major substructures in the inner halo -- the North-East Shelf\footnote{Sometimes called the ``Eastern Shelf'' \citep[see][]{mcconnachie:18}.} and the Western Shelf -- are thought to be due, respectively, to the second and third orbital wraps of debris from the Giant Stream progenitor, and are well-reproduced by modelling of this accretion event \citep[e.g.,][]{fardal:07,fardal:12,fardal:13}. In such models, the Giant Stream itself is composed of trailing debris from the first pass of the progenitor. \citet{mackey:10b} noted the paucity of globular clusters projected onto the Giant Stream outside $R_{\rm proj} = 25$\ kpc, given its ranking as the most luminous substructure in the M31 halo and the expectation that its progenitor was comparable in mass to the LMC \citep{fardal:13}. This could be explained if the progenitor system retained the majority of its clusters until the latter stages of its disruption, perhaps due to these objects being centrally concentrated within the satellite. Such behaviour is observed for the Sagittarius dwarf galaxy, presently being disrupted by the Milky Way, which still possesses four clusters coincident with its main body \citep[e.g.,][]{dacosta:95}\footnote{Although Sagittarius has notably {\it also} left a circum-Galactic stellar stream studded with globular clusters that have already been stripped from its main body \citep[e.g.,][]{bellazzini:03,law:10}, which is not obviously true for the Giant Stream.}.

In this case we might expect to find a number of globular clusters projected onto the North-East Shelf and the Western Shelf, and a quick inspection of the maps in Figure \ref{f:innermaps} reveals several such candidates.  Going one step further, if there {\it were} originally a number of centrally-located clusters within the progenitor system then these might plausibly form a co-moving group and thus provide a means of identifying its present location, which is thought to lie within the North-East Shelf \citep[e.g.,][]{fardal:13,sadoun:14}.  We defer further investigation along these lines to a future work -- although precise radial velocities are now available for the majority of globular clusters in the inner parts of M31 \citep[e.g.,][]{caldwell:11,strader:11,caldwell:16}, this exercise requires a detailed and careful comparison to the various Giant Stream models due to the complexity of the kinematics in the two shelf regions. 

\subsection{Quantifying the cluster-substructure correlation}
\label{ss:correlation}
In \citet{mackey:10b} we tested the significance of the association between globular clusters and field substructures in the M31 outer halo.  By examining the typical density of the stellar halo locally around each globular cluster, we showed that the likelihood that the apparent cluster-substructure association could be due to the chance alignment of clusters scattered according to a smooth underlying distribution was low -- well below $1\%$ system-wide, and less than $3\%$ for each of the North-West Stream, the South-West Cloud, and the Stream C/D overlap region individually. 

However, the methodology employed in our analysis was in several ways non-optimal, mainly due to the limitations of the available data at the time. For example, the PAndAS footprint covered less than half its final area; local stellar densities were inferred from a smoothed two-dimensional histogram rather than calculated directly from star counts; no allowance was made for the declining mean stellar density with projected radius, meaning the global analysis was likely more strongly influenced by measurements in the range $R_{\rm proj} \sim 30-50$\ kpc compared to those at larger radii; no correction for contamination was made save for the subtraction of the visible south-north gradient from the density histogram; and there were still systematic offsets at the few-percent level in the photometry from field-to-field within the PAndAS mosaic.

With the availability of the final calibrated PAndAS point-source catalogue spanning the full survey footprint, as well as the contamination model of \citet{martin:13}  and the complete globular cluster catalogue described in Section \ref{s:data}, we are now in a position to re-examine the significance of the cluster-substructure correlation with a far superior methodology. Our analysis is based on the premise that, if globular clusters preferentially project onto streams or overdensities, then the surface density of M31 halo stars locally around each cluster ought to be systematically higher than the typical surface density observed at a comparable galactocentric radius. By quantifying how different the observed distribution of local densities around globular clusters is from the expected distribution, we can formally assess the significance of the correlation between clusters and field substructures.

To maintain readability, we reserve a detailed discussion of our methodology for Appendix \ref{a:method33}. In brief, we determined the surface density of metal-poor M31 halo stars in a circular aperture of radius $r=10\arcmin$ around each of the $92$ globular clusters with $R_{\rm proj} > 25$\ kpc, corrected for foreground contamination using the model of \citet{martin:13}, and with possible contributions from M31 satellite dwarfs and/or other nearby clusters excised. We then repeated this calculation for $1000$ randomly-selected locations in each of $135$ $1$-kpc-wide circular annuli centred on M31, spanning the range $20 \leq R_{\rm proj} \leq 155$\ kpc, in order to empirically determine the underlying density distribution for comparison.

\begin{figure*}
\begin{center}
\includegraphics[height=70mm,clip]{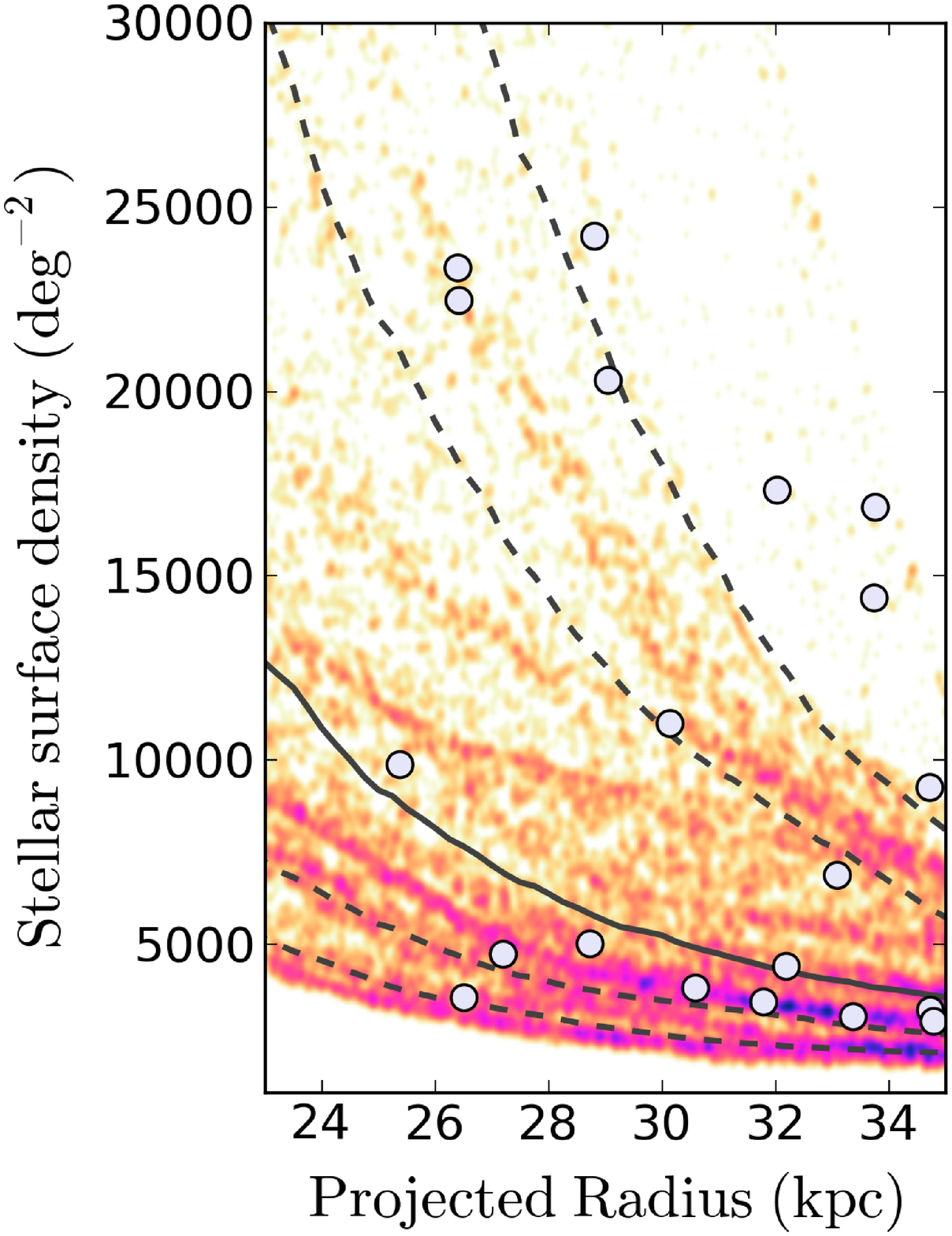}
\hspace{1mm}
\includegraphics[height=70mm,clip]{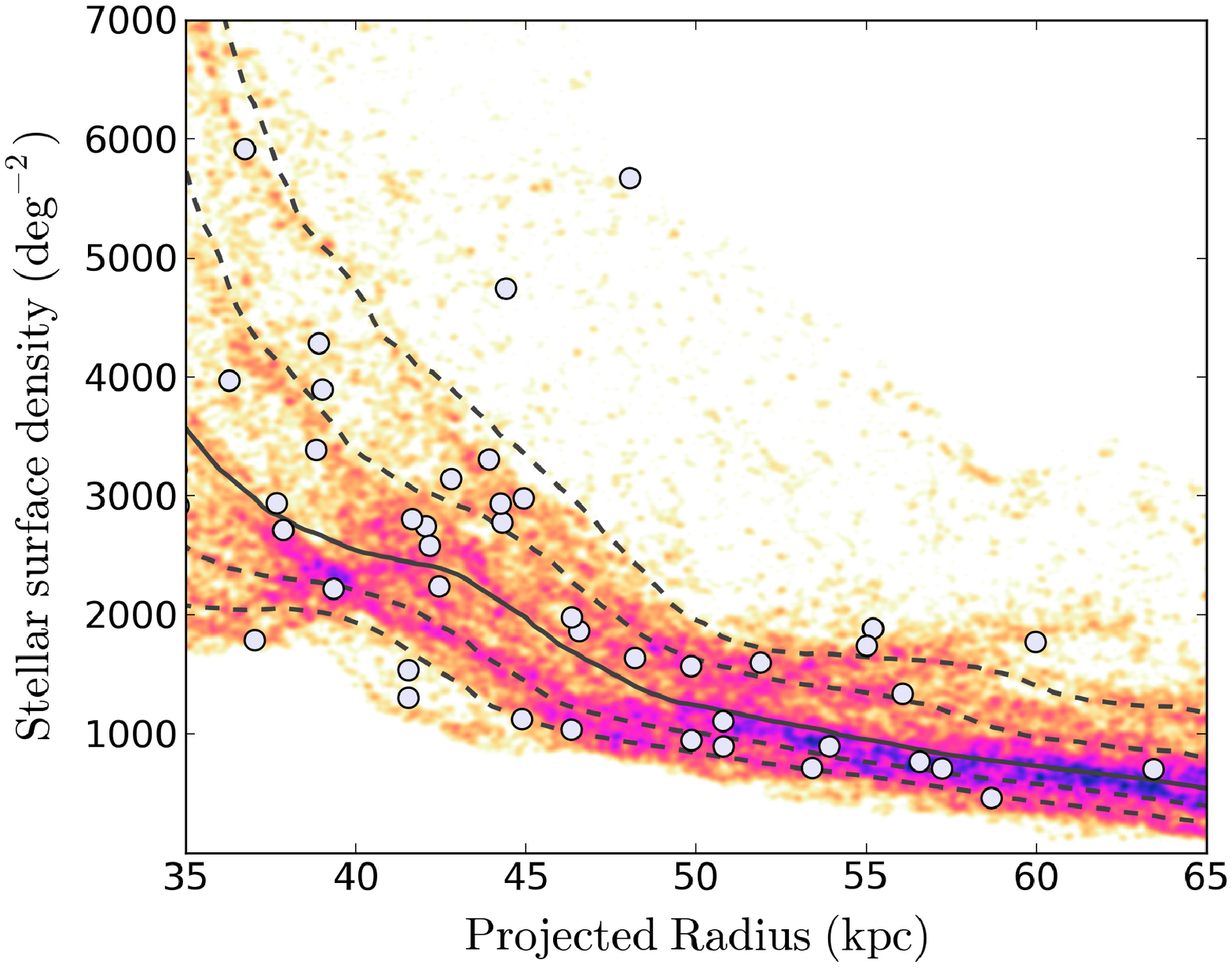}\\
\vspace{2mm}
\includegraphics[height=65mm,clip]{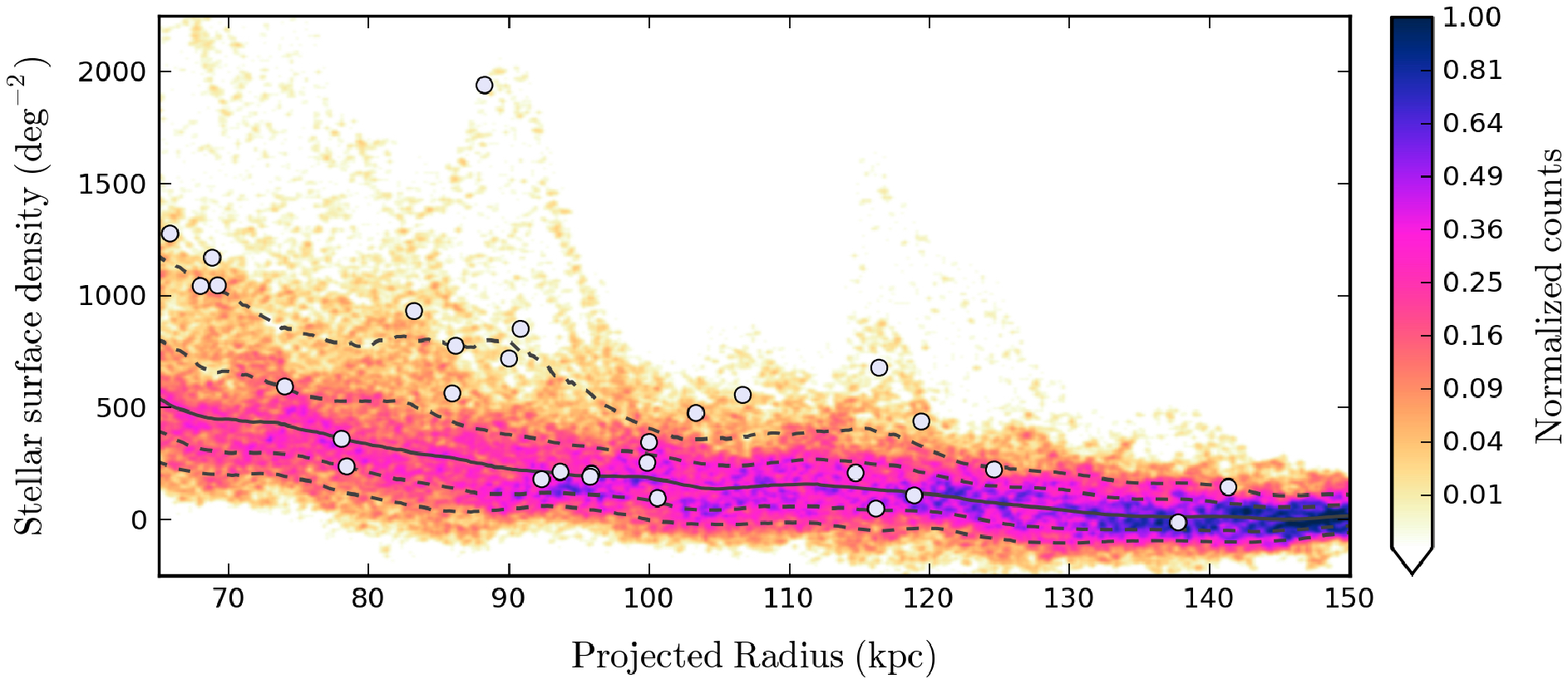}
\caption{Distribution of surface density in the M31 halo for stars with $-2.5 \la [$Fe$/$H$] \la -1.1$ (i.e., ``metal-poor'' stars), as a function of projected galactocentric radius within the PAndAS survey footprint. The full radial span has been split into three panels for clarity. Each panel has a different range on the $y$-axis, and the colour-map is normalised to the pixel with the highest number of counts; we use a non-linear (square-root) scaling to enhance the visibility of the tails of the distribution. The solid black contour shows the median of the distribution as a function of radius; the dashed contours show, from top to bottom, the $10\%$, $25\%$, $75\%$ and $90\%$ bands (i.e., at any given radius the $10\%$ band has ten percent of the randomly generated locations sitting at higher surface density). Measurements for the $92$ outer halo globular clusters are marked with light grey points.}
\label{f:densities}
\end{center}
\end{figure*}
 
Our results are displayed in Figure \ref{f:densities}. This shows the distribution of surface density in the M31 metal-poor stellar halo as a function of projected galactocentric radius, with individual measurements for the $92$ outer halo globular clusters overplotted.  The complexity of the halo is evident at all radii, with numerous filamentary features visible in each of the three panels. To guide the eye, we mark contours indicating the median of the distribution as a function of radius (solid line), and the $10\%$, $25\%$, $75\%$ and $90\%$ bands (dashed lines, top to bottom).  These density percentile bands are defined in terms of the fraction of the distribution lying above them -- for example, at any given radius, ten percent of the randomly generated locations have higher local surface densities than the value of the $10\%$ contour.

At very large galactocentric distances the median of the distribution approaches zero.  This is partly due to the intrinsic sparsity of the M31 halo at these radii, but also partly because, as we previously noted, the \citet{martin:13} contamination model was by necessity derived using the outermost reaches of the PAndAS survey area at $R_{\rm proj} \ga 120$\ kpc and thus includes a small but non-zero halo component. Fortunately our analysis depends on the {\it spread} of the distribution at given radius rather than its absolute level -- any oversubtraction due to the contamination model may affect the level but does not alter the spread.

Even a cursory inspection of the positions of the globular clusters in relation to the various contour lines in Figure \ref{f:densities} reveals that many objects sit above the $25\%$ line, and a substantial number even sit above the $10\%$ line. This is direct confirmation of our impression from the metal-poor halo map in Figure \ref{f:gcmaps} that cluster positions preferentially tend to correlate with the locations of stellar streams and overdensities. To quantify the association further, we assign a density percentile value, $\zeta_{\rm MP}$, to each globular cluster -- i.e., the fraction of the underlying metal-poor density distribution at a commensurate radius that sits above the local density measured for the cluster in question. We define the ``commensurate radius'' as being within $\pm 1$\ kpc of that for a given cluster, although our results are not strongly sensitive to the width of this interval. Values of $\zeta_{\rm MP}$ for individual clusters are reported in Appendix \ref{a:data}.  

\begin{figure}
\begin{center}
\includegraphics[width=78mm,clip]{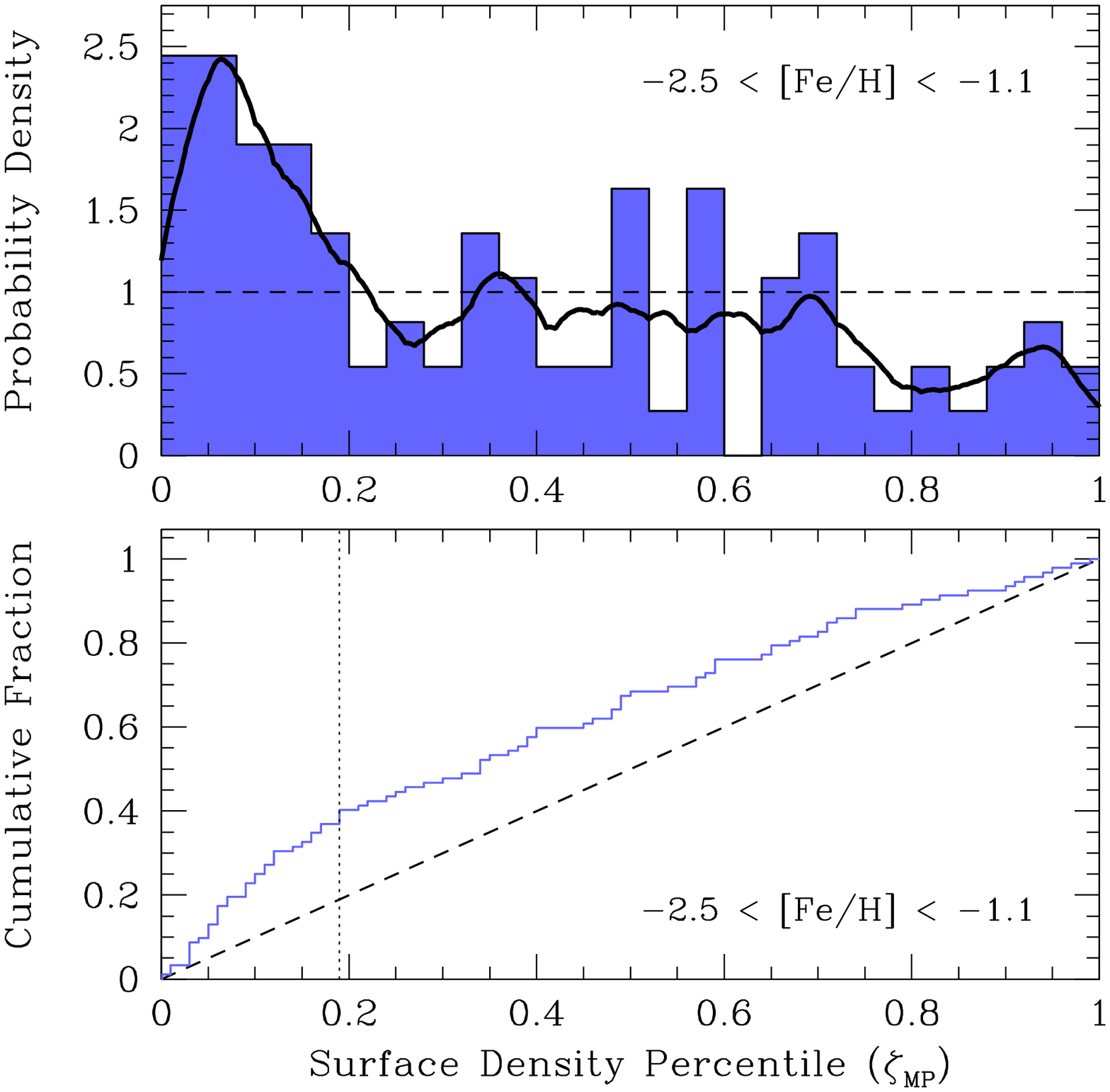}
\caption{Distribution of $\zeta_{\rm MP}$ for the $92$ M31 globular clusters with $R_{\rm proj} > 25$\ kpc. The upper panel shows a histogram, and a smoothed curve derived via a kernel density estimator with an Epanechnikov kernel. The scaling is such that the area under both the histogram and the curve is unity. The lower panel shows the data as a cumulative distribution. In both panels the dashed line shows the null hypothesis (uniform distribution) discussed in the text. The data are strongly concentrated at small values of $\zeta_{\rm MP}$, indicating a clear preference for globular clusters to sit on or near over-dense locations in the metal-poor stellar halo. In the lower panel the vertical dotted line indicates the location of the greatest separation between the measured distribution and the uniform distribution.}
\label{f:densdistmp}
\end{center}
\end{figure}
 
In Figure \ref{f:densdistmp} we construct the distribution of $\zeta_{\rm MP}$ for the $92$ M31 globular clusters with $R_{\rm proj} > 25$\ kpc.  The upper panel shows a histogram of these values, while the lower panel shows their cumulative distribution. It is evident that nearly half of the clusters have local densities in the top quartile of the observed distribution, while one-quarter have local densities in the top decile. To assess this pattern more formally, we adopt a null hypothesis \citep[as in][]{mackey:10b} that the M31 cluster system is smoothly arranged within the halo, such that there is no correlation with the underlying stellar populations.  Under this assumption the cluster positions would effectively be random, meaning that the expected distribution of density percentile values should be uniform.  Both panels in Figure \ref{f:densdistmp} show that this is not the case -- the observed distribution for M31 outer halo globular clusters is strongly peaked to small values of $\zeta_{\rm MP}$, indicating a clear preference for globular clusters to sit on or near over-dense locations in the metal-poor stellar halo.  To estimate the significance of this observation we use a simple Kolmogorov-Smirnov (K-S) test.  The greatest separation between the cumulative distribution of $\zeta_{\rm MP}$ for our globular cluster ensemble and that expected for our null hypothesis is $0.212$ at a percentile value of $\zeta_{\rm MP} = 0.19$; the probability that the two distributions were drawn from the same parent distribution is only $0.04\%$.

\begin{figure*}
\begin{center}
\includegraphics[height=70mm,clip]{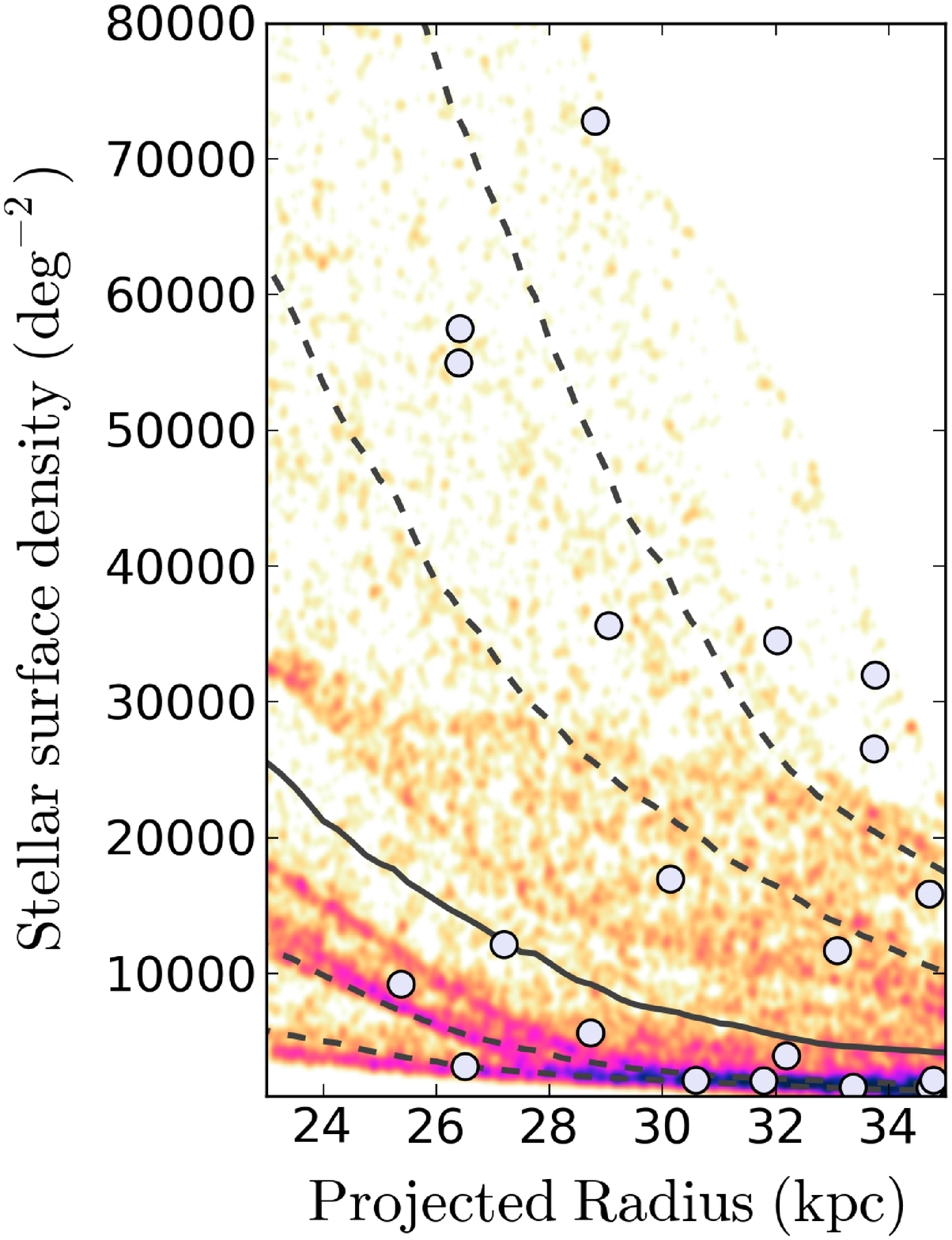}
\hspace{1mm}
\includegraphics[height=70mm,clip]{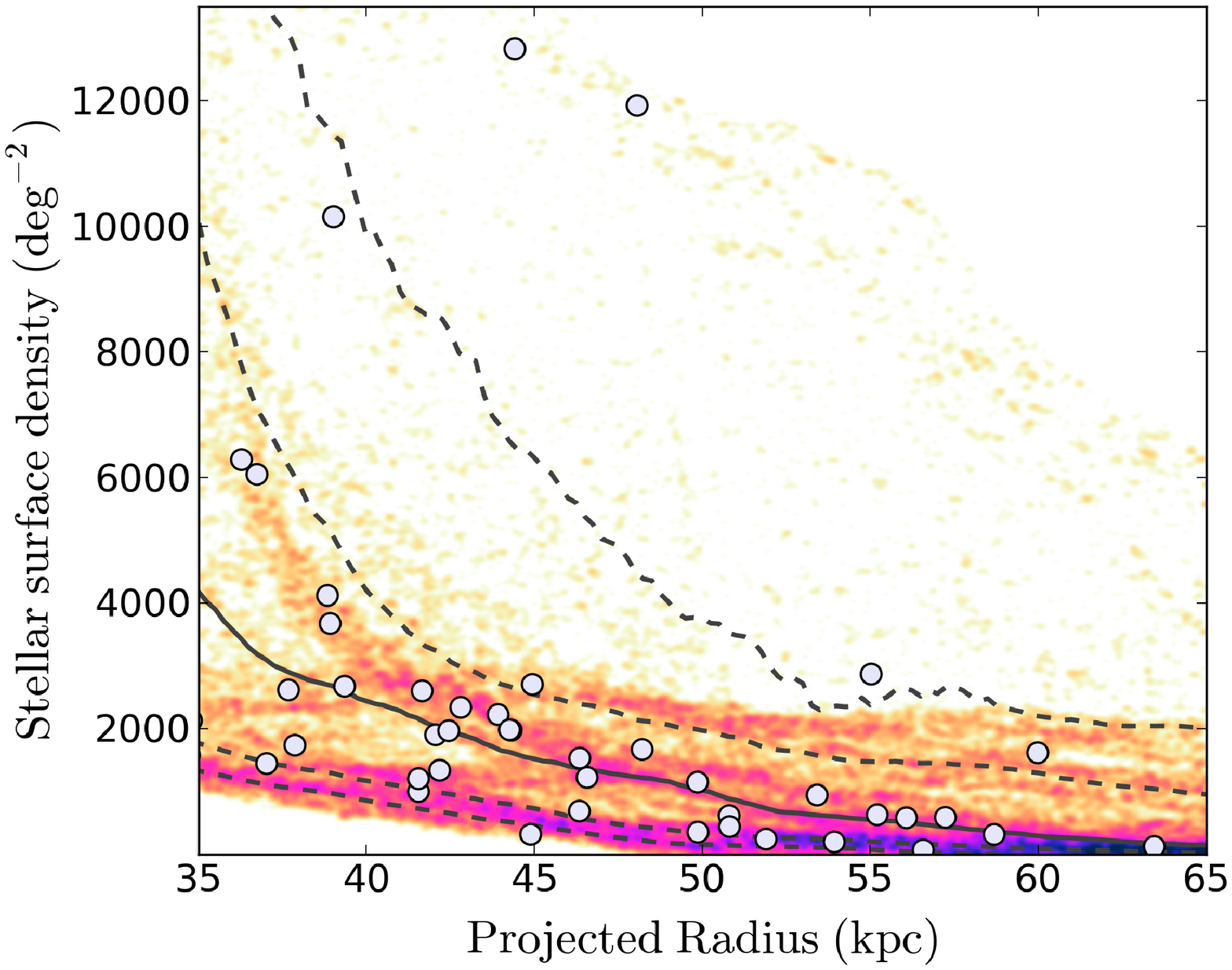}\\
\vspace{2mm}
\includegraphics[height=65mm,clip]{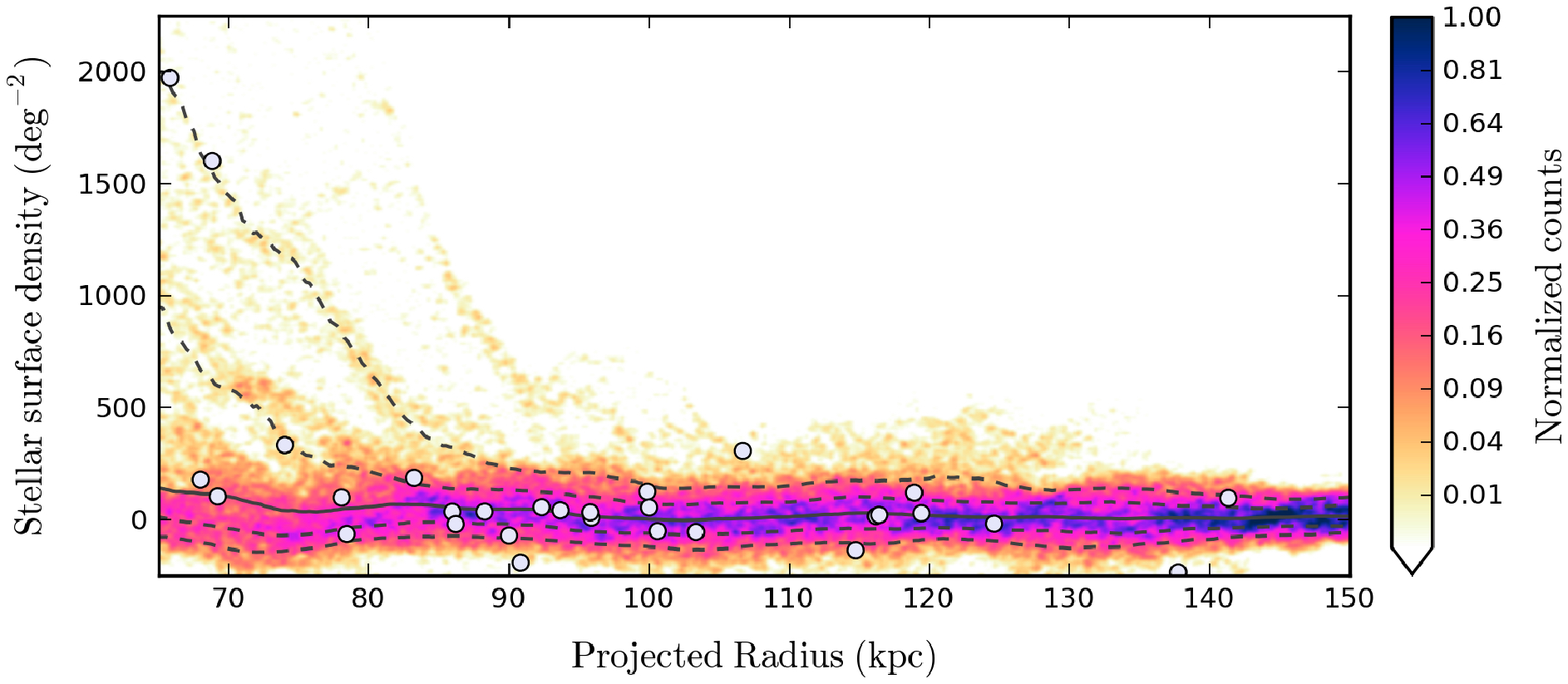}
\caption{Same as Figure \ref{f:densities}, but now for M31 halo stars with $-1.1 \la [$Fe$/$H$] \la 0.0$ (i.e., ``metal-rich'' stars). As before, the solid black contour shows the median of the distribution as a function of radius, while the dashed contours show, from top to bottom, the $10\%$, $25\%$, $75\%$ and $90\%$ bands, and the measurements for our $92$ outer halo globular clusters are marked with light grey points.}
\label{f:densitiesmr}
\end{center}
\end{figure*}

It is interesting to examine the globular cluster cumulative distribution in Figure \ref{f:densdistmp} in more detail. This distribution splits into three distinct regions -- that below $\zeta_{\rm MP} \approx 0.25$, featuring the apparent strong excess of clusters over the number expected in the case of the null hypothesis; that above $\zeta_{\rm MP} \approx 0.75$, which seems to show a deficit of clusters compared to the prediction for the null hypothesis; and that in between these two limits, which shows an approximately linear increase with a slope comparable to that predicted for the null hypothesis (i.e., where the separation between the two cumulative distributions remains approximately constant). 

We can examine the significance of the excess at small $\zeta_{\rm MP}$ by noting that, in the case of the null hypothesis, the probability distribution for observing a given number of clusters within a certain percentile range $(\zeta_1,\zeta_2)$ is binomial. Here, the number of ``trials'', $n$, is the number of clusters in the sample (i.e., $n = 92$), the number of ``successes'', $k$, is the number of clusters falling within $(\zeta_1,\zeta_2)$, and the probability of success, $p$, is the width of this region (i.e., $p = \zeta_2 - \zeta_1$). The likelihood of observing at least $k$ clusters in the range $(\zeta_1,\zeta_2)$ is given by:
\begin{equation}
P(X \ge k) = \sum\limits_{i=k}^{n} \frac{n!}{i!(n-i)!} p^i (1-p)^{n-i}
\label{e:binomial}
\end{equation}
For our globular cluster distribution, there are $k=41$ clusters with $\zeta_{\rm MP} \le 0.25$. This is substantially above the expected number of $k = 23$ in the case of the null hypothesis (where $\zeta_{\rm MP}$ is distributed uniformly), and indeed according to Equation \ref{e:binomial} the probability of observing at least $41$ clusters with such small values of $\zeta_{\rm MP}$ is tiny, at $0.004\%$.  This strongly reinforces the conclusion we drew from the result of the K-S test. The excess of clusters holds even to much smaller values of $\zeta_{\rm MP}$ -- repeating the test for the $k = 23$ clusters observed to have $\zeta_{\rm MP} \le 0.10$ returns a probability of $0.003\%$. 

Moving to the other end of the scale, how significant is the apparent deficit of clusters with large percentile values?  There are $11$ objects with $\zeta_{\rm MP} \ge 0.75$. We use Equation \ref{e:binomial} to calculate the probability of observing this number or fewer by noting that an equivalent test is to determine the probability of observing at least $k=81$ clusters with $p = 0.75$. The outcome is $0.16\%$, indicating that not only do the outer halo clusters in M31 preferentially associate with regions of high stellar density, they also tend to avoid regions of low stellar density. 

\begin{figure}
\begin{center}
\includegraphics[width=78mm,clip]{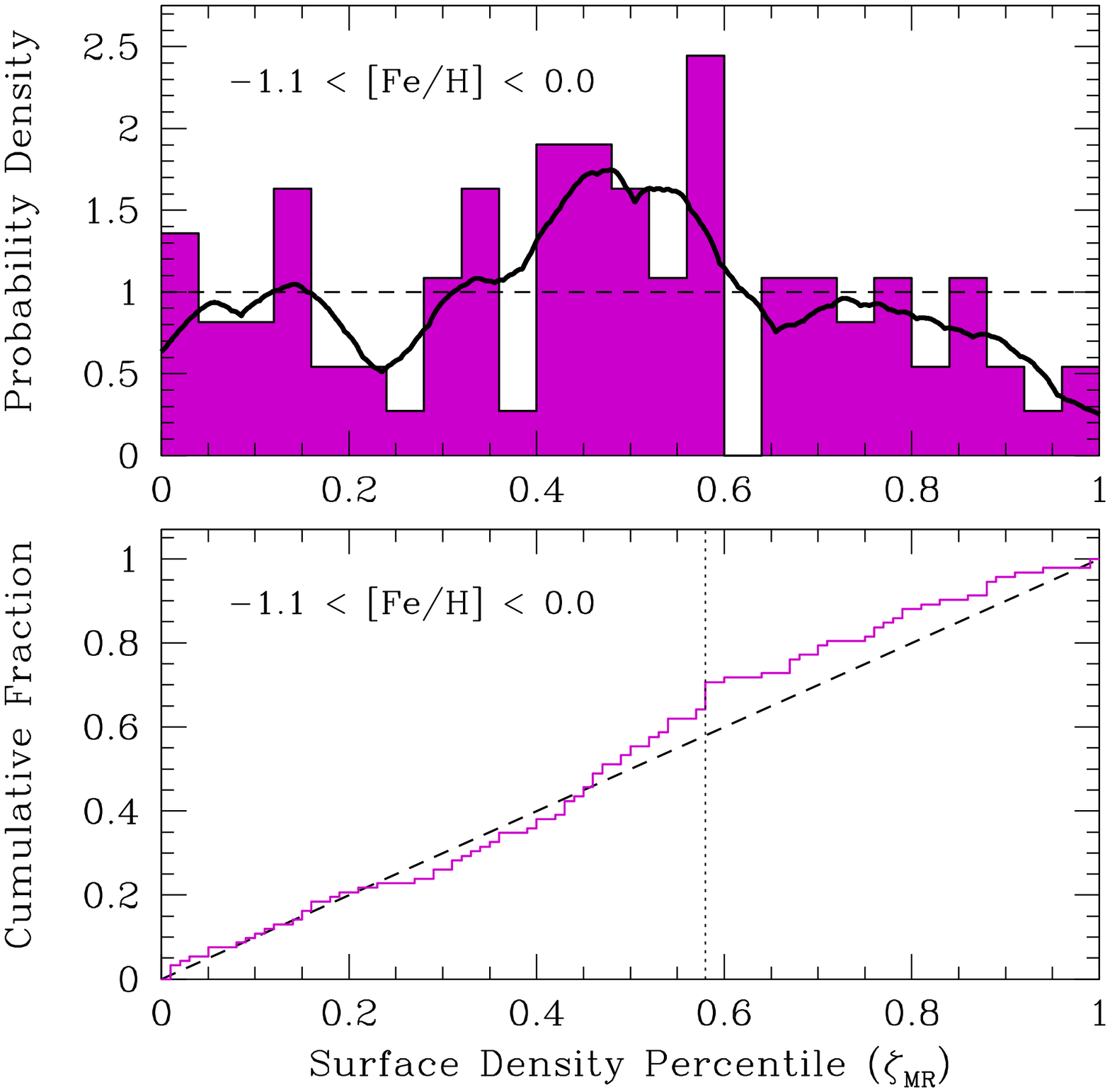}
\caption{Same as Figure \ref{f:densdistmp}, but now showing the distribution of $\zeta_{\rm MR}$. As before, the dashed line shows the expectation for the null hypothesis (a uniform distribution), while the vertical dotted line in the lower panel indicates the location of the greatest separation between the measurements and this uniform distribution. In contrast to the situation for  $\zeta_{\rm MP}$, here the data match the uniform distribution much more closely; the strong peak at small values of $\zeta$ evident in the metal-poor plot is clearly absent.}
\label{f:densdistmr}
\end{center}
\end{figure}

For completeness we repeated the full measurement procedure using stars with $-1.1 \la [$Fe$/$H$] \la 0.0$ -- i.e., those falling within the CMD box labelled ``MR'' in Figure \ref{f:cmd} -- even though there is little in Figure \ref{f:gcmaps} to suggest a strong correlation between globular clusters and the locations of the few metal-rich substructures visible in the M31 halo.  Our numerical results reinforce this impression. Figure \ref{f:densitiesmr} shows the distribution of metal-rich surface density in the M31 stellar halo as a function of projected galactocentric radius, while Figure \ref{f:densdistmr} shows the distribution of metal-rich $\zeta_{\rm MR}$ values for the $92$ M31 globular clusters with $R_{\rm proj} > 25$\ kpc. The strong peak at small values of $\zeta_{\rm MP}$ evident in the metal-poor distribution is clearly absent, and indeed the cumulative distribution of $\zeta_{\rm MR}$ rather closely follows that expected for the null hypothesis. Unsurprisingly a K-S test cannot formally separate the two -- the chance that they were drawn from the same parent distribution is $\approx 10\%$.

\section{Properties of globular cluster subsystems in the M31 outer halo}
\label{s:properties}
Our analysis so far is valid in a global statistical sense. However, the availability of the local density parameter for each individual cluster in our sample also now offers the opportunity to more robustly identify and study subsets of objects that are, and are not, associated with stellar substructures in the outskirts of M31. This is of interest because the M31 periphery is the {\it only} location where there are sufficient data available for both the globular cluster system and the field halo to enable such a classification. Whilst many studies of globular cluster subgroups have been undertaken in the Milky Way system \citep[e.g.,][]{searle:78,zinn:93,mackey:04,mackey:05,forbes:10}, by necessity these have used the properties of the clusters themselves to determine the classification -- a good example being the supposedly-accreted ``young halo'' population, members of which have red horizontal branches (taken as a proxy for younger ages) at given metallicity. Here we are able, for the first time, to attempt the reverse approach -- uniformly identifying accreted clusters by the fact that they are clearly associated with an underlying halo substructure, and then exploring the properties of the subsystems so defined. For this exercise we utilise the data compilation described in Appendix \ref{a:data} and presented in Table \ref{t:fulldata}.

\subsection{Classification}
\label{ss:classification}
Full details of our classification scheme are provided in Appendix \ref{ss:class}. We split our sample of $92$ globular clusters with $R_{\rm proj} > 25$\ kpc into three groups. ``Substructure'' clusters exhibit strong spatial and/or kinematic evidence for a link with a halo substructure, while ``non-substructure'' clusters possess no such evidence. Clusters with weak or conflicting evidence for an association fall into an ``ambiguous'' category. We carefully consider all the available information for each given object when making our classification. Simply having a small value of $\zeta_{\rm MP}$ is not, by itself, sufficient to identify a ``substructure'' cluster; nor, in many cases, is the kinematic information uniquely decisive.  While \citet{veljanoski:14} previously used their radial velocity measurements to explore the association between a subset of clusters and the most prominent stellar substructures in the M31 halo, here we have added a formal measurement, through the calculation of $\zeta_{\rm MP}$, of the proximity of each given cluster to overdensities in the field (whether or not these are named and/or recognized as discrete features).

We identify $32$ clusters that have a high likelihood of being associated with an underlying field substructure, and $35$ that show no evidence for such an association. In $25$ cases the available data are ambiguous. The majority of these objects have a small value of $\zeta_{\rm MP}$ but exhibit no additional evidence for a substructure association. However, there are also several examples where a cluster has close proximity to a large stellar feature or kinematically-identified cluster grouping, but possesses an inconsistent velocity measurement. 

Our results imply that between $\approx 35 - 62\%$ of globular clusters at $R_{\rm proj} > 25$\ kpc exhibit properties consistent with having been accreted into the M31 halo. This is lower than the $\sim 80\%$ inferred by \citet{mackey:10b}. However, these authors did not examine the complete M31 halo but rather only a region mostly to the south and south-west of the M31 centre. As discussed in Section \ref{ss:profile} this region -- also studied by \citet{huxor:11} -- is rather unrepresentative in terms of the number of clusters it contains; it also appears somewhat enhanced in terms of ``substructure'' and ``ambiguous'' objects. 

\begin{figure*}
\begin{center}
\includegraphics[width=58mm,clip]{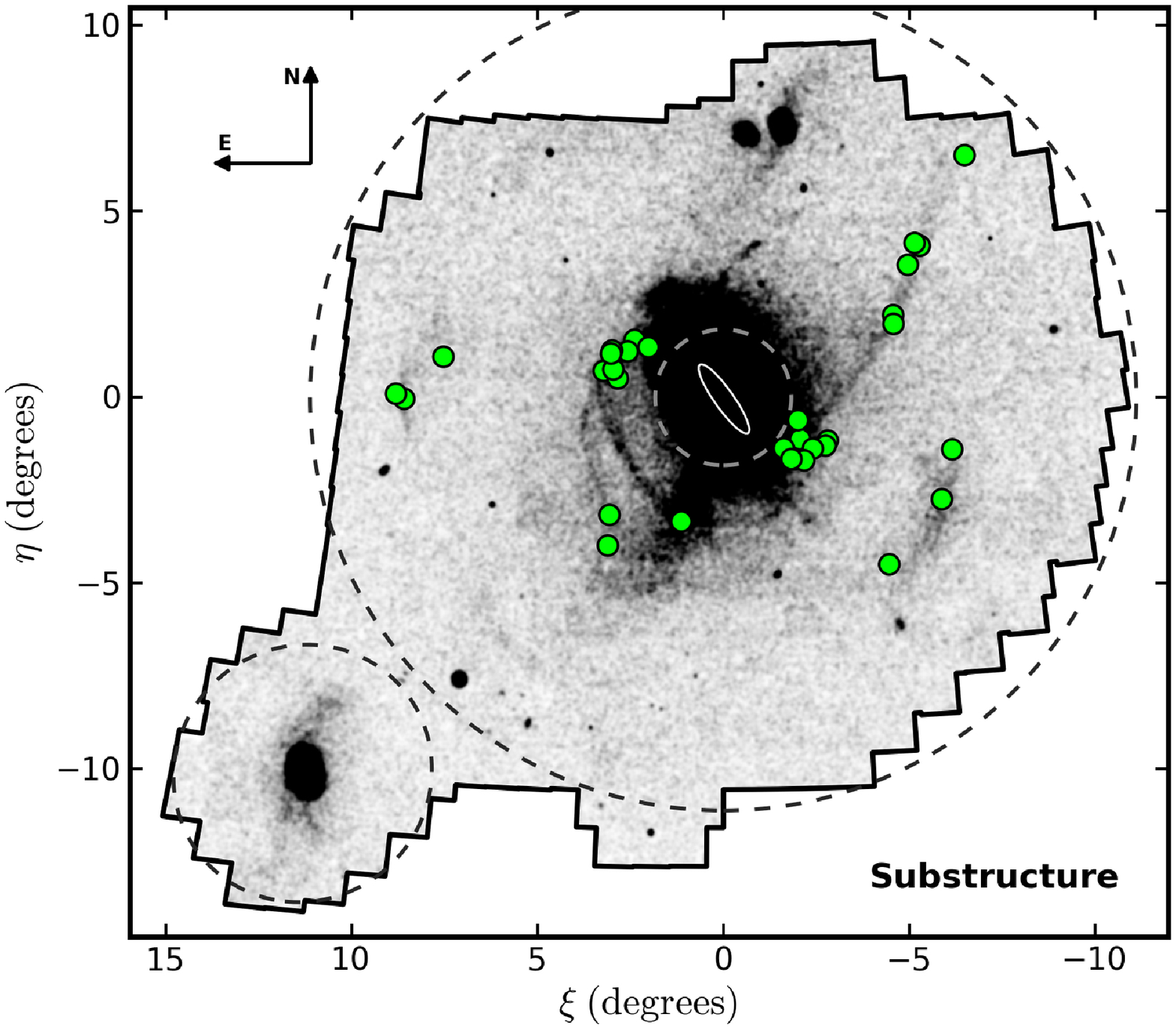}
\hspace{0mm}
\includegraphics[width=58mm,clip]{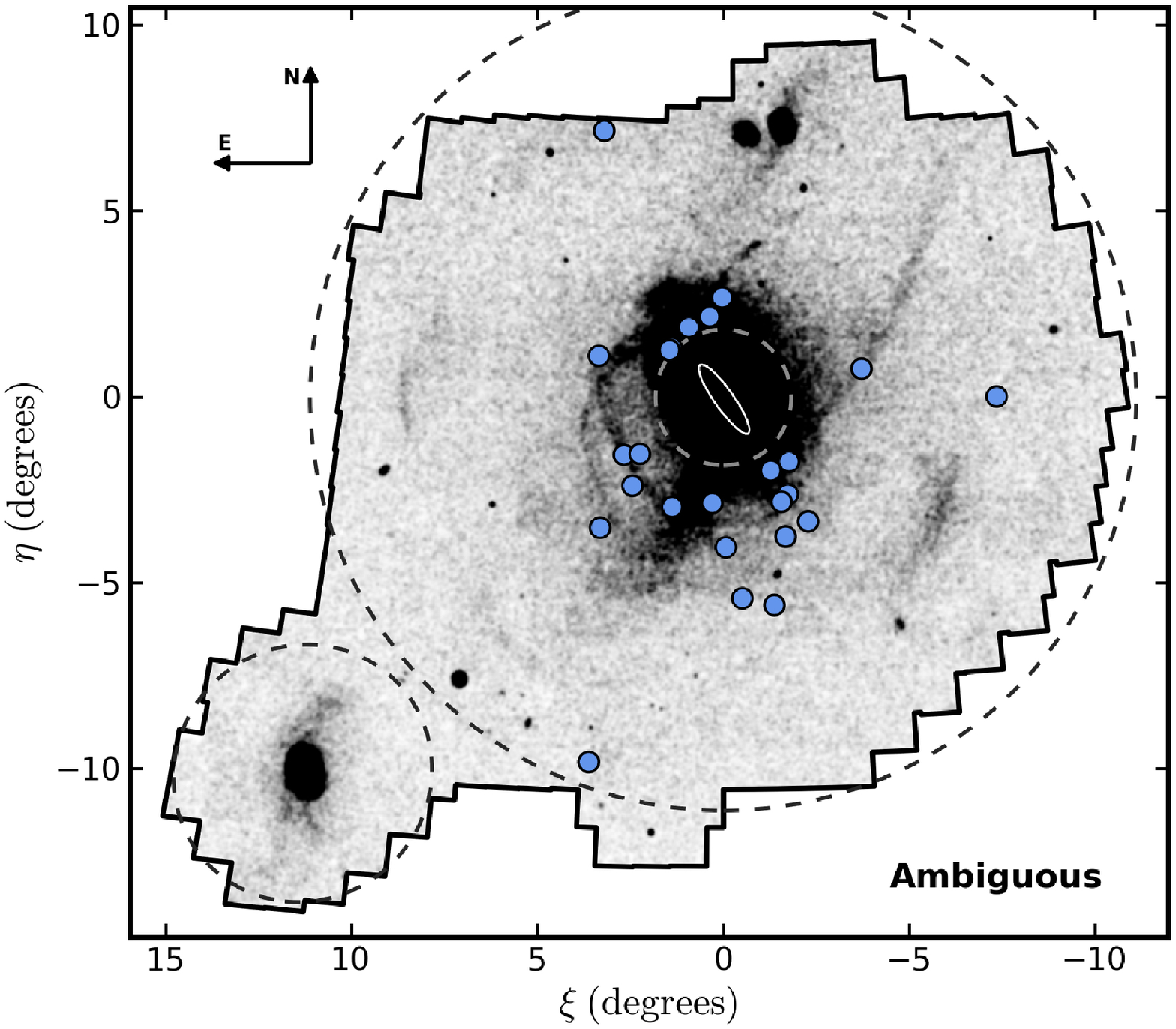}
\hspace{0mm}
\includegraphics[width=58mm,clip]{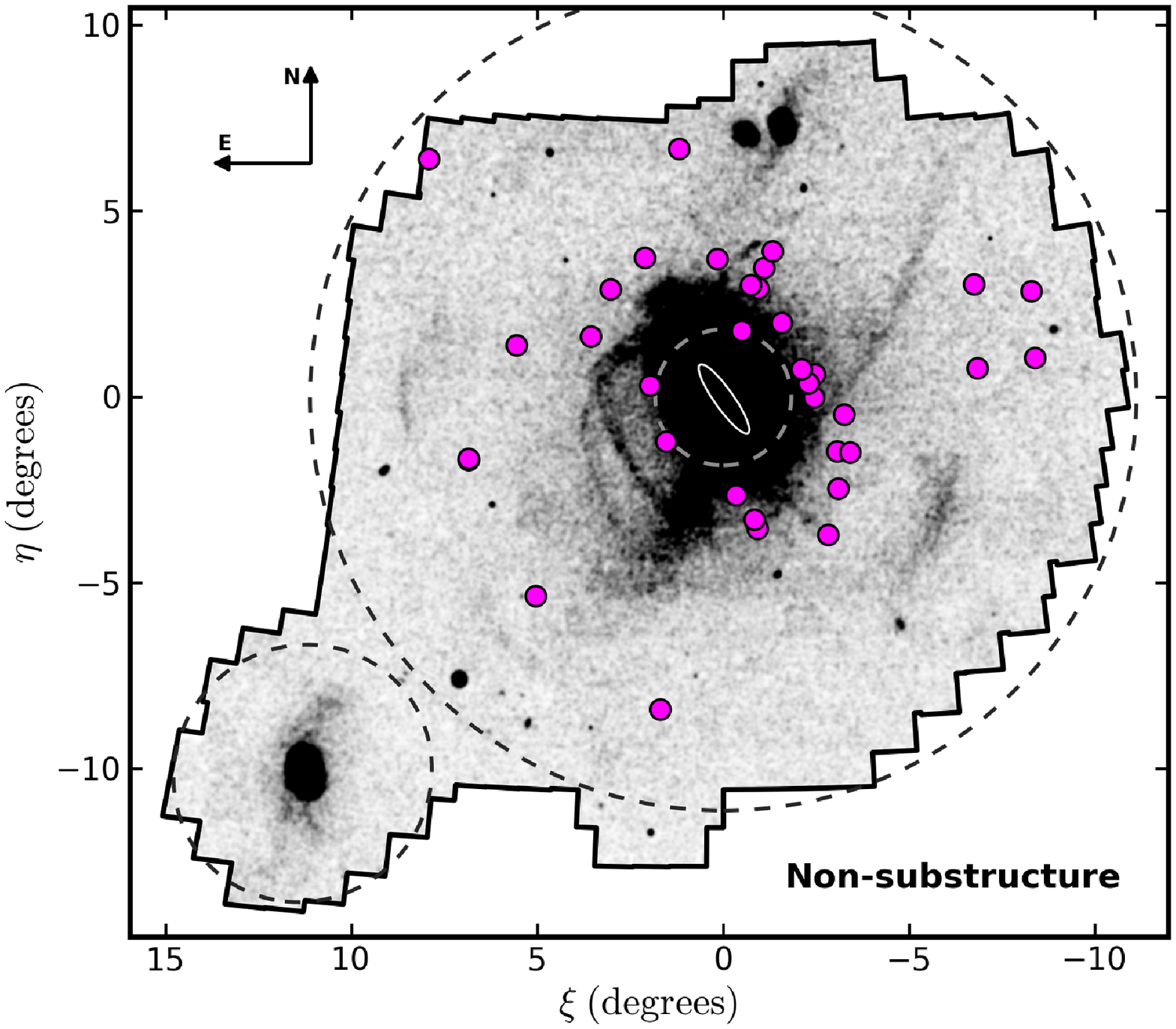}
\caption{Spatial distribution for clusters in each of the three classes introduced in Section \ref{ss:classification}, overplotted on the metal-poor field halo map from Figure \ref{f:maps}.}
\label{f:subsysmap}
\end{center}
\end{figure*}

Figure \ref{f:subsysmap} shows the spatial distribution of the clusters in each of the three classes, overplotted on the metal-poor field halo map from Figure \ref{f:maps}. As expected, the ``substructure'' members exhibit significant clustering and a tight correlation with various of the main stellar streams in the M31 halo, while the ``non-substructure'' objects are more dispersed.  There is a hint that inside $R_{\rm proj} \approx 50$\ kpc the clusters in this sub-group possess a somewhat flattened distribution oriented similarly to the M31 disk; however there are too few members to support this assertion with robust statistics.

\begin{figure}
\begin{center}
\includegraphics[width=78mm,clip]{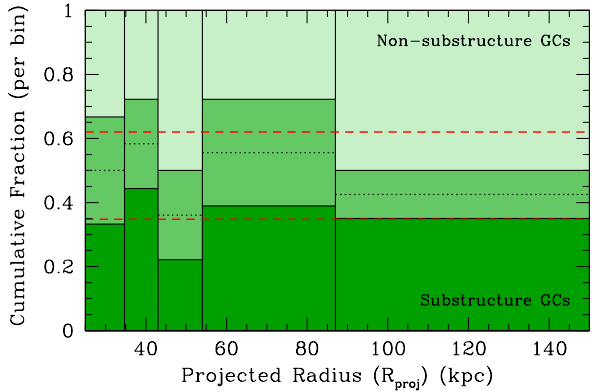}
\caption{The fraction of clusters populating each of the three classes, as a function of projected galactocentric radius. The ``ambiguous'' subgroup occupies the region between the ``substructure'' clusters (lower) and ``non-substructure'' clusters (upper) on the diagram. Each radial bin holds $\approx 20\%$ of the cluster population outside $R_{\rm proj} > 25$\ kpc (specifically, the inner four bins contain $18$ clusters each, while the outermost bin has $20$ clusters). The horizontal dashed red lines indicate the split between the three subgroups for the overall sample. The horizontal black dotted lines indicate, per bin, the centre of the ``ambiguous'' class.}
\label{f:fraction}
\end{center}
\end{figure}

In Figure \ref{f:fraction} we plot the fraction of clusters falling in the three different subgroups as a function of projected galactocentric radius. Although there are mild bin-to-bin fluctuations, the division of clusters between the ``substructure'' and ``non-substructure'' subsystems appears essentially constant with radius when averaged over all position angles. There are comparatively few ``ambiguous'' clusters at radii beyond $\sim 80$\ kpc; this may simply reflect the less complex nature of the M31 halo at large galactocentric distances, leading to more confident classification.

\subsection{Radial density profiles}
\label{ss:rpsubsys}
\begin{figure}
\begin{center}
\includegraphics[width=78mm,clip]{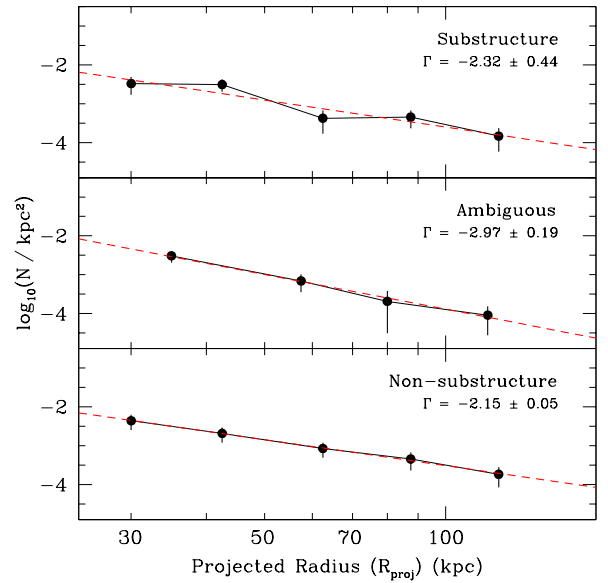}
\caption{Radial surface density profiles for the three subgroups of outer halo globular cluster systems defined in Section \ref{ss:classification}. The red dashed lines show the best-fit power-laws. All points have Poissonian error bars.}
\label{f:profilesubsys}
\end{center}
\end{figure}

Figure \ref{f:profilesubsys} shows radial surface density profiles for each of the three cluster subsystems. These were computed precisely as were the profiles in Section \ref{ss:profile}. It is evident from Figure \ref{f:profilesubsys} that the profile for substructure clusters exhibits much greater point-to-point fluctuations than does that for non-substructure clusters, which is remarkably smooth. The irregularity of the substructure profile agrees with naive expectation (and, indeed, our observations in Section \ref{ss:profile}) -- by definition, substructure clusters ought to be grouped both spatially and kinematically. However, it does not automatically follow that the non-substructure objects should possess a completely featureless decline with radius; that they appear to do so tells us something interesting about the nature of this population. 

Additional insight can be gained from power-law fits to the profiles. For the non-substructure profile we measure a power-law index of $\Gamma = -2.15 \pm 0.05$. This is an excellent match to the field halo profiles from \citet{ibata:14}, who measured $\Gamma = -2.08 \pm 0.02$ and $\Gamma = -2.13 \pm 0.02$ for {\it substructure-masked} populations with $-2.5 < [$Fe$/$H$] < -1.7$ and $-1.7 < [$Fe$/$H$] < -1.1$, respectively. Similarly, \citet{gilbert:12} found $\Gamma = -2.2 \pm 0.3$ for a {\it ``substructure-removed''}  sample from their large-scale spectroscopic survey. Given how closely the behaviour of the non-substructure component of the outer M31 cluster system mirrors that of the apparently smooth metal-poor component of the field halo, it is strongly tempting to link the two. 

For the substructure cluster profile we obtain $\Gamma = -2.32 \pm 0.44$. The much larger uncertainty associated with this measurement reflects the substantial point-to-point scatter in the profile. It is more difficult to interpret this measurement, as \citet{ibata:14} do not examine any substructure-only profiles. However, as discussed in Section \ref{ss:profile} they do provide unmasked profiles (i.e., including both the substructured and smooth halo components), finding $\Gamma = -2.30 \pm 0.02$ for the stars with $-2.5 < [$Fe$/$H$] < -1.7$, and $\Gamma = -2.71 \pm 0.01$ for stars with $-1.7 < [$Fe$/$H$] < -1.1$. Both these values are consistent with our measurement, given the uncertainties. The metal-rich field halo population, with $-1.1 < [$Fe$/$H$] < 0.0$, exhibits a much steeper radial fall-off with $\Gamma = -3.72 \pm 0.01$.

For completeness Figure \ref{f:profilesubsys} also shows a radial density profile for our ``ambiguous'' class of clusters; however, it is not clear that this is physically meaningful. Given the mild decline in the fraction of ambiguous clusters at large galactocentric radii, as in Figure \ref{f:fraction}, it is not too surprising that the power-law index for this profile is steeper than for the other two classes, at $\Gamma = -2.97 \pm 0.19$. Qualitatively, the amplitude of the point-to-point fluctuations in the profile falls somewhere between that for the substructure clusters and that for the non-substructure clusters. This makes sense, given the composite nature of the ``ambiguous'' class.

\subsection{Luminosity distributions}
\label{ss:lumfunc}
\begin{figure}
\begin{center}
\includegraphics[width=78mm,clip]{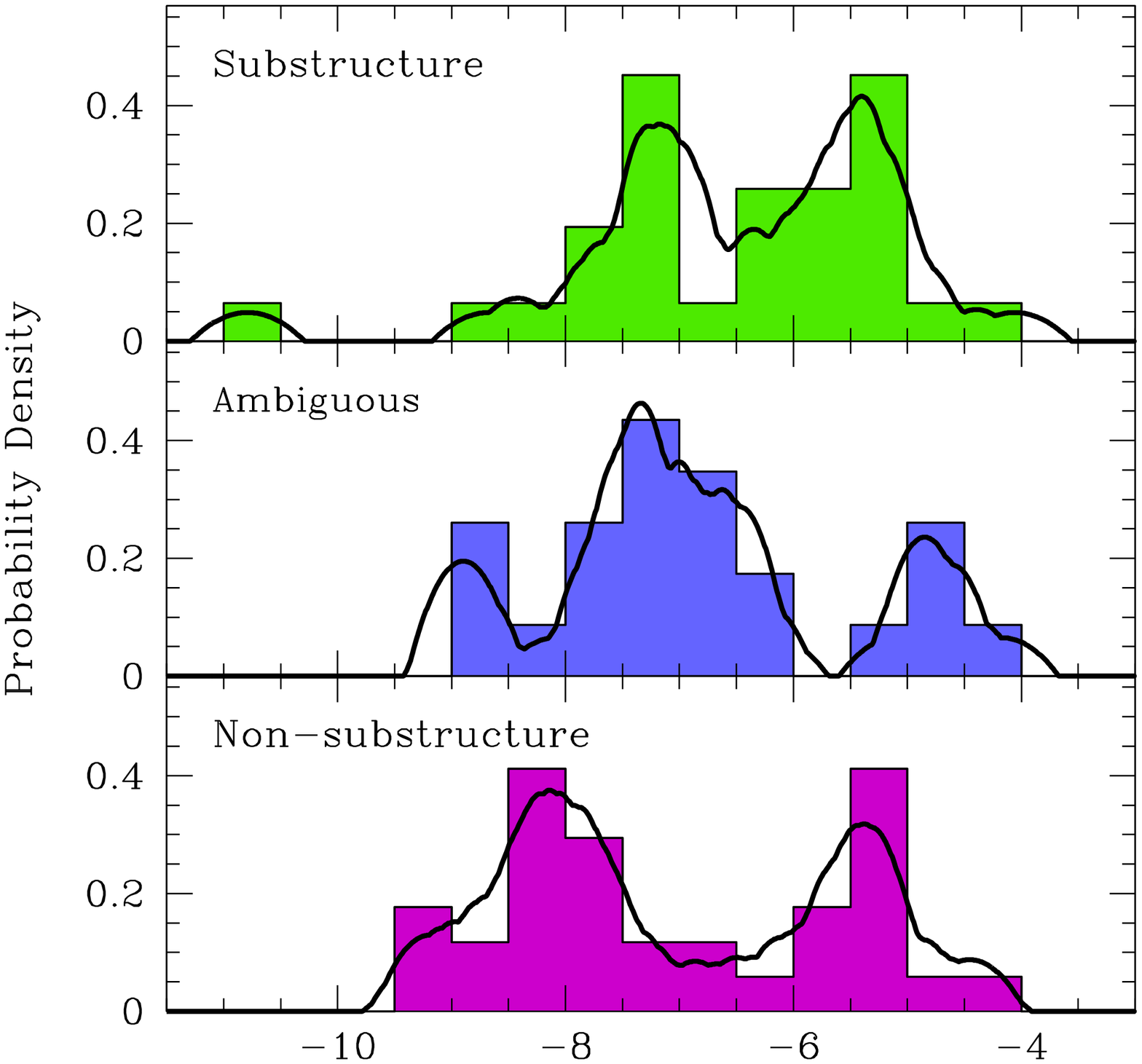}\\
\vspace{-1mm}
\includegraphics[width=78mm,clip]{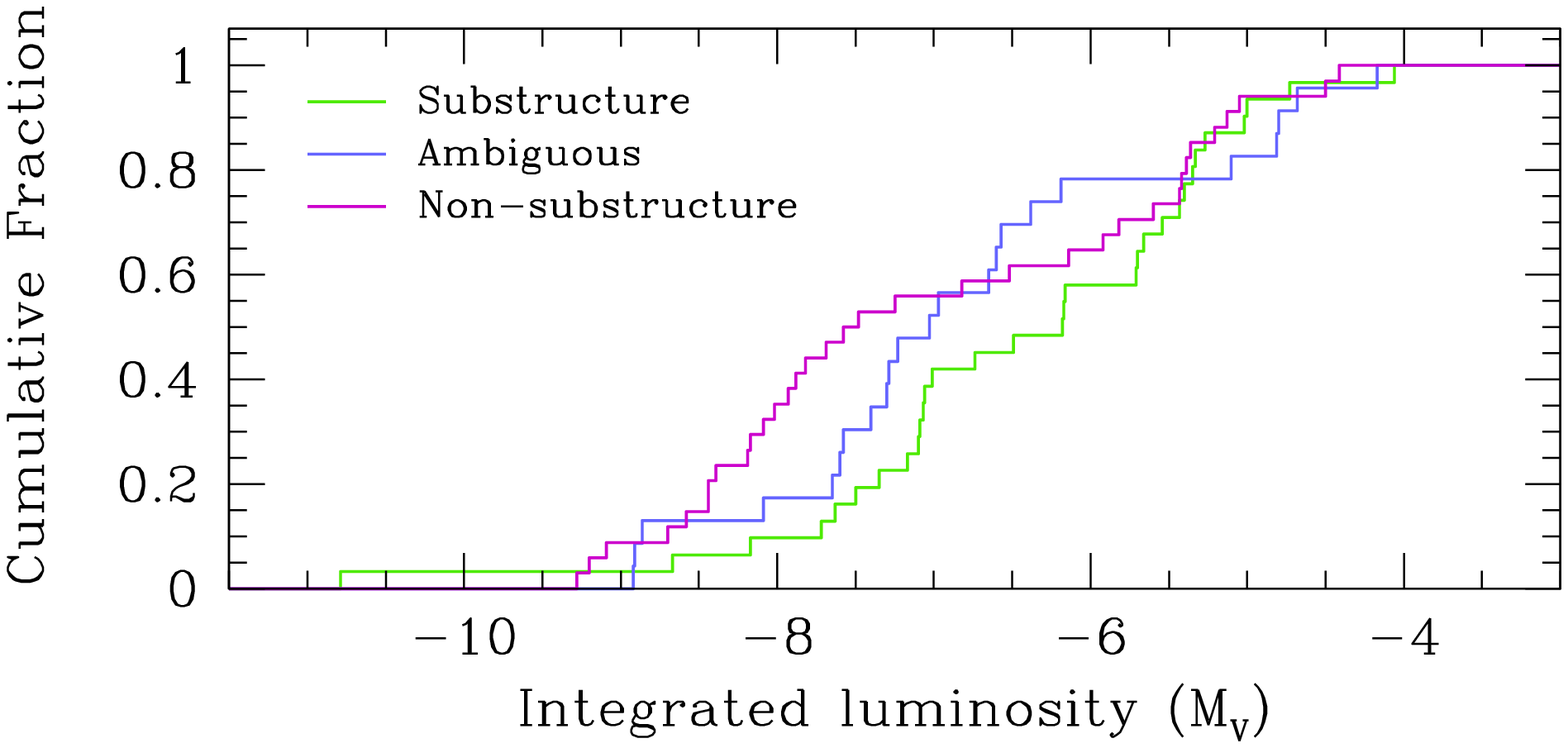}
\caption{Luminosity functions for the three globular cluster subsystems. In the upper three panels these are plotted as histograms, and as smoothed curves derived via kernel density estimation with an Epanechnikov kernel. The scaling is such that the area under the curves is unity. The lower panel shows cumulative luminosity distributions. Due to missing luminosity data for a few objects (see Appendix \ref{ss:lumcol}), the total sample considered here comprises $88$ clusters: $31$, $23$, and $34$ for the three sub-systems respectively.}
\label{f:subsysmv}
\end{center}
\end{figure}

Figure \ref{f:subsysmv} displays the luminosity function for each globular cluster subsystem. All three distributions possess very similar shapes. The most striking characteristic is their bimodal nature, with one peak near the canonical value of $M_V \sim -7.5$ and a second with comparable amplitude at much fainter $M_V \approx -5.4$. This property of the globular cluster population in the outer halo of M31 has previously been noted and discussed by \citet{huxor:14}; it is intriguing to see here that it is not restricted to a particular class of object.

The most significant difference between the substructure and non-substructure distributions is the location of the brighter peak. For the substructure clusters (and, indeed, those in the ambiguous class) the bright peak falls near $M_V \approx -7.25$. On the other hand, the peak for the non-substructure clusters is nearly a magnitude brighter at $M_V \approx -8.15$. This discrepancy is significant: (i) as noted in Appendix \ref{a:data}, the per-object uncertainty on the luminosity measurement for the bright, compact clusters that comprise this portion of the distribution is about $\pm 0.1$\ mag, much smaller than the apparent separation of the peaks; and (ii) the cumulative distributions plotted in the lower panel in Figure \ref{f:subsysmv} exhibit their strongest separation at $M_V\sim -7.7$, directly between the peaks -- a K-S test delivers a probability of only $\approx 2\%$ that the two sets of data were drawn from the same parent distribution.

The unusually bright luminosity function peak appears to be a characteristic peculiar to the outer halo non-substructure clusters -- a number of previous studies have measured the luminosity function peak to be at $M_V \approx -7.65$ for the M31 metal-poor globular cluster population as a whole (i.e., for a sample where the vast majority sits well inside $R_{\rm proj} = 25$\ kpc), in good agreement with that found for the metal-poor Milky Way population \citep[see, e.g.,][]{dicris:06,rejkuba:12,huxor:14}. We discuss the implications of the observed luminosity function differences between our cluster subsystems in more detail in Section \ref{s:discussion}.

\subsection{Size distributions}
\begin{figure}
\begin{center}
\includegraphics[width=78mm,clip]{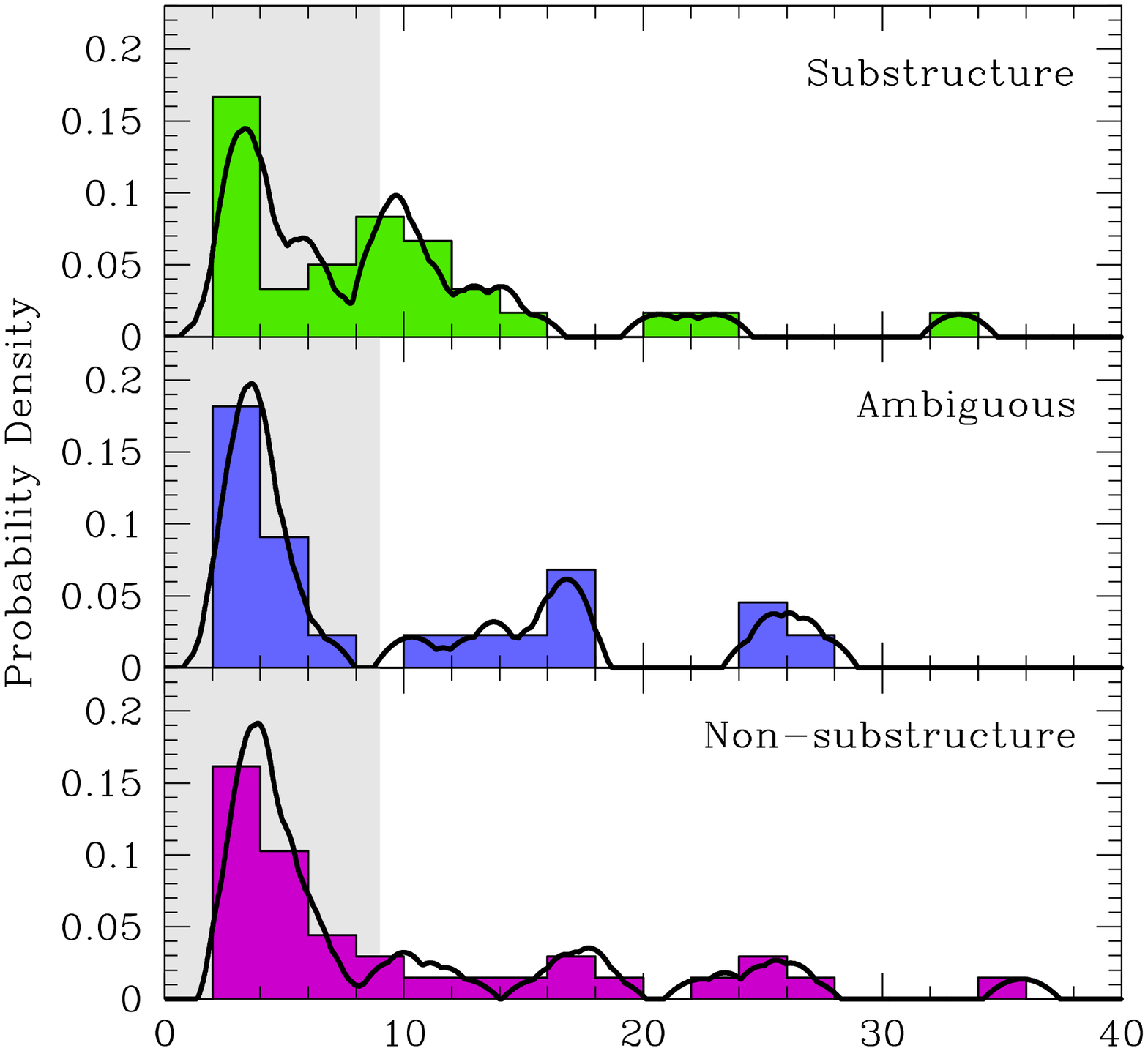}\\
\vspace{-1mm}
\includegraphics[width=78mm,clip]{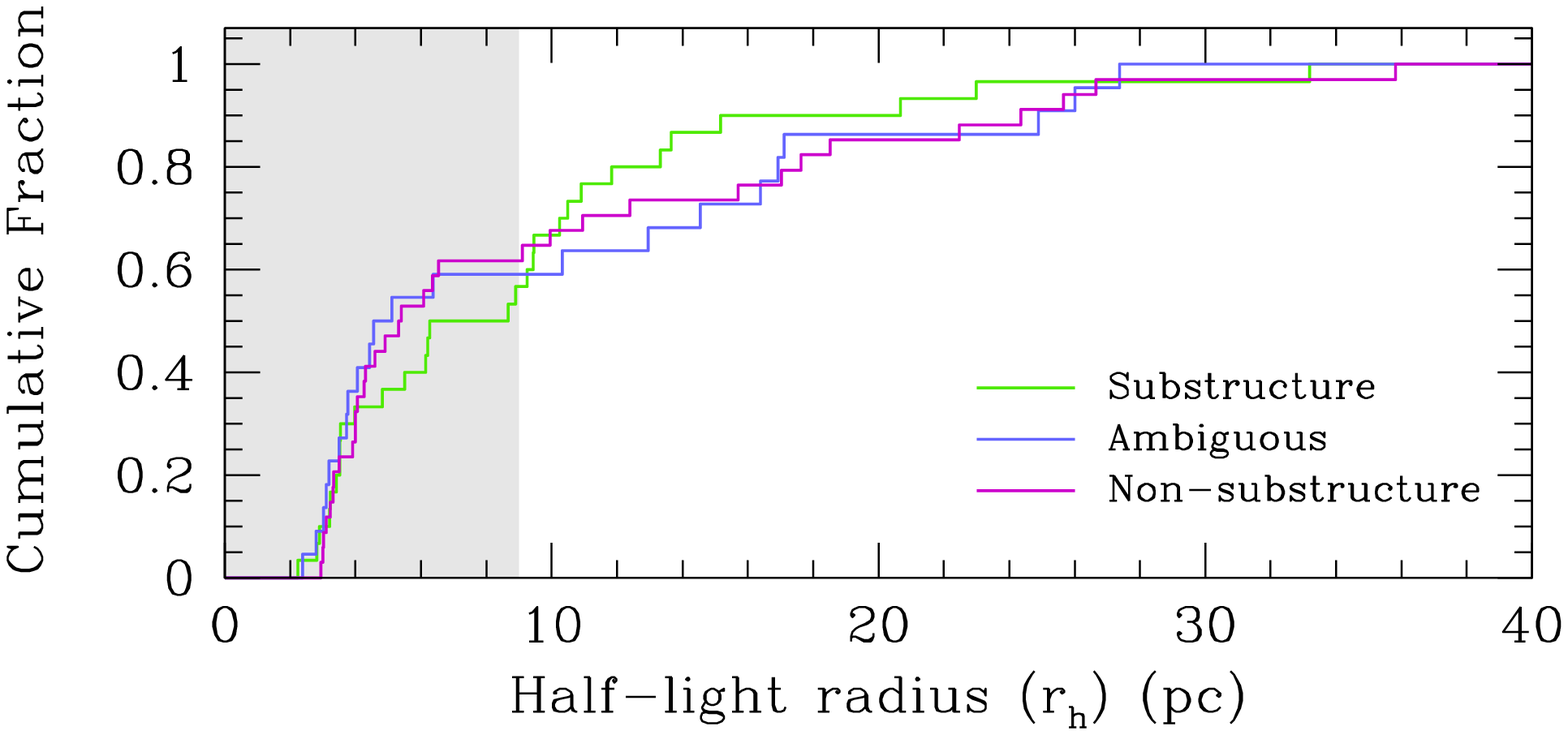}
\caption{Size distributions for the three globular cluster subsystems. In the upper three panels these are plotted as histograms, and as smoothed curves derived via kernel density estimation with an Epanechnikov kernel. The scaling is such that the area under the curves is unity. The lower panel shows cumulative size distributions. Due to missing size data for a few objects (see Appendix \ref{ss:rh}), the total sample considered here comprises $86$ clusters: $30$, $22$, and $34$ for the three sub-systems respectively. In all panels the grey shaded region indicates the range over which the measured cluster sizes require a correction for the effects of atmospheric seeing; this region should be given low weight when considering the various distributions.}
\label{f:subsysrh}
\end{center}
\end{figure}

Figure \ref{f:subsysrh} shows the distribution of half-light radii for our three cluster subgroups.  Again, all three distributions possess very similar shapes. The shaded region indicates the range of sizes over which an empirical correction has been applied to each measured half-light radius to account for the effects of atmospheric seeing, as described in Section \ref{ss:rh}. The {\it shapes} of the distributions in this region should not be trusted; however the proportion of each subgroup that falls below the limiting $r_h = 9$\ pc is unaffected by the correction and appears remarkably consistent across the three samples at $\approx 60\%$. 

As noted by several previous studies \citep[e.g.,][]{huxor:05,mackey:06,huxor:11,huxor:14}, the outer halo globular cluster population in M31 includes many very extended objects with half-light radii as large as $r_h \approx 35$\ pc. Figure \ref{f:subsysrh} shows that these extended clusters are not concentrated in a single subgroup -- both the substructure and non-substructure classes include this type of object in roughly equal proportions.  The main difference between the two distributions is the apparent presence of a mild excess of clusters with sizes $\sim8-12$\ pc in the substructure group; however the cumulative distributions plotted in the lower panel reveal that the significance of this difference is low, especially given the typical individual measurement uncertainties of $\approx 10\%$ in $r_h$. 

\subsection{Colour distributions}
\label{ss:gccol}
\begin{figure}
\begin{center}
\includegraphics[width=78mm,clip]{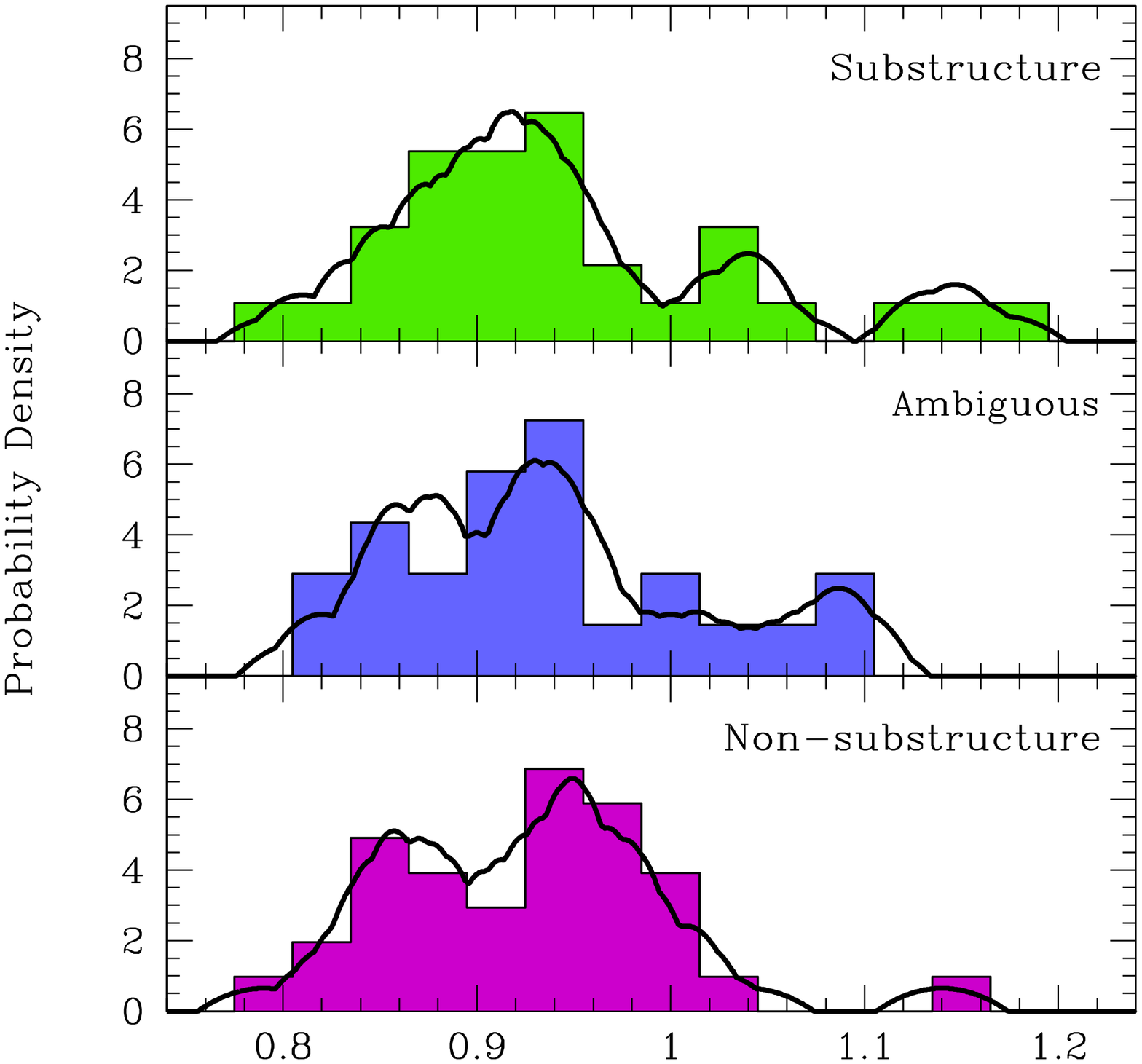}\\
\vspace{-1mm}
\includegraphics[width=78mm,clip]{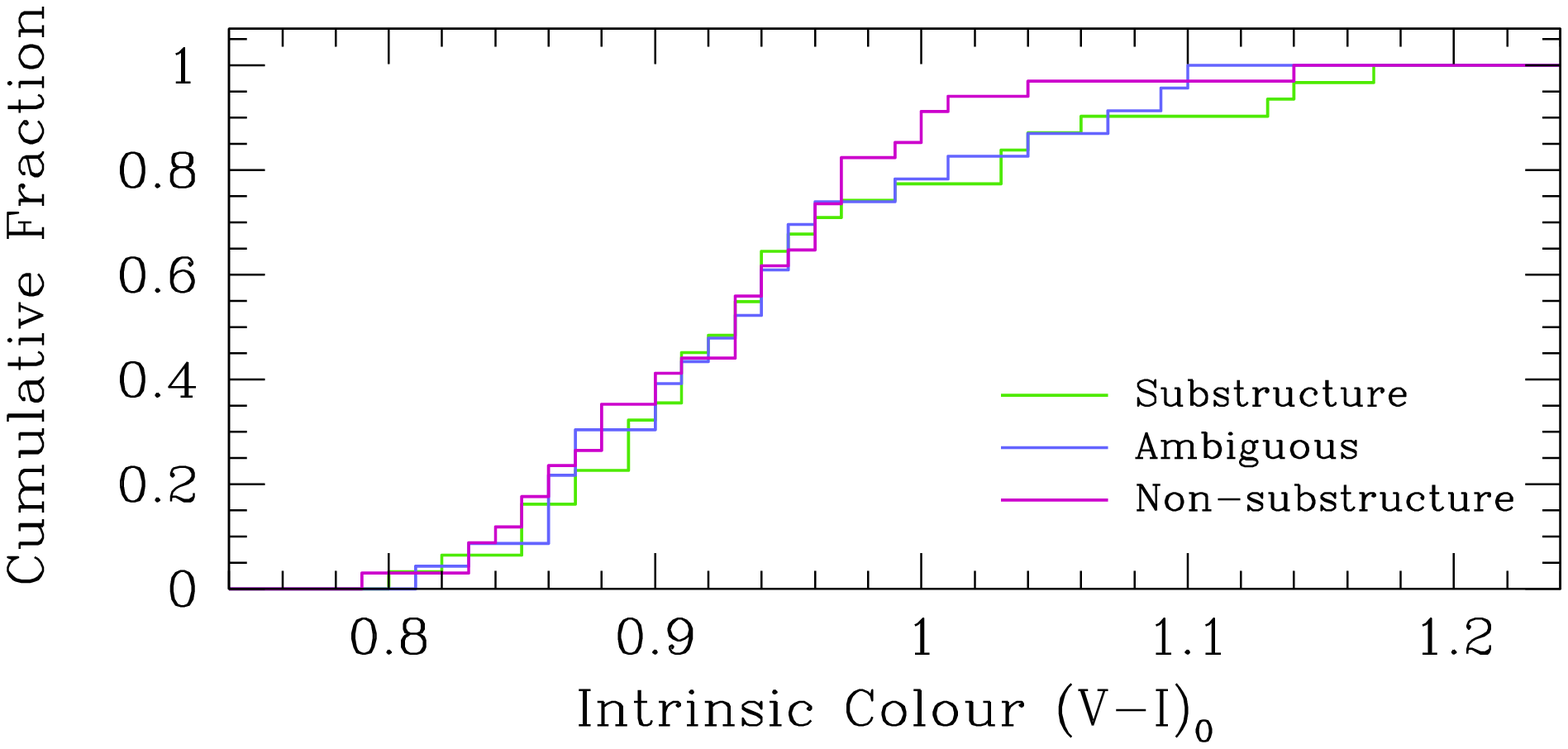}
\caption{Colour distributions for the three globular cluster subsystems. In the upper three panels these are plotted as histograms, and as smoothed curves derived via kernel density estimation with an Epanechnikov kernel. The scaling is such that the area under the curves is unity. The lower panel shows cumulative colour distributions. Due to missing photometric data for a few objects (see Appendix \ref{ss:lumcol}), the total sample considered here comprises $88$ clusters: $31$, $23$, and $34$ for the three sub-systems respectively.}
\label{f:subsysvmi}
\end{center}
\end{figure}

In Figure \ref{f:subsysvmi} we construct $(V-I)_0$ colour distributions for the three cluster subsystems. Once again these are very similar to each other. Indeed, for colours bluer than $(V-I)_0 \approx 1.0$ the cumulative distributions plotted in the lower panel show that the subsystems are effectively indisinguishable. At redder colours, the substructure group appears to harbour a handful of objects extending to $(V-I)_0 \approx 1.2$ that are largely absent in the non-substructure group. Very few of these red objects have been studied in detail, so their nature is not immediately obvious.  It is likely that their colours reflect a somewhat higher metallicity than the bulk of the outer halo population -- see, for example, Figure 5 in \citet{georgiev:09}, which shows the $(V-I)$ colour distributions for globular clusters in dwarf galaxies within $12$\ Mpc along with evolutionary tracks for single stellar population models with $[$Fe$/$H$] = -2.25$ and $-1.65$. 

Reinforcing this notion are the observations of \citet{colucci:14} and \citet{sakari:15}, who derived elemental abundance estimates for nine outer halo globular clusters from high-resolution integrated spectra. We observe a good correlation between the integrated colours and the spectroscopically-derived metallicities for the objects in this sample. For PA-06, PA-53, PA-54, PA-56, MGC1, and G2, the spectroscopic metallicities fall in the range $-2.1 \la [$Fe$/$H$] \la -1.6$; these clusters all have $(V-I)_0 < 0.9$. For H10 and H23 the spectroscopic metallicities are between $-1.4 \la [$Fe$/$H$] \la -1.1$, and these clusters have $0.9 < (V-I)_0 < 1.0$. The most metal-rich cluster in the sample is PA-17, with $[$Fe$/$H$] \approx -0.9$ and a correspondingly red $(V-I)_0 = 1.14$. Overall, this suggests that the substructure group includes the majority of objects populating the metal-rich tail of the cluster metallicity distribution at $R_{\rm proj} \ge 25$\ kpc.

\section{Discussion}
\label{s:discussion}
The availability of the complete PAndAS data set, providing contiguous coverage of the M31 stellar halo to approximately half the virial radius \citep{mcconnachie:18}, has allowed us to undertake a comprehensive investigation of the links between the remote field star populations and the globular cluster system in this galaxy. We have focused on the ``outer halo'' region spanning $R_{\rm proj} = 25-150$\ kpc, where the globular cluster census is essentially complete \citep{huxor:14}, and our understanding of the field populations is not confusion-limited \citep[e.g.,][]{ibata:14}. 

\subsection{Relationship between clusters and the field halo}
\label{ss:discuss0}
Our halo maps and radial density profiles robustly demonstrate that the globular clusters outside $25$\ kpc in M31 are overwhelmingly associated with the metal-poor stellar halo -- i.e., that portion with $[$Fe$/$H$] \la -1.1$. Notably, this cut contains only the minority fraction of halo luminosity over the range $25-150$\ kpc \citep[$\sim15-30\%$, depending on the assumed stellar ages -- see][]{ibata:14}. Within this metal-poor halo the constituent stellar populations split approximately evenly into a heavily substructured component, and an apparently much smoother diffuse component \citep{ibata:07,ibata:14,mcconnachie:09,mcconnachie:18,gilbert:12}. Similarly, we have shown that the remote globular cluster population in Andromeda plausibly comprises a composite system where some fraction is robustly associated with the field substructures at high statistical significance, and another fraction appears to behave rather like the diffuse smooth-halo component. The properties of these two groups will be discussed in detail in the following sub-sections.

Quantitative evidence for a strong association between a subset of globular clusters and underlying halo substructures in the outskirts of M31 comes from Section \ref{ss:correlation}, where we demonstrated that clusters in our remote sample preferentially project onto over-dense regions in the metal-poor field halo with very high significance\footnote{Specifically, we recall that (i) a simple K-S test rejects the possibility of no correlation between clusters and field overdensities with a probability $>99.95\%$, and (ii) in the case of no correlation, the probability distribution for observing a given number of clusters within a certain range of local density percentiles is binomial such that the likelihood of observing at least the $41$ clusters in our sample possessing $\zeta_{\rm MP} \le 0.25$ would be just $0.004\%$.}. This does not mean that {\it all} globular clusters fall onto metal-poor substructures; it simply states that {\it many more} clusters have high local densities of metal-poor stars {\it than would be expected if the clusters were randomly scattered throughout the halo}. Of course, this cluster-substructure association does not just manifest in a purely spatial sense. \citet{veljanoski:13a,veljanoski:14} established very clearly that kinematic correlations are evident amongst some groups of clusters that sit in close proximity to prominent features in the field halo, and even amongst some groups of clusters for which no underlying over-density is apparent. Also relevant are the handful of studies that provide the ``missing link'' between clusters and the field: a velocity measurement for a halo substructure that matches the kinematics for nearby clusters possessing high local densities \citep{collins:09,bate:14,mackey:14}.

Perhaps surprisingly, the opposite appears to be true for the {\it metal-rich} component of the outer M31 halo (i.e., that portion with $[$Fe$/$H$] \ga -1.1$): we find no evidence for a statistically significant correlation between clusters and metal-rich field overdensities in our survey region. Again, this does not mean that {\it no} globular clusters belong to metal-rich substructures; it simply states that the number of clusters possessing high local densities of metal-rich halo stars {\it is not substantially in excess of the number that would be expected if the clusters and field substructures were decoupled}. 

It is possible that this observation can be traced to the particular circumstances of the M31 halo. Although the metal-rich cut contains the majority of the stellar luminosity outside $25$\ kpc, \citet{ibata:14} have shown that this component is overwhelmingly dominated by the debris from a {\it single} accretion event, which produced the Giant Stream. Models of this event generally agree that the progenitor system, at least as massive as the Large Magellanic Cloud, fell into M31 on a highly radial orbit such that its first pericentric passage came within a few kpc of the galactic centre \citep[e.g.,][]{fardal:06,fardal:08,fardal:13,mori:08,sadoun:14}. The Giant Stream represents the trailing material stripped during this first pericentric pass; the remainder of the progenitor is thought to reside almost exclusively {\it inside} $R_{\rm proj} = 25$\ kpc, forming the North-East Shelf and Western Shelf from successive orbital passages. In this case, it is entirely plausible that most or all of the globular clusters that were members of the accreted system are now located in the inner halo of M31, especially if they were relatively tightly bound to the incoming satellite\footnote{In Appendix \ref{ss:class} we identify just a single cluster, PA-37, that is plausibly associated with the Giant Stream outside $25$\ kpc.}. If this is true, it would suggest that the apparent lack of clusters associated with the metal-rich portion of the outer halo is mainly due to the specifics of the Giant Stream accretion event combined with the restricted radial span of our analysis ($25\le R_{\rm proj} \le 150$\ kpc), rather than representing a more general property of metal-rich halo populations.

We conclude by noting that, while good quality metallicity measurements exist for a small fraction of our remote globular cluster sample \citep[e.g.,][]{mackey:06,mackey:07,alvesbrito:09,colucci:14,sakari:15}, in general there is insufficient information presently available to robustly compare the metallicity distributions of clusters and the underlying field halo. However, as discussed in Section \ref{ss:gccol}, integrated colour measurements for our clusters strongly suggest that a very significant majority are metal poor with $[$Fe$/$H$]\la -1.1$ (with the bulk possessing $[$Fe$/$H$]\la -1.5$). Even the reddest clusters in the sample are unlikely to be much more metal-rich than $[$Fe$/$H$]\approx -0.9$.

\subsection{The substructured cluster population}
\label{ss:discuss1}
In order to move from the global statistical analysis presented in Section \ref{ss:correlation} and discussed above, to a more intricate investigation of the M31 accretion history as traced by globular clusters, we attempted to robustly identify the subgroup of clusters responsible for the excess signal at high local (metal-poor) stellar densities and for the instances of correlated kinematics described by \citet{veljanoski:14}. Full details are provided in Section \ref{s:properties} and Appendix \ref{a:data}. We found that $\approx 35\%$ of the remote cluster system ($32$ objects) can be unambiguously associated with substructure in the M31 halo, while another $27\%$ ($25$ objects) show some indication for such an association (these constitute the ``ambiguous'' class). The total fraction falling into the ``substructured'' cluster population could therefore be as high as $\approx 62\%$. 

\subsubsection{Comparison with metal-poor field substructures}
From their three-dimensional fits to the masked PAndAS data, \citet{ibata:14} find that $58\%$ of the total number of halo stars in the range $-1.7 < [$Fe$/$H$] < -1.1$ are in the substructured component, and that this decreases to $42\%$ of more metal-poor halo stars with $-2.5 < [$Fe$/$H$] < -1.7$. These estimates are entirely consistent with that derived from our globular cluster sample. In terms of luminosity, from Tables $4$ and $5$ in \citet{ibata:14} we calculate\footnote{We provide full details here, as similar calculations will be relevant throughout this Section. Assuming an age of $13$ Gyr for halo stars, Table $4$ in \citet{ibata:14} shows that the total luminosity in the range $-2.5 < [$Fe$/$H$] < -1.7$ is $0.09\times10^9 L_\odot$, and in the range $-1.7 < [$Fe$/$H$] < -1.1$ is $0.17\times10^9 L_\odot$. Similarly, their Table $5$ shows that the smooth halo luminosity in the range $-2.5 < [$Fe$/$H$] < -1.7$ is $0.08\times10^9 L_\odot$, and in the range $-1.7 < [$Fe$/$H$] < -1.1$ is $0.11\times10^9 L_\odot$. The simplest method of estimating the substructure fraction is simply to add the luminosities, calculate the fraction in the smooth halo, and take the complement. This returns a value of $26.9\%$ in the substructured component. However, as noted by \citet{ibata:14}, it is not strictly correct to add luminosities across separate metallicity intervals in their Table $5$, because different substructure masks are used per interval. As an alternative, we note that the substructure fraction is $11.1\%$ for $-2.5 < [$Fe$/$H$] < -1.7$, and $35.3\%$ for $-1.7 < [$Fe$/$H$] < -1.1$. Taking the mean, weighted by the total luminosities listed in Table $4$ (for which the values {\it are} comparable across metallicity intervals) returns an overall fraction of $27.0\%$. Hence it seems that directly adding across the metallicity intervals in Table $5$ is an acceptable approximation for these two metal-poor bins. This is consistent with Figure $10$ in \citet{ibata:14}, which shows that the substructure masks for the two bins are in fact very similar. Lastly, we note that \citet{ibata:14} also provide luminosity estimates for a stellar age of $9$ Gyr. Repeating our calculation under this assumption returns a substructure fraction of $31.0\%$.} that just under $30\%$ is in the substructured component across the range $-2.5 < [$Fe$/$H$] < -1.1$.  This is slightly lower than our minimum globular cluster fraction.

It is instructive to calculate the specific frequency of the substructured population of globular clusters. This parameter, first introduced by \citet{harris:81}, is commonly used to connect the total luminosity $M^T_V$ of a host galaxy with the number of globular clusters $N_{GC}$ that it hosts:
\begin{equation}
S_N = N_{GC} \times 10^{0.4(M^T_V + 15)}
\label{eq:sn}
\end{equation}
The distribution of specific frequency with host galaxy luminosity exhibits a characteristic U-shape, with $S_N\approx 1$ and very little scatter at $M^T_V \sim -18$, and much higher mean values and larger scatter at the bright ($M^T_V \la -21$) and faint ($M^T_V \ga -15$) extremes \citep[e.g.,][]{miller:07,peng:08,georgiev:10,harris:13,beasley:16,lim:18}. In particular, the specific frequencies for dwarf spheroidal and nucleated dwarf elliptical galaxies with $M^T_V\sim-13$ can be as high as $\approx 10-30$ (this is seen locally for the Fornax dwarf which, with $N_{GC}=5$ and $M^T_V\sim-13.4$, has $S_N\approx22$).

Based on the luminosites taken from Tables $4$ and $5$ of \citet{ibata:14} across the range $-2.5 < [$Fe$/$H$] < -1.1$, as detailed above, we infer $M^T_V=-14.7$ for the metal-poor substructure component of the outer M31 halo\footnote{Note that this value is slightly different from the integrated magnitudes listed by \citet{ibata:14}. Maintaining consistency with our previous calculation, we take the total $13$ Gyr luminosity for the substructure component to be $0.07\times10^9 L_\odot$ and assume an absolute solar magnitude $M^\odot_V = 4.83$.}. This implies $S_N=42$ for our group of $32$ robust substructure clusters, extending to a maximum $S_N=75$ if {\it all} $25$ ``ambiguous'' clusters are included. These values are a factor of $\sim2-5$ higher than observed for typical nearby dwarf spheroidal systems.

An alternative way of viewing this problem is by considering the individual globular cluster populations of the several most luminous metal-poor halo substructures, using the memberships assigned in Appendix \ref{ss:class} and the luminosities compiled by \citet{mcconnachie:18}: the North-West Stream has $M^T_V=-12.3$ and $6-7$ globular clusters such that $S_N=70-85$; the South-West Cloud has $M^T_V=-11.3$ and $3-5$ clusters such that $S_N=90-150$; the East Cloud has $M^T_V=-10.7$ and $2-3$ clusters such that $S_N=105-155$; and Streams C and D, added together, have $M^T_V=-13.6$ and $11-15$ clusters such that $S_N=40-55$. 

How can we reconcile these specific frequencies with the much lower values measured for the type of dwarf galaxies usually assumed to be the progenitors of the M31 halo streams? This issue has previously been considered in the context of the East Cloud and the South-West Cloud by \citet{mcmonigal:16a}, who noted that the estimated mean metallicities for these substructures, at $[$Fe$/$H$]\approx-1.3$, imply much higher progenitor luminosities $M^T_V\approx-13.5$ according to the luminosity-metallicity relation \citep{kirby:11}. In this case the observed high specific frequencies would reflect the almost complete destruction of these systems, with only $\sim15\%$ of their stellar content now located in large halo substructures. The original specific frequencies would have been $\sim10-30$, consistent with dwarf spheroidals observed in the local Universe at the present day.

A significant problem with this picture is that it is difficult to hide the missing portions of the destroyed progenitors. It seems unlikely that the debris is scattered in many low-luminosity substructures with surface brightnesses falling below the PAndAS faint detection limit -- i.e., the apparently smooth halo -- because (i) the amount of material required to account for the high cluster specific frequency is larger than the total luminosity in the smooth halo component at metallicities $-2.5 < [$Fe$/$H$] < -1.1$ measured by \citet{ibata:14}; (ii) the smooth halo is typically more metal-poor than the substructured component at given galactocentric radius \citet{ibata:14}; and (iii) as we discuss below, the smooth halo component may well also possess its {\it own} complement of globular clusters that would serve to keep the overall specific frequency high. It is also difficult to argue that the missing material now resides within $R_{\rm proj} = 25$\ kpc, as this would require rather efficient separation of clusters and field stars during the accretion process to produce the observed specific frequencies.

It is relevant that M31 possesses a number of other peculiar characteristics that have recently led to a number of authors advancing ``major merger'' scenarios, in which a single dominant accretion involving a galaxy with stellar mass $M_\star \approx 2.5\times10^{10}M_\odot$ (i.e., a $\sim 4$:$1$ merger event) occurred within the last few ($\la 5$) Gyr \citep[e.g.,][]{hammer:18,dsouza:18}. Modelling by these authors has shown that such an event is likely to be preferred cosmologically, and can apparently account for various observed properties of the M31 disk, the high metallicity and complexity of the inner halo, the overall halo profile, the Giant Stream, the similar metallicities of the outer halo streams, and perhaps even the existence of M32.

Returning to the M31 globular cluster system, the ``excess clusters'' problem outlined above could find a natural resolution if most, or all, of the significant remote-halo substructures are due to a single luminous progenitor \citep[e.g.,][]{mcconnachie:18}\footnote{Notably, both \citet{veljanoski:14} and \citet{ferguson:16} have also articulated the idea that one or two large accretion events might explain the coherent rotation signal measured for the outer halo cluster system \citep[see][]{veljanoski:13a,veljanoski:14}.}. A merging galaxy with $M_\star \approx 2.5\times10^{10}M_\odot$ has $M^T_V\sim -20.5$ and could easily accommodate the entire accreted outer halo globular cluster population in M31: the $32-57$ objects identified here would imply $S_N\approx0.2-0.4$, whereas the typical specific frequency for galaxies of this luminosity is $S_N\approx 1$. Indeed, it would be expected that such a progenitor {\it also} contributed significantly to the inner halo cluster population in M31, as implied by our arguments regarding the Giant Stream in Section \ref{ss:discuss0} above. Recent modelling by \citet{hughes:19} has shown that the ages and metallicites of globular clusters associated with halo streams correlate with the mass and infall time of their progenitor systems. Detailed information on the substructure clusters identified here therefore provides an important avenue for testing the feasibility of the major merger scenario in future.

\subsubsection{Typical properties of substructure clusters}
Our analysis in Section \ref{s:properties} revealed that substructure clusters are (i) generally quite compact with $65\%$ having $r_h < 10$ pc, but with the distribution exhibiting a tail extending to at least three times this size; and (ii) mostly blue (metal-poor) with $75\%$ having $0.8 < (V-I)_0 < 0.95$, but with the distribution exhibiting a tail extending to quite red colours $(V-I)_0 \sim 1.2$ that is inferred to be comprised of clusters with metallicities as high as $[$Fe$/$H$] \sim -0.9$. Colour-magnitude diagrams have been published for a handful of substructure members. These are HEC12 \citep[EC4 in][]{mackey:06}, H24 \citep[GC9 in][]{mackey:07}, PA-07 and PA-08 \citep{mackey:13a}, PA-56 \citep{sakari:15}, and PA-57 and PA-58 \citep{mcmonigal:16a}. While most are clearly typical metal-poor globular clusters, a few have features indicative of objects that are several Gyr younger than the oldest clusters seen in the Milky Way \citep[these are PA-07, PA-08, and PA-58 -- see][]{mackey:13a,mcmonigal:16a}. 

The presence of both a somewhat metal-enhanced subset and a somewhat younger subset of clusters in the substructure class (apparently with significant overlap) is consistent with the accretion at late times of progenitors that had the chance to undergo extended star and cluster formation including significant chemical evolution. Indeed, \citet{hughes:19} have demonstrated that higher metallicities and younger ages are a generic property of more-recently accreted clusters (especially those originating in higher-mass satellites). A specific local example is provided by the Sagittarius dwarf, which is currently disintegrating in the Milky Way's halo. Sagittarius possesses several younger metal-rich cluster members (e.g., Terzan 7, Palomar 12, and Whiting 1), and indeed its cluster system appears to exhibit a strikingly different age-metallicity distribution than the bulk of the Galactic globular cluster population \citep[e.g.,][]{marinfranch:09,forbes:10,dotter:11,leaman:13}.

The double-peaked luminosity function is a striking characteristic of the remote-halo globular cluster population in M31 that has, to date, received surprisingly little attention. The origin of the bimodal shape is unknown; however, the fact that it is clearly seen for the substructure clusters is particularly interesting because these objects are unambiguously known to have formed in other systems. It is notable that the luminosity functions shown by \citet{mackey:05} for the supposedly-accreted ``young halo'' globular clusters in the Milky Way, and for the globular cluster populations of the four largest dwarf satellites of the Milky Way (Fornax, Sagittarius, and the Large and Small Magellanic Clouds), are also plausibly bimodal, with $\approx 25\%$ of both distributions sitting at magnitudes fainter than $M_V \sim -6$. The remote M31 clusters populating the faint peak of the luminosity distribution are generally rather diffuse with a median $r_h\sim15$\ pc \citep{huxor:14}, in stark contrast to the much more compact sizes ($r_h\sim 3$\ pc) seen for the clusters populating the brighter peak. The bimodal luminosity function could therefore be a consequence of an ``extended'' mode of cluster formation that has been suggested to occur preferentially in the relatively benign tidal environments found in lower-mass galaxies as compared to larger galaxies \citep[e.g.,][]{elmegreen:08,dacosta:09}.

\subsection{The non-substructured cluster population}
\label{ss:discuss2}
As a result of our classification efforts, we were also able to identify a subset of clusters that exhibit no persuasive evidence for association with halo substructure -- that is, they do not have local densities of field stars in the top quartile observed at a commensurate radius, they do not possess a velocity similar to that seen for any nearby substructure, and they are not a member of an obvious kinematic group. We found that $\approx 38\%$ of the remote cluster system ($35$ objects) fall into this category. While it is not a great leap to link the clusters in the ``substructure'' class with the relatively recent accretion of one or more dwarf galaxies into the M31 halo, the nature of what we have called the ``non-substructure'' group is less immediately obvious. For example, this subset could easily include clusters that should have been assigned to the substructure class, but are associated with features that happen to be fainter than the PAndAS detection limit. 

\subsubsection{Comparison with the metal-poor smooth halo}
A good starting point in our examination of the non-substructure group is its radial surface density profile. As noted in Section \ref{ss:rpsubsys}, not only does this profile follow a completely featureless power-law, it possesses an essentially {\it identical} slope to that observed by \citet{ibata:14} for the apparently smooth component of the metal-poor field halo. Quantitatively, for the non-substructure clusters we measure a power-law index of $\Gamma = -2.15 \pm 0.05$, while \citet{ibata:14} obtained $\Gamma = -2.08 \pm 0.02$ and $\Gamma = -2.13 \pm 0.02$ for their substructure-masked populations with $-2.5 < [$Fe$/$H$] < -1.7$ and $-1.7 < [$Fe$/$H$] < -1.1$, respectively. 

Given how closely the non-substructure group of clusters appears to mirror the properties of the metal-poor smooth halo, it is strongly tempting to link the two. After all, both sets consist of what remains after all identifiable traces of substructure have been removed, and there is no reason to expect, {\it a priori}, that the residual cluster population should possess any particular power-law slope, nor, even, that its profile need be particularly smooth. 

\citet{ibata:14} provide extensive discussion on the nature of the metal-poor smooth halo component. They identify it with the hot kinematic component detected spectroscopically by \citet{gilbert:12}, and, while noting that it could plausibly consist of many extremely low surface brightness structures that are presently undetectable, conclude that even within the limitations imposed by PAndAS the observed degree of spatial homogeneity means that it was most likely built up at very early times from a large number of low-luminosity satellites \citep[cf.][]{johnston:08}. Other studies have shown that stars, and presumably clusters, that were formed {\it in situ} and then scattered out of the disk due to merger activity can also provide a substantial contribution to the smooth halo \citep[e.g.,][]{zolotov:09,font:11,kruijssen:15}. However, this component is thought to be almost entirely confined to the inner $\approx 20-30$\ kpc, and is hence not expected to be important over the radial range considered here.

{\it If} the clusters in our non-substructure group are indeed part of the remote metal-poor smooth halo in M31, then they provide additional insight regarding its origin. Once again, the specific frequency of the population is instructive. From Table $5$ in \citet{ibata:14}, the total luminosity of the smooth halo with $-2.5 < [$Fe$/$H$] < -1.1$ is $\approx 0.19\times10^9 L_\odot$, corresponding to $M^T_V=-15.8$. This would imply a specific frequency $S_N\sim19$ for the associated globular cluster system. Following \citet{cole:17} and \citet{mcconnachie:18}, the lowest-luminosity Local Group dwarfs hosting multiple globular clusters are Fornax ($M^T_V = -13.4$) and Sagittarius ($M^T_V \approx -13.5$)\footnote{Although Sagittarius might initially have been as luminous as $M^T_V \sim -15$ \citep{ostholt:12}}; significant cluster systems are also found in NGC 147 ($M^T_V = -14.6$) and 185 ($M^T_V = -14.8$) \citep[e.g.,][]{veljanoski:13b}, and NGC 6822 ($M^T_V = -15.2$) \citep[e.g.,][]{veljanoski:15}. It would require only $\sim5-6$ systems of luminosity comparable to Fornax, or $\sim 2$ of luminosity comparable to NGC 147 and 185, to build the entire metal-poor smooth halo measured by \citet{ibata:14}\footnote{Note that this is not a strictly fair comparison as a substantial fraction of the luminosity in all of these galaxies represents more metal-rich populations due to their extended star-formation histories.}. However, it may then be difficult to explain the observed lack of substructure even if the accretions occurred at early times.

At lower luminosities the evidence for Local Group dwarfs hosting globular clusters is more ambiguous, but single faint clusters appear to be present in each of Andromeda I \citep[$M^T_V=-11.7$;][]{caldwell:17}, the Pegasus dIrr \citep[$M^T_V=-12.2$;][]{cole:17}, Eridanus II \citep[$M^T_V=-7.1$;][]{koposov:15,crnojevic:16}, and Andromeda XXV \citep[$M^T_V=-9.7$;][]{cusano:16}. Further afield, more examples of faint dwarfs hosting multiple and/or luminous globular clusters are known \citep[e.g.,][]{dacosta:09,georgiev:09}. Nonetheless, such systems are rare; if the metal-poor smooth halo in M31 was indeed built up from a large number of low-luminosity accretions then it appears likely that the globular clusters associated with this component arrived with only a small fraction of the progenitor systems. 

It is worth emphasising that in this scenario, essentially the {\it entire} M31 globular cluster system outside $25$\ kpc is comprised of accreted objects. This observation is qualitatively consistent with the predictions of various models for the formation and assembly of globular cluster systems in large galaxies like the Milky Way and Andromeda \citep[e.g.,][]{prieto:08,muratov:10,griffen:10,renaud:17,pfeffer:18,kruijssen:18}. Moreover, recent work by \citet{andersson:18} has shown that accretion from dwarf galaxies is able to populate even the outermost regions of an M31-like halo ($R\sim200$\ kpc) with globular clusters.

\subsubsection{Typical properties of non-substructure clusters}
Our analysis in Section \ref{s:properties} showed that, in general, the typical properties of clusters in the non-substructure group differ only mildly from those observed for the substructure group. The luminosity function is again strongly double-peaked, and the distribution of sizes includes $\approx 35\%$ of objects with half-light radii larger than $10$\ pc. If the double-peaked luminosity function and the presence of diffuse low-luminosity clusters are good indicators of an {\it ex situ} origin as inferred for the substructure group, then this constitutes additional evidence that the non-substructure members also formed in now-destroyed dwarfs. The integrated colours for the non-substructure group are almost exclusively ($\approx 95\%$) bluer than $(V-I)_0 = 1.0$, consistent with the notion that the host systems were chemically primitive when they were accreted, presumably because the accretions occurred at early times\footnote{The one very red object in the non-substructure group is PA-17. This cluster was observed spectroscopically at high resolution by \citet{sakari:15}, who found that its abundance patterns were indicative of formation in an LMC-like progenitor. It is therefore likely that PA-17 is a misclassified object that would belong more naturally in the substructure class (but does not obviously exhibit any of that class's defining characteristics).}. 

One key difference between the substructure and non-substructure groups identified in Section \ref{s:properties} is the location of the main (classical) peak in the luminosity function, which sits nearly a magnitude brighter for the non-substructure clusters than for the substructure objects. Since the shape of the luminosity function for ancient clusters is generally assumed to be the result of various physical effects that lead to mass loss and cluster disruption, this discrepancy may offer insight into either (i) the different types of progenitor systems that originally hosted the substructure and non-substructure clusters, or (ii) the different durations over which various erosive processes (particularly those induced externally, such as gravitational shocks) were important. It is also possible that the brighter peak in the non-substructure group could be due to the presence of a higher number of stripped nuclear star clusters, which typically appear more luminous than would be inferred given standard evolutionary processes \citep[e.g.,][]{kruijssen:12}. 

\section{Summary \& Conclusions}
\label{s:summary}
In this paper we have investigated the links between the globular cluster system and the field halo in M31 at projected radii $R_{\rm proj} = 25-150$\ kpc, utilising the final point-source catalogue from the Pan-Andromeda Archaeological survey \citep{mcconnachie:18} together with our essentially complete census of the cluster population in this region. This represents the first global, quantitative such study in an L$^*$ galaxy. Our main results are as follows:
\begin{enumerate}
\item{We identify $92$ globular clusters spanning the range $25 \le R_{\rm proj}  \la 150$\ kpc in M31. This is a factor of $\approx 7$ higher than the number known in the Milky Way over a roughly commensurate region (i.e., $30 \la R \la 190$\ kpc)\footnote{Here we assume that on average the deprojected radius $R = 4/\pi \times R_{\rm proj}$.}.  Clusters are found to the very edge of the PAndAS footprint, suggesting that the system likely extends to even larger radii. This notion is reinforced by the fact that several M31 clusters are known to have 3D galactocentric distances $R\sim 200$\ kpc \citep[e.g.,][]{mackey:10a,mackey:13b}.\vspace{1mm}}
\item{The radial density profile for M31 globular clusters exhibits a large bump at radii $R_{\rm proj} \approx 30-50$\ kpc but otherwise declines as a power-law with index $\Gamma = -2.37 \pm 0.17$ over the range $25-150$\ kpc, or $\Gamma = -2.15 \pm 0.10$ if the bump is excluded. This is similar to the behaviour of the {\it metal-poor field halo} ($[$Fe$/$H$] < -1.1$) in Andromeda as observed by \citet{ibata:14}, indicating that the globular clusters outside $25$\ kpc in M31 are overwhelmingly associated with this component even though, with $M^T_V \approx -16.3$, it contains only the minority fraction of the total halo luminosity ($\sim 15-30\%$ depending on the assumed age).\vspace{1mm}}
\item{By mapping the spatial density of metal-poor giants together with the positions of the globular clusters, we qualitatively confirm the apparent association between clusters and stellar substructures at $R_{\rm proj} \ge 25$\ kpc first noted in the south and west by \citet{mackey:10b}. In contrast, the metal-rich map ($[$Fe$/$H$] > -1.1$), which is dominated by debris from the accretion event that produced the Giant Stellar Stream, exhibits no such association. This is likely due to the particular circumstances of the event itself, rather than reflecting a general property of L$^*$ galaxy halos. Inside $R_{\rm proj} \approx 25$\ kpc the complexity of the field is so great that, regardless of metallicity, it is impossible to draw any robust links between clusters and substructures using spatial information only.\vspace{1mm}}
\item{By calculating the surface density of metal-poor halo stars in the vicinity of each remote globular cluster and comparing to the azimuthal distribution at an equivalent radius, we show that the positions of clusters correlate with overdensities in the stellar halo at greater than $99.95\%$ significance. That is, many more clusters exhibit high local densities of metal-poor stars than would be expected if the positions of the clusters and field substrcutures were completely decoupled -- nearly half of clusters with $R_{\rm proj} > 25$\ kpc have local densities in the top quartile of the observed distribution, while one-quarter have local densities in the top decile. On the other hand, a similar calculation for the metal-rich halo indicates no statistical preference for such an association.\vspace{1mm}}
\item{We utilise the calculated local densities of metal-poor halo stars, in combination with previously-measured radial velocities, to identify two cluster subsets: one containing objects that are robustly associated with halo substructures, and one containing objects that exhibit no evidence for any association. These groups are labelled as ``substructure'' and ``non-substructure'', respectively. A third class, ``ambiguous'', indicates clusters for which there is weak and/or conflicting evidence for a substructure association. All classifications are listed in Table \ref{t:fulldata}.\vspace{1mm}}
\item{We find that at a minimum, $\approx 35\%$ of remote clusters fall into the ``substructure'' category; however, the fraction could be as high as $\sim 60\%$ if all of the ``ambiguous'' clusters are also assumed to be substructure objects. It is straightforward to see that these clusters have arrived in the halo of M31 via the relatively recent accretion and destruction of their parent dwarfs. The radial density profile for this set exhibits large point-to-point scatter and has a power-law decline of index $\Gamma = -2.32 \pm 0.44$, very similar to the unmasked metal-poor halo profiles measured by \citet{ibata:14}.\vspace{1mm}}
\item{The substructure clusters are generally metal-poor ($[$Fe$/$H$] \la -1$) although their integrated colours extend to quite red values, likely indicating that $\sim 15\%$ are metal-richer and/or younger objects. This suggests that the host system(s) had time to undergo extended star formation and chemical enrichment, consistent with the idea that they were accreted at late times \citep[cf.][]{hughes:19}. The luminosity function for this cluster subset is strikingly bimodal, which may reflect the origin of these clusters in a relatively benign tidal environment. Puzzlingly, the specific frequency of substructure clusters relative to the metal-poor field overdensities is substantially higher than typically observed for present-day cluster-hosting dwarfs in the Local Group. A scenario where M31 underwent a major merger in the last few Gyr, as advocated by several groups \citep[e.g.,][]{hammer:18,dsouza:18}, may help explain this observation.\vspace{1mm}}
\item{The non-substructure clusters comprise at least $\approx 40\%$ of the remote halo population. Their radial surface density profile is markedly featureless and has a power-law index $\Gamma = -2.15\pm0.05$, precisely matching the profiles observed by \citet{ibata:14} for slices of the metal-poor {\it smooth} halo. We speculate that the non-substructure objects could be linked to this smooth halo component, which has $M^T_V\sim-15.8$. If so, then their properties (uniformly metal-poor, double-peaked luminosity function) are consistent with an origin in primitive dwarfs that were accreted into the M31 halo at very early times \citep[$\sim12$ Gyr ago -- e.g.,][]{johnston:08}. Most low luminosity dwarfs seen at the present day do not host clusters; hence, if, as suggested by \citet{ibata:14}, the smooth halo was formed by the destruction of many low-luminosity systems, then perhaps only a relatively small fraction donated globular clusters.}
\end{enumerate}
It is therefore plausible that the {\it entire} M31 globular cluster system outside $25$\ kpc has been accumulated via the accretion of cluster-bearing dwarf satellites over a Hubble time of growth. Precise measurements of the properties of these clusters -- in particular their metal abundances, ages, and line-of-sight distances -- hold the enticing prospect of helping quantitatively unravel the assembly history of Andromeda.

\section*{Acknowledgments}
ADM holds an Australian Research Council (ARC) Future Fellowship (FT160100206), and acknowledges support from ARC Discovery Project DP150103294. We would like to thank the anonymous referee for their thorough reading of this manuscript and for their helpful suggestions for improvement.

This paper is based on observations obtained with MegaPrime/MegaCam, a joint project of CFHT and CEA/DAPNIA, at the Canada-France-Hawaii Telescope (CFHT) which is operated by the National Research Council (NRC) of Canada, the Institute National des Sciences de l'Univers of the Centre National de la Recherche Scientifique of France, and the University of Hawaii.

This paper is also based in part on observations obtained at the Gemini Observatory, which is operated by the Association of Universities for Research in Astronomy, Inc., under a cooperative agreement with the NSF on behalf of the Gemini partnership: the National Science Foundation (United States), the National Research Council (Canada), CONICYT (Chile), Ministerio de Ciencia, Tecnolog\'{i}a e Innovaci\'{o}n Productiva (Argentina), and Minist\'{e}rio da Ci\^{e}ncia, Tecnologia e Inova\c{c}\~{a}o (Brazil). These observations were obtained via programmes GN-2014B-Q-26 and GN-2015B-Q-17.

\appendix

\section{Gemini observations of globular cluster candidates in the M31 outer halo}
\label{a:gemini}
This Appendix provides details of our efforts to classify candidate globular clusters in the outer halo of M31 with a view to obtaining as complete a catalogue as possible for the present paper. 

At the same time as we were undertaking our PAndAS search, \citet{dtz:13,dtz:14} identified a total of $100$ sources in the Sloan Digital Sky Survey (SDSS) possessing properties consistent with those expected for M31 globular clusters. Of these, $81$ fall at $R_{\rm proj} > 25$ kpc. In \citet{huxor:14} we confirmed that $11$ of the $93$ objects listed by \citet{dtz:13} are indeed globular clusters: $8$ of these appeared independently in our PAndAS catalogue, while $3$ fell inside our minimum search radius of $R_{\rm proj} = 25$\ kpc. We also ruled out $48$ of the \citet{dtz:13} candidates as background galaxies. The remaining $34$ objects from \citet{dtz:13} could not be classified -- either they were at very large galactocentric radii ($\sim 150 < R_{\rm proj} < 230$\ kpc) and hence beyond the edge of the PAndAS footprint, or they fell in small gaps in the PAndAS mosaic. \citet{dtz:14}, which was not published at the time we wrote \citet{huxor:14}, presented an additional $6$ high confidence clusters and one candidate.  Three of the objects (SDSS-A, SDSS-B, and SDSS-C) appear independently in our PAndAS catalogue (as PA-14, PA-17, and PA-21, respectively) and are certain globular clusters, while the candidate (SDSS-F) is unambiguously a background galaxy in the PAndAS imaging.

More recently, \citet{dtz:15} extended their search to the entire SDSS footprint with an improved cluster selection methodology, in order to identify possible intergalactic globular clusters in the Local Group. This included a complete reanalysis of their previous two M31 catalogues, resulting in a final sample of $22$ M31 globular clusters including $12$ at radii beyond $R_{\rm proj} = 25$ kpc\footnote{In total, this list is comprised of the $14$ objects discussed in \citet{huxor:14}, an additional three from \citet{dtz:14}, and five new discoveries at $R_{\rm proj} < 25$\ kpc (where we did not search with PAndAS).}. Notably this work ruled out almost all of the remaining remote candidates from \citet{dtz:13,dtz:14} that we had been unable to assess in \citet{huxor:14} -- the final list of \citet{dtz:15} contains only two objects at $R_{\rm proj} > 25$\ kpc that were not also in our catalogue. 

Prior to the publication of \citet{dtz:15}, we obtained short-exposure ``snapshot'' images of $16$ candidate clusters with the Gemini Multi-Object Spectrograph (GMOS) at the Gemini North telescope on Mauna Kea, Hawaii. These observations were initially aimed at improving the classification of objects identified as possible M31 clusters in \citet{huxor:14} and \citet{dtz:13,dtz:14} so as to maximise the completeness of our outer halo catalogue, but also ultimately serve as an independent check of the reanalysis by \citet{dtz:15}. Our list of targets can be seen in Table \ref{t:snapshot} and includes all but two unclassified candidates from \citet{dtz:13,dtz:14} with $25 < R_{\rm proj} \le 150$\ kpc, plus the strongest SDSS candidates at larger radii. The observations were carried out in queue mode as part of programmes GN-2014B-Q-26 and GN-2015B-Q-17 (PI: Mackey).  Twelve of the objects were imaged on 2014 July 18, three (PA-Cand-2, SDSS-C70, and dTZZ-21) were observed on 2014 July 29, and the remaining one (PA-55) on 2015 July 28.  All data were collected during clear conditions and under good seeing ($0.5\arcsec - 0.7\arcsec$).

\begin{table}
\centering
\caption{Classifications from GMOS imaging for the $16$ candidate clusters.}
\begin{tabular}{@{}lccccc}
\hline \hline
Name & Alt. & \multicolumn{2}{c}{Position (J2000.0)} & $R_{\rm proj}$ & Type \\
 & Name & RA & Dec & (kpc) & \\
\hline
SDSS-C1     & $...$ & $00\,00\,37.9$ & $+32\,25\,04$ & $166.7$ & Galaxy \\
dTZZ-05     & SDSS-D & $00\,36\,08.6$ & $+39\,17\,30$ & $32.0$ & Cluster \\
SDSS-C12   & $...$ & $00\,37\,37.5$ & $+25\,08\,45$ & $220.8$ & Galaxy \\
PA-Cand-1 & $...$ & $00\,44\,58.4$ & $+40\,21\,38$ & $13.7$ & Galaxy \\
SDSS-C29   & $...$ & $00\,47\,32.6$ & $+28\,03\,56$ & $180.9$ & Galaxy \\
SDSS-C43   & $...$ & $01\,03\,15.6$ & $+29\,28\,34$ & $170.9$ & Galaxy \\
SDSS-C46   & $...$ & $01\,06\,06.6$ & $+27\,45\,30$ & $195.9$ & Galaxy \\
SDSS-C49   & $...$ & $01\,07\,24.8$ & $+29\,07\,34$ & $179.6$ & Galaxy \\
PA-Cand-2 & $...$ & $01\,07\,53.9$ & $+48\,22\,42$ & $114.6$ & Galaxy \\
SDSS-C57   & $...$ & $01\,18\,09.8$ & $+29\,14\,09$ & $191.6$ & Galaxy \\
PAndAS-55 & $...$ & $01\,19\,20.4$ & $+46\,03\,12$ & $111.5$ & Galaxy \\
dTZZ-21     & SDSS-G & $01\,28\,49.2$ & $+47\,04\,22$ & $137.8$ & Cluster \\
SDSS-C70   & $...$ & $01\,36\,14.1$ & $+45\,37\,35$ & $145.2$ & Galaxy \\
SDSS-17     & $...$ & $23\,41\,50.0$ & $+44\,50\,07$ & $159.4$ & Galaxy \\
SDSS-C74   & $...$ & $23\,46\,49.9$ & $+45\,14\,50$ & $149.1$ & Galaxy \\
SDSS-C75   & $...$ & $23\,48\,40.9$ & $+39\,37\,45$ & $142.2$ & Galaxy \\
\hline
\label{t:snapshot}
\end{tabular}
\end{table}

\begin{figure*}
\begin{center}
\includegraphics[width=110mm,clip]{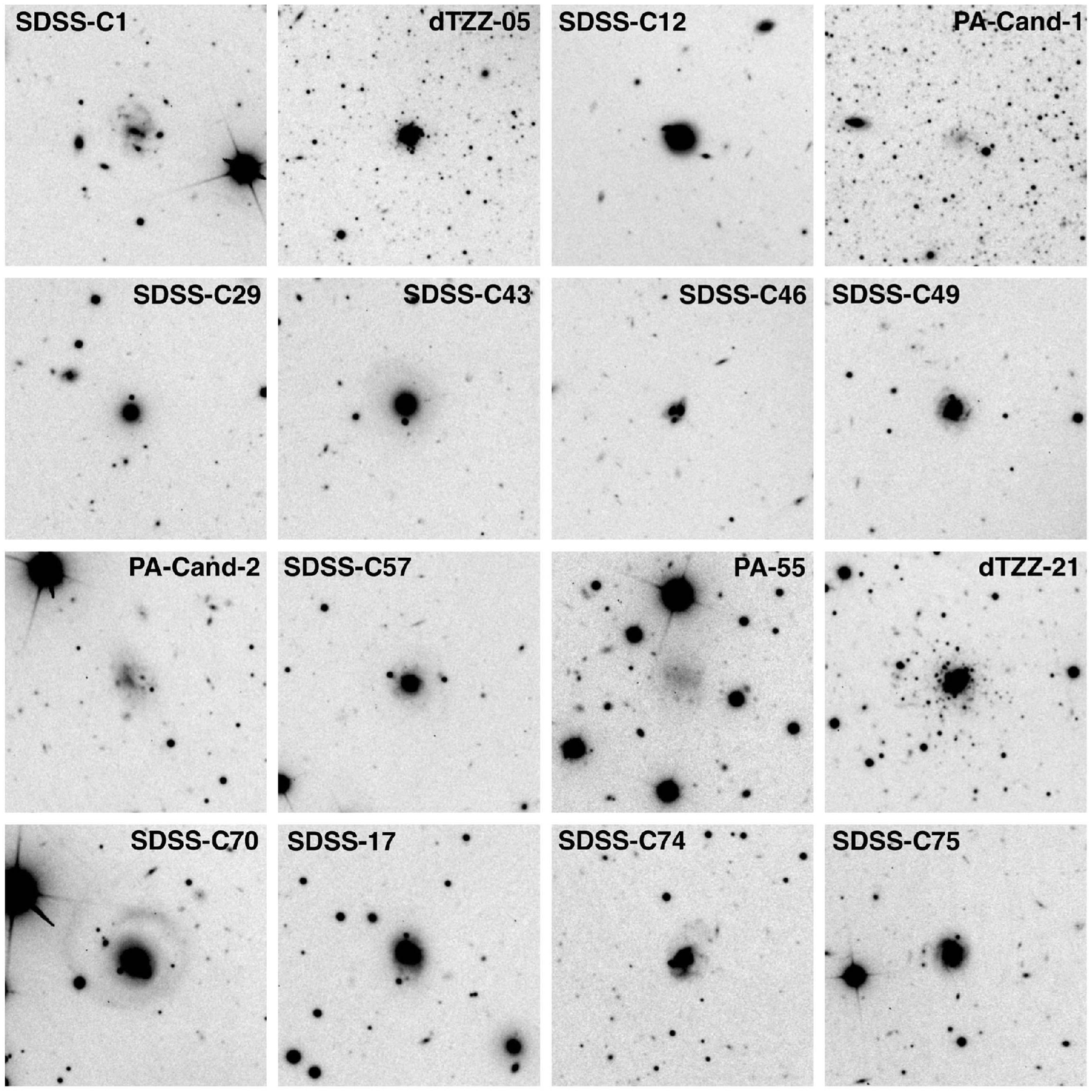}
\caption{GMOS $i^\prime$-band images of the $16$ targets listed in Table \ref{t:snapshot}. Each thumbnail is $1\arcmin \times 1\arcmin$ and oriented such that north is up and east is to the left. It is evident that only two of these objects are genuine globular clusters (dTZZ-05 and dTZZ-21); the remainder are background galaxies.}
\label{f:snapshot}
\end{center}
\end{figure*}

For a given object we obtained two frames of exposure duration $145$s each, with a $5\arcsec$ dither to fill in the GMOS inter-CCD gaps. All imaging was conducted through the GMOS $i^\prime$ filter.  We reduced the data using standard procedures in the GMOS software package in {\sc iraf}.  Bias and flat-field images were applied with the {\sc gireduce} task, the three CCD frames in a given exposure were mosaicked into a single frame with {\sc gmosaic}, and then the two frames for a given object were stacked together using {\sc imcoadd}.  

Images of our $16$ targets, taken as $1\arcmin \times 1\arcmin$ cut-outs from the final reduced GMOS frames, are shown in Figure \ref{f:snapshot}.  It is evident from simple visual inspection that $14$ of the objects are background galaxies. This includes the two low confidence candidates identified in \citet{huxor:14}, as well as PA-55 (which exhibits a misleading morphology in the PAndAS imaging because it fell partially in a MegaCam chip gap). However, two of the GMOS targets are {\it bona fide} globular clusters -- these are dTZZ-05 ($=\,$SDSS-D) and dTZZ-21 ($=\,$SDSS-G), both initially identified by \citet{dtz:14} and then reconfirmed by \citet{dtz:15}.  dTZZ-21 is of particular interest, sitting at the second-largest projected galactocentric radius ($R_{\rm proj} = 137.8$\ kpc) of all known clusters in M31. Its existence provides further evidence of the huge spatial extent of the M31 globular cluster system, and suggests that a handful more members are likely to be found beyond the edge of the PAndAS footprint (see also Section \ref{ss:profile}). Overall, our Gemini imaging is entirely consistent with the reanalysis of \citet{dtz:15}, and reinforces their conclusion of that none of the remaining unclassified remote candidates listed by \citet{dtz:13,dtz:14} are likely to be M31 globular clusters \citep[see also][]{mackey:16}.

\section{Quantifying the cluster-substructure correlation in the M31 outer halo: detailed methodology}
\label{a:method33}
This Appendix contains a complete description of the procedure employed in Section \ref{ss:correlation}. We begin by determining the surface density of M31 halo stars in the vicinity each of the $92$ globular clusters with $R_{\rm proj} > 25$\ kpc. To do this we consider a circular aperture of radius $r = 10\arcmin$ around each target.  Selecting an appropriate size for this region is a trade-off between remaining truly ``local'' to a given globular cluster, and ensuring a sufficient number of halo stars are present to avoid large random fluctuations in the measured densities. The latter issue is of particular concern at galactocentric distances beyond $R_{\rm proj} \approx 50$\ kpc, where the natural decline in halo density means that stars in the field can be very sparsely distributed.  A degree of experimentation revealed that aperture radii in the range $5\arcmin \la r \la 15\arcmin$ were acceptable.  Anything smaller than $r \sim 5\arcmin$ resulted in many cases where too few halo stars were present, while for radii larger than $r \sim 15\arcmin$ the effective smoothing length was too large, leading to a noticeable loss of resolution (cf. Figure \ref{f:maps} where the maps were smoothed with a Gaussian kernel of $\sigma = 2.5\arcmin$).

At the assumed distance of M31 our preferred aperture radius $r = 10\arcmin$ corresponds to a physical radius of roughly $2.3$\ kpc. This matches quite well the typical sizes of many of the dwarf spheroidal satellite galaxies of M31\footnote{That is, several times the measured half-light radii for these systems \citep[see, e.g.,][]{mcconnachie:12,martin:16}.}, as well as the widths of several of the narrower halo streams (e.g., the North-West Stream, the East Cloud, Stream D, and Stream Cp).  Although we adopt this selection radius throughout the following calculations, we note that our conclusions hold irrespective of the value chosen within $5\arcmin \la r \la 15\arcmin$.

We used the number of stars lying in a given aperture to determine the local density. For the first phase of our analysis we included only stars with $-2.5 \la [$Fe$/$H$] \la -1.1$ -- that is, those falling within the CMD box labelled ``MP'' in Figure \ref{f:cmd}.  Since we are interested in the possible correlation between clusters and overdensities of stars in the M31 field halo, it is important to ensure our density calculations are not biased by other populations that are present in the PAndAS point source catalogue.  These include stars belonging to the globular clusters themselves, stars belonging to the bound dwarf satellites of M31 (of which there are now in excess of 30 known systems), point sources residing in large background galaxies (often their own globular cluster systems), unresolved background galaxies, and foreground members of the Milky Way.

To exclude any stars belonging to the target M31 globular clusters, many of which are partially resolved in the PAndAS imaging, we excised a small circular area of radius $r_{\rm cen} = 53\arcsec$ from the centre of each aperture.  This corresponds to a physical radius $r_{\rm cen} = 200$\ pc, which is significantly larger than the tidal radii for all M31 globular clusters for which such measurements exist from high-resolution space-based imaging \citep[e.g.,][]{barmby:07,tanvir:12}.  The only M31 cluster known to extend beyond this size is  MGC1 \citep{martin:06}, and even in this extreme case the {\it vast} majority of cluster members lie within $200$\ pc of its centre \citep{mackey:10a}.  A few globular clusters in our sample lie closer to each other than $r = 10\arcmin$ on the sky; in these cases we also excised the portion of the circle of radius $r_{\rm cen} = 53\arcsec$ belonging to the neighbouring system that overlapped with the aperture for the target cluster, and then corrected the area of the aperture accordingly. 

To make sure stars belonging to M31 dwarf galaxies were not included in the density calculations we cross-matched a list of these systems against our globular cluster catalogue to identify any cases where the $10\arcmin$ aperture for a given target encroached on an area of radius $3r_h$ about an M31 dwarf.  Here $r_h$ is the half-light radius of the dwarf in question, as listed by \citet{martin:16}.  We found only one such example -- the cluster H11, which sits $8.7\arcmin$ from And XVII\footnote{Although the radial velocity of H11 indicates that it is unlikely to be physically associated with the dwarf \citep{veljanoski:14}.}. For this object we excluded stars lying in the overlapping section, and again corrected the area of the aperture appropriately. 

We followed precisely the same procedure for large background galaxies hosting substantial groupings of point sources.  The most prominent examples of such systems were catalogued during the dwarf galaxy search of \citet{martin:13} -- we compiled a list from their Table 1 and cross-matched this against our globular cluster catalogue using a conservative exclusion radius of $3\arcmin$ (comparable to that for the smallest M31 dwarf galaxies described above). 

Finally, we used the \citet{martin:13} contamination model (i.e., Equation \ref{e:contam}) to determine what fraction of the apparent density of stars in a given aperture is due to the density of non-M31 sources lying inside the ``MP'' box on the CMD at the location $(\xi,\eta)$ of the cluster in question. Since the \citet{martin:13} model varies smoothly and gradually with position in the PAndAS footprint, our implicit assumption that the local density of contaminants is constant over a span of $20\arcmin$ is justified. 

Having measured the surface density of M31 halo stars with $-2.5 \la [$Fe$/$H$] \la -1.1$ in the local vicinity of each globular cluster in our catalogue, we repeated the calculation for a large number of locations sampling the PAndAS footprint in order to determine the underlying distribution of stellar surface densities as a function of projected galactocentric radius. We first split the PAndAS survey area into circular annuli centred on M31, with thickness $1$\ kpc and radii in the range $20 \le R_{\rm proj} \le 155$\ kpc. Inside each annulus we randomly generated $1000$ locations with a uniform spatial distribution, and at each location we determined the local surface density of the M31 metal-poor field halo within a circular aperture of radius $r = 10\arcmin$. As previously, we were careful to exclude non-halo populations from our star counts.  Any random location falling within the exclusion zone of a dwarf galaxy or a background system of point sources was re-generated; for all legitimate locations we excised any portion of an aperture that overlapped with the exclusion zone of a globular cluster, dwarf galaxy, or large background system.  We used the \citet{martin:13} contamination model to correct each measured density for the contribution of non-M31 sources.

Note that our procedure of generating a fixed {\it number} of apertures per annulus leads to a spatially variable sampling rate. While we could, in terms of our final results, equally have kept the {\it density} of points constant with radius, the algorithm we adopted allows simple visualisation of the distribution of local stellar densities as a function of galactocentric distance (i.e., Figure \ref{f:densities} below) without the necessity of applying any radially-dependent normalisation. To determine the number of locations required per annulus we simply ensured that the sampling density in the annulus spanning the greatest physical area within the (irregular) PAndAS footprint met some minimum requirement; in all other annuli the sampling density was, by definition, higher. This particular annulus occurs at $R_{\rm proj} = 130$\ kpc, and with $1000$ random positions the sampling density is such that the aperture for any given location would typically encompass $\approx 25$ other locations in the list.

\section{Full globular cluster catalogue used in this work}
\label{a:data}
The complete catalogue of remote M31 globular clusters used in this work is presented in Table \ref{t:fulldata}. It consists of $92$ objects with projected galactocentric radii $R_{\rm proj} > 25$\ kpc. For each cluster we list the metal-poor and metal-rich density percentile values ($\zeta_{\rm MP}$ and $\zeta_{\rm MR}$) calculated in Section \ref{ss:correlation}.  We also provide ancillary photometric and kinematic data, plus a classification, that together form the basis of our analysis in Section \ref{s:properties}. Below we provide information on our various data sources, and our classification scheme.

\subsection{Names and positions}
As described in Section \ref{ss:outercat}, our catalogue consists of $52$ objects discovered as part of the PAndAS globular cluster survey by \citet{huxor:14} (all clusters possessing names beginning with ``PAndAS''), plus $32$ found in pre-PAndAS surveys by our group: \citet{huxor:05,huxor:08} (all clusters possessing names beginning with ``H'' or ``HEC''), and MGC1 from \citet{martin:06}. Another $6$ come from several earlier works as compiled by \citet{galleti:04} in Version 5 of the RBC (G001, G002, G339, G353, EXT8, and B517), while the final two come from the SDSS catalogue of \citet{dtz:15} (dTZZ-05 and dTZZ-21). Coordinates for $88$ objects are taken from \citet{huxor:14}, where they were derived as part of the uniform photometric measurements conducted for all M31 outer halo clusters known at that time and imaged by PAndAS. As reported in that paper, typical uncertainties in these positions are a few tenths of an arcsecond. Four clusters are missing from the PAndAS measurements. For these objects we adopted coordinates from the following sources: \citet{huxor:08} for H9, the RBC for B517, and \citet{dtz:15} for dTZZ-05 and dTZZ-08. The galactocentric radii and position angles listed in Table \ref{t:fulldata} were calculated assuming that M31 has its centre at $00\,42\,44.3$ $+41\,16\,09.4$ as listed in the NASA Extragalactic Database (NED)\footnote{See \href{https://ned.ipac.caltech.edu/}{https://ned.ipac.caltech.edu/}}. As previously stated we assume the M31 distance modulus to be $(m-M)_0 = 24.46$ \citep{conn:12}, corresponding to a physical distance of $780$\ kpc and an angular scale of $3.78$\ pc per arcsecond.

\subsection{Luminosities and colours}
\label{ss:lumcol}
We adopted luminosity and colour estimates for $86$ clusters from the uniform photometric measurements conducted by \citet{huxor:14} using PAndAS imaging. As described in detail by those authors, instrumental MegaCam $g$- and $i$-band magnitudes were first determined for each cluster by constructing curves-of-growth; these measurements were then transformed to the standard Johnson-Cousins $V$ and $I$ passbands, and dereddened using $E(B-V)$ values from \citet{schlegel:98} and the coefficients of \citet{schlafly:11}. This process provides the intrinsic $(V-I)_0$ colours reported in Table \ref{t:fulldata}; the absolute luminosities $M_V$ were obtained by subtracting our assumed distance modulus from the integrated $V_0$ magnitudes. 

\citet{huxor:14} demonstrated typical uncertainties of $\approx 0.1$ mag for the clusters in their sample with high quality photometry (i.e., with quality flags of ``A'' or ``B''), by comparing their luminosity measurements to those available from {\it Hubble Space Telescope} imaging for a small subset of objects. For compact clusters ($r_h \la 10$\ pc), they further identified an additional mean systematic offset of $\approx +0.1$ mag, in the sense that the PAndAS photometry under-estimates the luminosity compared to the {\it HST} measurement. For very extended clusters ($r_h \sim 30$\ pc) this systematic difference could be as large as $\sim +0.5$ mag.

As noted above, four clusters are missing from the PAndAS measurements (H9, B517, dTZZ-05, and dTZZ-21); in addition, two others (PA-41 and PA-51) are too close to the edge of an image to allow useful photometry. We were able to track down luminosity and colour estimates from the literature for each of these objects except PA-51. For H9 and B517 we adopted the SDSS photometry from \citet{peacock:10}. Measurements for H9 are also available in \citet{huxor:08}; although these come from imaging in rather different passbands, they are consistent with the Peacock et al. results. For dTZZ-05, dTZZ-21, and PA-41 we used the SDSS photometry reported by \citet{dtz:14,dtz:15}. For all these clusters we converted the SDSS photometry to Johnson-Cousins $V$ and $I$ using the photometric transformation defined by Lupton on the SDSS web pages\footnote{\href{http://www.sdss.org/dr13/algorithms/sdssUBVRITransform/\#Lupton2005}{www.sdss.org/dr13/algorithms/sdssUBVRITransform/\#Lupton2005}}.

\citet{huxor:14} flagged four of the clusters in our sample as having poor quality photometry, typically due to one or more bright stars falling in close proximity to the target. These objects are PA-09, PA-15, PA-42, and PA-54. Since no better photometry is available in the literature we retain the Huxor et al. results for completeness; however we mark these with asterisks in Table \ref{t:fulldata}, and these data were not used for the analysis presented in Section \ref{s:properties}.

Finally, two clusters (MGC1 and PA-48) have been studied in detail with resolved imaging and found to have substantially different line-of-sight distances than the generic M31 distance modulus assumed above. For these objects we adopted the individually-measured luminosities and distance moduli from \citet{mackey:10a} for MGC1, and \citet{mackey:13b} for PA-48. Uncertainties for these two clusters are $\la 0.1$ mag.

\subsection{Cluster sizes}
\label{ss:rh}
For $70$ clusters in our sample we adopted the half-light radii ($r_{\rm h}$) determined from the PAndAS curves-of-growth by \citet{huxor:14}. Typical uncertainties for these measurements are $\approx 10\%$. As before, we flagged the measurements for PA-09, PA-15, PA-42, and PA-54 as being of lower quality, and excluded these from the analysis in Section \ref{s:properties}. Fifteen clusters have more precise size measurements derived from {\it Hubble Space Telescope} imaging -- these are G1, G2, G339, and G353 \citep{barmby:07}, plus H1, H4, H5, H10, H23, H24, H27, HEC7, HEC12, and B514 \citep{tanvir:12}, and PA-48 \citep{mackey:13b}. Typical uncertainties for these clusters are smaller than $\approx 5\%$. We also adopted the high quality measurement by \citet{mackey:10a} for MGC1, and the estimates from SDSS photometry by \citet{dtz:15} for dTZZ-05, dTZZ-21, and PA-41. We were unable to locate size data for three objects in our sample (H9, B517, and PA-51). 

\citet{huxor:14} show that seeing issues affect the fidelity of their cluster size measurements for objects with (true) half-light radii smaller than $r_{\rm h} \approx 9$\ pc. By comparing size measurements derived from PAndAS imaging with those derived from {\it HST} imaging, and utilising a set of mock measurements on simulated cluster images, they show that the ratio between the true and observed half-light radius for objects in their sample is well approximated by (see their Figures 9 and 10):
\begin{equation}
\frac{r_{\rm h,obs}}{r_{\rm h,true}} =\begin{cases}
   -0.059\,r_{\rm h,true} + 1.535 & \text{if $r_{\rm h,true} < 9$\ pc}\\
   1.000 & \text{if $r_{\rm h,true} \geq 9$\ pc}.
\end{cases}
\end{equation}
We used this to define an empirical correction to the half-light radii for all $38$ objects in our sample for which we adopted the measurments of \citet{huxor:14}, and for which $r_{\rm h} < 9$\ pc:
\begin{equation}
r_{\rm h,true} = 13.000 - 8.475 \left(2.356 - 0.236\,r_{\rm h,obs} \right)^{1/2} \,\,.
\end{equation}
Image-to-image changes in the seeing profile lead to an intrinsic cluster-to-cluster variation in the quality of this correction. However, as described in Section \ref{ss:pandas}, the rms scatter in the seeing across the PAndAS survey as a whole is relatively small ($0\farcs 10$ in both filters). We decided to also apply the above correction to the ground-based size measurements adopted for MGC1 (from Gemini/GMOS) and dTZZ-05, dTZZ-21, and PA-41 (SDSS). While the image quality for the MGC1 observation is comparable to the mean PAndAS image quality, SDSS in general has much poorer seeing such that the corrected half-light radii are likely still too large.  However, the discrepancy is not as bad as if we had left these measurements uncorrected.  Even so, we are careful not to give much weight to size data for clusters with $r_{\rm h} < 9$\ pc in Section \ref{s:properties}.

We converted all the adopted half-light radii to parsecs by assuming the usual M31 distance modulus, except for MGC1 and PA-48 as described in \ref{ss:lumcol} above. 

\subsection{Radial velocities}
\label{ss:rv}
\citet{veljanoski:14} obtained radial velocity measurements for $76$ clusters in our sample, and these constitute the majority of the values assumed in the present work. However, we update the velocities for $15$ of these objects with higher precision measurements as follows. For G1 and G2, we take the values listed in the RBC \citep[see][]{galleti:06}; for G1 the RBC velocity is within $1\sigma$ of that from \citet{veljanoski:14}, while for G2 the values are discrepant at the $\approx 2\sigma$ level. We adopt the velocities for PA-06, PA-17, PA-53, PA-54, PA-56, and H23 from the high-resolution spectroscopy of \citet{sakari:15}; for H23, PA-06, and PA-56 the \citet{sakari:15} measurements are within $1\sigma$ of the values reported by \citet{veljanoski:14}, while for PA-17 and PA-54 the difference is $\la 1.3\sigma$, and for PA-53 it is $1.8\sigma$. For H10, H27, and MGC1 we use the velocities derived from the high-resolution spectra of \citet{alvesbrito:09}; for MGC1 and H27 the \citet{alvesbrito:09} measurements are identical to those from \citet{veljanoski:14} (but with smaller uncertainties), while for H10 the measurements differ by $< 1\sigma$\footnote{Note that \citet{sakari:15} also observed H10, and their derived velocity sits in between those of \citet{alvesbrito:09} and \citet{veljanoski:14}.}. For PA-07 and PA-08, we adopt the velocities reported by \citet{mackey:13a} -- again, these sit within $1\sigma$ of the \citet{veljanoski:14} measurements. 

For two clusters -- PA-13 and PA-15 -- we use velocities derived from new spectroscopic observations. While \citet{veljanoski:14} obtained measurements for both these objects, their velocity for PA-13 has large uncertainties of $\pm 45$ km$\,$s$^{-1}$, while PA-15 sits very close to a bright star and we feared that its spectrum could have been contaminated due to the position angle of the slit. Our new data were obtained using the DEIMOS instrument on the 10m Keck II telescope in longslit mode during the night of 2013 September 11 (program 2013B-Z297D, PI: Mackey). Basic reduction was undertaken using {\sc iraf} following the procedures outlined in Section 2.2 of \citet{veljanoski:14}, while the radial velocity measurements were made using the Ca {\sc ii} triplet as in Section 2.4 of that paper.

Finally, we searched the literature for velocity measurements of objects not included in the \citet{veljanoski:14} sample but found only one example -- G353, for which we take the RBC value \citep[again, see][]{galleti:06}.

The velocities described above are listed in Table \ref{t:fulldata}; these ($V_{\rm helio}$) are all reported in the heliocentric frame. As in \citet{veljanoski:14}, we also computed the M31-centric velocity, $V_{\rm M31}$, using the following procedure. We first converted the velocities from the heliocentric to Galactocentric frame to eliminate the effect of the solar motion:
\begin{equation}
V_{\rm gal} = V_{\rm helio} + 251.24 \sin(l) \cos(b) + 11.1 \cos(l) \cos(b) + 7.25 \sin(b)
\end{equation}
where $l$ and $b$ are the Galactic latitude and longitude. This relation originally comes from \citet{courteau:99}, but we utilise values of the solar motion from \citet{schonrich:10} and \citet{mcmillan:11}. Next, we calculated the M31-centric velocity of each cluster via:
\begin{equation}
V_{\rm M31} = V_{\rm gal} - V_{\rm M31,r} \cos(\rho) - V_{\rm M31,t} \sin(\rho) \cos(\phi - \theta_{\rm t})
\end{equation}
as per \citet{vandermarel:08}. Here $V_{\rm M31,r}$ is the systemic radial velocity of M31 (i.e., taken along the line-of-sight to its centre) while $V_{\rm M31,t}$ is the systemic tangential motion, which occurs in a direction on the sky given by the position angle $\theta_{\rm t}$ (east of north). The position angle of the cluster with respect to the centre of M31 is $\phi$, and its angular separation is $\rho$. \citet{veljanoski:14} show that the third term in the above relation, involving the systemic transverse motion, is sufficiently unimportant for all the clusters in our sample that it can be set to zero\footnote{More specifically, they argue that the formal uncertainties on the individual components of the best available transverse velocity measurement for M31, made by \citet{vandermarel:12} using HST, are sufficiently large ($\approx 30$ km s$^{-1}$ each), and the magnitude of the motion so small (with a $1\sigma$ confidence region of $V_{\rm M31,t} \le 34.3$ km s$^{-1}$) that including this term in the calculation would introduce significantly larger random uncertainties into the final M31-centric velocities than ignoring it entirely. Recent results from Gaia DR2, although less accurate than the HST measurements, support this assertion \citep{vandermarel:18}. Note also that all clusters sit at $\rho \la 10\degr$, such that $\sin(\rho) \la 0.2$.}. For the systemic radial velocity of M31 we assume the value measured by \citet{vandermarel:08}: $-301 \pm 1$\ km s$^{-1}$ in the heliocentric frame, corresponding to $V_{\rm M31,r} = -109 \pm 4$ km s$^{-1}$ in the Galactocentric frame.

\subsection{Classification}
\label{ss:class}
We classify the clusters in our sample based on the strength of the evidence for an association (or not) with known substructures in the M31 halo. We define three classes: ``substructure'' clusters exhibit strong spatial and/or kinematic evidence for a link with a substructure, while ``non-substructure'' clusters possess no such evidence. Clusters with weak or conflicting evidence for an association fall into an ``ambiguous'' category. Our main aim in developing this scheme is to try and isolate relatively clean subsets of ``substructure'' and ``non-substructure'' objects in order to study the number and radial distribution of these clusters, and search for differences between their typical properties. \citet{veljanoski:14} have already discussed the association between clusters and the most prominent stellar substructures in the M31 halo, based on their radial velocity measurements and the proximity of clusters to the various streamlike features. Here, through the calculation in particular of $\zeta_{\rm MP}$, we are able to formally quantify the latter. 

We first consider the three best-established examples of a kinematic and spatial link between clusters and an underlying structure in the outer M31 field halo:
\begin{itemize}
\item{{\it North-West Stream:} Seven clusters project directly on top of this narrow radial stream extending $\ga 100$\ kpc to the north-west of M31. These are: PA-04, PA-09, PA-10, PA-11, PA-12, PA-13, and PA-15. All seven of these objects have $\zeta_{\rm MP} \leq 0.16$. \citet{veljanoski:14} showed that the first five in the list also exhibit strongly correlated radial velocities that confirm their membership of the stream; however their measurements for PA-13 and PA-15 were ambiguous. The velocities derived from our new Keck spectra for these two clusters (as described in \ref{ss:rv} above) suggest that PA-13 fits the observed kinematic trend while PA-15 does not. Hence we classify PA-04, PA-09, PA-10, PA-11, PA-12, and PA-13 as ``substructure'' objects, while we consider PA-15 to be ``ambiguous'' due to its conflicting velocity and small $\zeta_{\rm MP}$.\vspace{1.5mm}}
\item{{\it South-West Cloud:} Three clusters fall onto this diffuse overdensity $\sim 90$\ kpc to the south-west of M31: PA-07, PA-08, and PA-14. Each has high $\zeta_{\rm MP} \leq 0.17$. \citet{mackey:14} showed that all three possess correlated radial velocities that precisely match the velocity of the stellar substructure itself \citep[see also][]{mackey:13a,bate:14,veljanoski:14}. We hence consider PA-07, PA-08, and PA-14 to be ``substructure'' clusters.\vspace{1.5mm}}
\item{{\it East Cloud:} This faint arc sitting $\sim 120$\ kpc due east of M31 has three coincident clusters: PA-56, PA-57, and PA-58. All three have very high local densities with $\zeta_{\rm MP} \leq 0.06$. \citet{veljanoski:14} demonstrated that PA-57 and PA-58 possess quite similar velocities that are well separated from the M31 systemic velocity, and that both are therefore very likely members of the substructure \citep[see also][]{mcmonigal:16a}. Here we note that adding PA-56 defines a clear trend between velocity and position angle, much as observed for the clusters lying on the South-West Cloud by \citet{veljanoski:14}. On this basis, and its very small $\zeta_{\rm MP} = 0.03$, we also consider PA-56 to be a ``substructure'' cluster.}
\end{itemize}
In Section \ref{ss:correlation} we showed that a very significant excess of clusters in our sample have $\zeta_{\rm MP}$ in the top quartile of observed values at given $R_{\rm proj}$. All $23$ of the clusters sitting on the three substructures discussed above fit this picture, with all possessing $\zeta_{\rm MP} \leq 0.17$. Based on these observations, when classifying the remainder of our sample we typically consider objects with $\zeta_{\rm MP} \leq 0.25$ to exhibit strong evidence for projecting onto a substructure or halo overdensity (named or not).

We next examine two high-density cluster groups investigated kinematically by \citet{veljanoski:14}:
\begin{itemize}
\item{{\it Stream C/D overlap:} Nine objects congregate in a small region of sky where the three arc-like substructures named Stream Cp, Stream Cr, and Stream D overlap in projection. \citet{veljanoski:14} showed that eight of these clusters split naturally into two kinematically cold subgroups: one consisting of H24, PA-41, PA-43, PA-45, and PA-46, and another consisting of B517, PA-44, and PA-47. Their analysis is, by itself, sufficient to classify these eight clusters as ``substructure'' objects -- the occurrence of subgroups clustered so closely in both position and velocity space is extremely unlikely in the case of a smooth pressure-supported halo. However, it is also informative to consider the local field-star densities for these objects. Three of them (PA-43, PA-47, and B517) have $\zeta_{\rm MP} \la 0.20$, comparable to the values exhibited by the clusters projected onto the three large substructures discussed above. A further four (H24, PA-41, PA-45, and PA-46) have slightly lower local densities in the range $0.25 \la \zeta_{\rm MP} \la 0.35$, while PA-44 has $\zeta_{\rm MP} = 0.79$. It is worth noting that for the objects in this region of sky, which have $R_{\rm proj} \la 40$\ kpc, the $\zeta_{\rm MP}$ values are rather biased by the dominant presence of the Giant Stellar Stream. Masking this feature would lower the observed values of $\zeta_{\rm MP}$ for clusters in the Stream C/D overlap region by $\approx 0.1$ -- hence, in general, the local field-star densities for the majority of these objects are quite similar to those for the three well-established examples above, where the clusters sit at much larger projected radii. The ninth cluster on the Stream C/D overlap, PA-49, has a velocity intermediate between the two cold kinematic subgroups identified by \citet{veljanoski:14}, as well as a rather unremarkable $\zeta_{\rm MP} = 0.39$. We classify this object as ``ambiguous'' since we cannot confirm or rule out an association with field substructure in this region.\vspace{1.5mm}}
\item{{\it Association 2:} This is a tight grouping of eleven clusters first identified by \citet{mackey:10b}. It constitutes the single highest local density enhancement of globular clusters, relative to the azimuthal average at commensurate radius, seen in the M31 halo. \citet{veljanoski:14} showed that eight of these clusters split naturally into two kinematically cold subgroups: one consisting of H2, PA-18, PA-19, and PA-21, and another consisting of H7, PA-22, PA-23, and H8. As before, this analysis is, by itself, sufficient to classify these objects as ``substructure'' clusters. Intriguingly, six of the eight have moderate to large values of $\zeta_{\rm MP}$; the remaining two (PA-23 and H8) project onto the so-called G1 Clump \citep[e.g.,][]{reitzel:04,ibata:05,faria:07} and have very small $\zeta_{\rm MP}$. The three remaining clusters in this region are G1, G2, and dTZZ-05. All three project onto the G1 Clump and have very small $\zeta_{\rm MP} \leq 0.05$. G1 and G2 have similar velocities but are not part of either of the kinematic groups identified by \citet{veljanoski:14}, and, despite their location (and the nomenclature) are likely not associated with the G1 Clump \citep{faria:07}. dTZZ-05 was not discovered at the time of the \citet{veljanoski:14} work and has no radial velocity measurement. Due to the complexity of this region we classify both G2 and dTZZ-05 as ``ambiguous''. In the absence of additional information G1 would warrant a similar classification; however, this object has long been suspected to constitute the accreted core of a now-destroyed nucleated dwarf. This is supported by its extremely high mass \citep[$\sim10^7\,M_\odot$; e.g.,][]{ma:09}, its structural parameters \citep[e.g.,][]{ma:07}, the presence of a substantial internal metallicity dispersion \citep[e.g.,][]{meylan:01}, and the possible existence of an intermediate-mass black hole within its core radius \citep[e.g.,][]{gebhardt:02,gebhardt:05,pooley:06,ulvestad:07,kong:07,kong:10}\footnote{But see \citet{baumgardt:03} and \citet{millerjones:12} for contrary viewpoints.}. Given these data we classify G1 as a ``substructure'' object.}
\end{itemize}

Finally, we consider a variety of smaller-scale instances of possible kinematic and spatial links between clusters and underlying halo substructures:
\begin{itemize}
\item{{\it Stream C:} Three clusters sit on the anticlockwise extension of Streams Cp and Cr from the region discussed above, which continue to overlap in projection. \citet{veljanoski:14} showed that HEC12 shares a common velocity with Stream Cp \citep[see also][]{chapman:08,collins:09}; the likelihood of an association between the two is increased by our measurement of $\zeta_{\rm MP} = 0.05$ for this cluster. Similarly, \citet{veljanoski:14} demonstrated that HEC13 has a velocity matching that for stream Cr; notably we find $\zeta_{\rm MP} = 0.04$ and $\zeta_{\rm MR} = 0.09$ for this object. We hence classify both HEC12 and HEC13 as ``substructure'' clusters. H26 also sits in this region and has a small $\zeta_{\rm MP} = 0.06$. However, its velocity is quite different from that of either Stream Cp or Cr and we classify it as ``ambiguous''.\vspace{1.5mm}}
\item{{\it Stream D:} Three clusters fall on the anticlockwise extension of Stream D. This is another complex region where the interpretation of $\zeta_{\rm MP}$ is affected by the presence of the main body of the Giant Stellar Stream at comparable radius. Both HEC11 and PA-42 fall very close to the main trace of Stream D, which is reflected in their local density measurements -- these would fall close to our threshold of $\zeta_{\rm MP} \approx 0.25$ if not for the Giant Stream. However, neither cluster possesses a velocity close to that estimated for Stream D by \citet{chapman:08}. H23, on the other hand, has a velocity that matches that of the stream, but an extremely large $\zeta_{\rm MP}$. We conservatively classify all three clusters as ``ambiguous''.\vspace{1.5mm}}
\item{{\it Giant Stellar Stream:} This feature is particularly intriguing.  It is by far the most luminous substructure in the M31 halo, and yet appears significantly underabundant in clusters \citep[see e.g.,][]{mackey:10b}. There are three objects that project onto the stream: PA-37, H19, and H22. Unsurprisingly, all three have small values of $\zeta_{\rm MP}$ (and, indeed, $\zeta_{\rm MR}$). The velocity for PA-37 agrees well with the profile measured by \citet{ibata:04} given its position on the stream. We consider this a firm association and classify PA-37 as a ``substructure'' cluster.  However, the velocities for H19 and H22 show much poorer agreement; we therefore assign them an ``ambiguous'' classification.\vspace{1.5mm}}
\item{{\it South-West Cloud extension:} \citet{bate:14} showed that the South-West Cloud may well extend significantly clockwise towards the outermost portion of the Giant Stream. Two clusters, H10 and H15, sit projected onto this possible extension. Both possess velocities that plausibly fit the evolution with position angle shown by \citet{veljanoski:14} for the three confirmed South-West Cloud clusters PA-07, PA-08, and PA-14. The local density for H10 is low, $\zeta_{\rm MP} = 0.74$, but that for H15 is within our threshold at $\zeta_{\rm MP} = 0.24$, despite the low surface-brightness of the possible extension. Again, we conservatively classify both clusters ``ambiguous''.}
\end{itemize}

Having exhausted the named substructures and known kinematic groups in the M31 halo, we are left with two clear sets of clusters -- those with local densities higher than our threshold at $\zeta_{\rm MP} = 0.25$, and those with local densities below:
\begin{itemize}
\item{{\it High local density:} There are $13$ clusters possessing $\zeta_{\rm MP} \leq 0.25$, and which we therefore classify as ``ambiguous'' -- PA-03, PA-24, PA-26, PA-36, PA-39, PA-48, PA-50, B514, H9, H17, HEC7, G339, and G353. Notably, the position of PA-39 and G353 is consistent with a possible inward extension of the Stream C/D overlap region; however, neither cluster has a measured velocity so we are unable to confirm or refute such an association.\vspace{1.5mm}}
\item{{\it Low local density:} A total of $35$ clusters have $\zeta_{\rm MP} > 0.25$ and no kinematic evidence for an association substructure in the field halo. These are: PA-01, PA-02, PA-05, PA-06 PA-16, PA-17, PA-20, PA-25, PA-27, PA-30, PA-31, PA-33, PA-38, PA-40, PA-51, PA-52, PA-53, PA-54, dTZZ-21, EXT8, MGC1, H1, H3, H4, H5, H11, H12, H18, H25, H27, HEC1, HEC2, HEC3, HEC6, and HEC10. We assign each such cluster a ``non-substructure'' classification.}
\end{itemize}

In summary, we identify $32$ clusters that have a high likelihood of being associated with an underlying field substructure, and $35$ that show no evidence for such an association. In $25$ cases the available data are ambiguous and do not allow us to confirm or rule out a substructure association.

\bsp
\label{lastpage}

\begin{landscape}
\begin{table}
\centering
\caption{Data  for the sample of $92$ globular clusters with $R_{\rm proj} \ga 25$\ kpc used in this work.}
\begin{tabular}{lccccccccccccccccc}
\hline \hline
Name & \multicolumn{2}{c}{Coordinates (J2000.0)} & $R_{\rm proj}$ & PA & $\zeta_{\rm MP}$ & $\zeta_{\rm MR}$ & $M_V$ & Source & $r_{\rm h}$ & Source & $(V-I)_0$ & Source & $V_{\rm helio}$ & $V_{\rm M31}$ & Source & Class & Note \\
 & RA & Dec & (kpc) & (deg) & & & (mag) & & (pc) & & (mag) & & (km$\,$s$^{-1}$) & (km$\,$s$^{-1}$) & & & \\
\hline
PAndAS-01 & $23\,57\,12.0$ & $+43\,33\,08.3$ & $118.9$ & $289.0$ & $0.54$ & $0.18$ & $-7.48$ & H14 & $6.1$ & H14 & $0.83$ & H14 & $-333 \pm 21$ & $-11 \pm 21$ & V14 & N & $...$ \\
PAndAS-02 & $23\,57\,55.7$ & $+41\,46\,49.3$ & $114.7$ & $277.2$ &  $0.34$ & $0.94$ & $-6.82$ & H14 & $25.7$ & H14 & $0.90$ & H14 & $-266 \pm 4$ & $54 \pm 5$ & V14 & N & $...$ \\
PAndAS-03 & $00\,03\,56.4$ & $+40\,53\,19.2$ & $100.0$ & $270.2$ &  $0.16$ & $0.29$ & $-4.17$ & H14 & $27.4$ & H14 & $0.86$ & H14 & $...$ & $...$ & $...$ & A & ($\zeta_{\rm MP}$)\\
PAndAS-04 & $00\,04\,42.9$ & $+47\,21\,42.5$ & $124.6$ & $315.1$ &  $0.12$ & $0.60$ & $-7.09$ & H14 & $3.5$ & H14 & $0.89$ & H14 & $-397 \pm 7$ & $-75 \pm 8$ & V14 & S & NW Stream \\
PAndAS-05 & $00\,05\,24.2$ & $+43\,55\,35.7$ & $100.6$ & $294.3$ &  $0.72$ & $0.70$ & $-5.05$ & H14 & $17.0$ & H14 & $0.97$ & H14 & $-183 \pm 7$ & $136 \pm 8$ & V14 & N & $...$ \\
PAndAS-06 & $00\,06\,12.0$ & $+41\,41\,21.0$ & $93.7$ & $276.5$ &  $0.48$ & $0.47$ & $-8.02$ & H14 & $3.3$ & H14 & $0.87$ & H14 & $-341 \pm 1$ & $-24 \pm 3$ & S15 & N & $...$ \\
PAndAS-07 & $00\,10\,51.4$ & $+39\,35\,58.6$ & $86.0$ & $257.2$ &  $0.17$ & $0.58$ & $-5.00$ & H14 & $13.3$ & H14 & $0.89$ & H14 & $-433 \pm 8$ & $-121 \pm 9$ & M13 & S & SW Cloud \\
PAndAS-08 & $00\,12\,52.5$ & $+38\,17\,47.9$ & $88.3$ & $245.0$ &  $0.01$ & $0.52$ & $-5.40$ & H14 & $9.5$ & H14 & $1.03$ & H14 & $-411 \pm 4$ & $-101 \pm 5$ & M13 & S & SW Cloud \\
PAndAS-09 & $00\,12\,54.7$ & $+45\,05\,55.9$ & $90.8$ & $307.7$ &  $0.09$ & $0.99$ & $-6.75^*$ & H14 & $2.8^*$ & H14 & $0.83^*$ & H14 & $-444 \pm 21$ & $-127 \pm 21$ & V14 & S & NW Stream \\
PAndAS-10 & $00\,13\,38.7$ & $+45\,11\,11.1$ & $90.0$ & $308.9$ &  $0.12$ & $0.88$ & $-5.43$ & H14 & $15.2$ & H14 & $0.93$ & H14 & $-435 \pm 10$ & $-118 \pm 10$ & V14 & S & NW Stream \\
PAndAS-11 & $00\,14\,55.6$ & $+44\,37\,16.4$ & $83.2$ & $305.7$ &  $0.06$ & $0.21$ & $-6.74$ & H14 & $9.4$ & H14 & $0.87$ & H14 & $-447 \pm 13$ & $-131 \pm 13$ & V14 & S & NW Stream \\
PAndAS-12 & $00\,17\,40.1$ & $+43\,18\,39.0$ & $69.2$ & $295.9$ &  $0.09$ & $0.53$ & $-5.33$ & H14 & $13.7$ & H14 & $0.94$ & H14 & $-472 \pm 5$ & $-158 \pm 6$ & V14 & S & NW Stream \\
PAndAS-13 & $00\,17\,42.7$ & $+43\,04\,31.8$ & $68.0$ & $293.4$ &  $0.11$ & $0.45$ & $-6.49$ & H14 & $3.5$ & H14 & $0.85$ & H14 & $-468 \pm 11$ & $-154 \pm 11$ & M18 & S & NW Stream \\
PAndAS-14 & $00\,20\,33.9$ & $+36\,39\,34.5$ & $86.2$ & $224.9$ &  $0.10$ & $0.77$ & $-7.01$ & H14 & $10.9$ & H14 & $0.90$ & H14 & $-363 \pm 9$ & $-59 \pm 9$ & V14 & S & SW Cloud \\
PAndAS-15 & $00\,22\,44.1$ & $+41\,56\,14.2$ & $51.9$ & $281.8$ &  $0.16$ & $0.79$ & $-5.04^*$ & H14 & $4.8^*$ & H14 & $0.93^*$ & H14 & $-435 \pm 19$ & $-124 \pm 19$ & M18 & A & (NW Stream)\\
PAndAS-16 & $00\,24\,59.9$ & $+39\,42\,13.1$ & $50.8$ & $246.6$ &  $0.83$ & $0.64$ & $-8.44$ & H14 & $4.3$ & H14 & $0.97$ & H14 & $-490 \pm 15$ & $-183 \pm 15$ & V14 & N & $...$ \\
HEC1 & $00\,25\,33.9$ & $+40\,43\,38.9$ & $44.9$ & $261.9$ &  $0.94$ & $0.99$ & $-5.82$ & H14 & $15.7$ & H14 & $0.86$ & H14 & $-233 \pm 9$ & $75 \pm 9$ & V14 & N & $...$ \\
H1 & $00\,26\,47.8$ & $+39\,44\,46.2$ & $46.3$ & $244.6$ &  $0.91$ & $0.70$ & $-8.70$ & H14 & $3.5$ & T12 & $0.88$ & H14 & $-245 \pm 7$ & $61 \pm 8$ & V14 & N & $...$ \\
PAndAS-17 & $00\,26\,52.2$ & $+38\,44\,58.1$ & $53.9$ & $231.6$ &  $0.67$ & $0.79$ & $-8.17$ & H14 & $3.3$ & H14 & $1.14$ & H14 & $-260 \pm 1$ & $45 \pm 3$ & S15 & N & $...$ \\
H2 & $00\,28\,03.2$ & $+40\,02\,55.6$ & $41.6$ & $247.5$ &  $0.97$ & $0.76$ & $-7.50$ & H14 & $4.0$ & H14 & $0.91$ & H14 & $-519 \pm 16$ & $-213 \pm 16$ & V14 & S & Assoc. 2 \\
PAndAS-18 & $00\,28\,23.3$ & $+39\,55\,04.9$ & $41.6$ & $244.8$ &  $0.95$ & $0.68$ & $-5.35$ & H14 & $23.0$ & H14 & $0.94$ & H14 & $-551 \pm 18$ & $-245 \pm 18$ & V14 & S & Assoc. 2 \\
HEC2 & $00\,28\,31.5$ & $+37\,31\,23.5$ & $63.5$ & $217.4$ &  $0.37$ & $0.58$ & $-5.60$ & H14 & $12.4$ & H14 & $0.96$ & H14 & $-341 \pm 9$ & $-39 \pm 9$ & V14 & N & $...$ \\
H3 & $00\,29\,30.2$ & $+41\,50\,31.9$ & $34.7$ & $284.1$ &  $0.57$ & $0.78$ & $-6.52$ & H14 & $4.3$ & H14 & $1.01$ & H14 & $-86 \pm 9$ & $222 \pm 9$ & V14 & N & $...$ \\
H4 & $00\,29\,45.0$ & $+41\,13\,09.4$ & $33.4$ & $269.9$ &  $0.71$ & $0.88$ & $-7.82$ & H14 & $4.0$ & T12 & $0.93$ & H14 & $-368 \pm 8$ & $-61 \pm 9$ & V14 & N & $...$ \\
PAndAS-19 & $00\,30\,12.2$ & $+39\,50\,59.3$ & $37.9$ & $240.2$ &  $0.50$ & $0.67$ & $-4.73$ & H14 & $6.3$ & H14 & $0.91$ & H14 & $-544 \pm 6$ & $-239 \pm 7$ & V14 & S & Assoc. 2 \\
H5 & $00\,30\,27.3$ & $+41\,36\,19.5$ & $31.8$ & $279.3$ &  $0.68$ & $0.81$ & $-8.44$ & H14 & $9.1$ & T12 & $0.85$ & H14 & $-392 \pm 12$ & $-85 \pm 12$ & V14 & N & $...$ \\
B514 & $00\,31\,09.8$ & $+37\,54\,00.1$ & $55.2$ & $214.4$ &  $0.03$ & $0.46$ & $-8.91$   & H14 & $3.5$ & T12 & $0.95$ & H14 & $-471 \pm 8$   & $-170 \pm 9$  & V14 & A & ($\zeta_{\rm MP}$) \\
PAndAS-20 & $00\,31\,23.7$ & $+41\,59\,20.1$ & $30.6$ & $289.7$ &  $0.65$ & $0.88$ & $-5.43$ & H14 & $6.3$ & H14 & $1.00$ & H14 & $...$ & $...$ & $...$ & N & $...$ \\
PAndAS-21 & $00\,31\,27.5$ & $+39\,32\,21.8$ & $37.7$ & $232.1$ &  $0.46$ & $0.54$ & $-7.06$ & H14 & $2.9$ & H14 & $0.87$ & H14 & $-600 \pm 7$ & $-296 \pm 8$ & V14 & S & Assoc. 2 \\
H7 & $00\,31\,54.6$ & $+40\,06\,47.8$ & $32.2$ & $241.5$ &  $0.49$ & $0.57$ & $-7.17$ & H14 & $10.5$ & H14 & $0.92$ & H14 & $-426 \pm 23$ & $-122 \pm 23$ & V14 & S & Assoc. 2 \\
PAndAS-22 & $00\,32\,08.4$ & $+40\,37\,31.6$ & $28.7$ & $253.0$ &  $0.57$ & $0.58$ & $-6.18$ & H14 & $6.2$ & H14 & $1.06$ & H14 & $-437 \pm 1$ & $-132 \pm 3$ & V14 & S & Assoc. 2 \\
G001 & $00\,32\,46.5$ & $+39\,34\,40.6$ & $34.7$ & $229.1$ &  $0.05$ & $0.14$ & $-10.79$ & H14 & $3.2$ & B07 & $0.95$ & H14 & $-332 \pm 3$   & $-29 \pm 4$   & G06 & S & G1 \\
PAndAS-23 & $00\,33\,14.1$ & $+39\,35\,15.9$ & $33.7$ & $227.9$ &  $0.03$ & $0.03$ & $-5.02$ & H14 & $5.5$ & H14 & $1.17$ & H14 & $-476 \pm 5$ & $-173 \pm 6$ & V14 & S & Assoc. 2 \\
G002 & $00\,33\,33.8$ & $+39\,31\,19.0$ & $33.8$ & $225.7$ &  $0.01$ & $0.01$ & $-8.92$   & H14 & $3.2$ & B07 & $0.87$ & H14 & $-313 \pm 17$ & $-10 \pm 17$ & G06 & A & (Assoc. 2) \\
PAndAS-24 & $00\,33\,50.6$ & $+38\,38\,28.0$ & $42.8$ & $213.7$ &  $0.19$ & $0.39$ & $-4.68$ & H14 & $16.9$ & H14 & $0.91$ & H14 & $...$ & $...$ & $...$ & A & ($\zeta_{\rm MP}$) \\
PAndAS-25 & $00\,34\,06.2$ & $+43\,15\,06.7$ & $34.8$ & $321.9$ &  $0.65$ & $0.67$ & $-5.21$ & H14 & $5.4$ & H14 & $1.04$ & H14 & $...$ & $...$ & $...$ & N & $...$ \\
H8 & $00\,34\,15.4$ & $+39\,52\,53.2$ & $29.1$ & $229.9$ &  $0.10$ & $0.16$ & $-5.71$ & H14 & $11.8$ & H14 & $1.03$ & H14 & $-463 \pm 3$ & $-160 \pm 4$ & V14 & S & Assoc. 2 \\
H9 & $00\,34\,17.3$ & $+37\,30\,43.3$ & $56.1$ & $204.2$ &  $0.22$ & $0.47$ & $-6.97$ & P10 & $...$ & $...$ & $0.86$ & P10 & $-374 \pm 5$ & $-75 \pm 6$ & V14 & A & ($\zeta_{\rm MP}$) \\
PAndAS-26 & $00\,34\,45.1$ & $+38\,26\,38.1$ & $43.9$ & $209.1$ &  $0.12$ & $0.36$ & $-5.10$ & H14 & $6.4$ & H14 & $1.09$ & H14 & $...$ & $...$ & $...$ & A & ($\zeta_{\rm MP}$) \\
PAndAS-27 & $00\,35\,13.5$ & $+45\,10\,37.9$ & $56.6$ & $341.3$ &  $0.64$ & $0.86$ & $-7.69$ & H14 & $4.1$ & H14 & $0.93$ & H14 & $-46 \pm 8$ & $263 \pm 9$ & V14 & N & $...$ \\
H10 & $00\,35\,59.7$ & $+35\,41\,03.5$ & $78.4$ & $193.8$ &  $0.74$ & $0.83$ & $-8.86$ & H14 & $5.1$ & T12 & $0.95$ & H14 & $-358 \pm 2$ & $-63 \pm 4$ & AB09 & A & (SW Cloud) \\
dTZZ-05 & $00\,36\,08.6$ & $+39\,17\,30.0$ & $32.0$ & $213.0$ &  $0.05$ & $0.05$ & $-7.03$ & dTZ14 & $3.0$ & dTZ14 & $0.86$ & dTZ14 & $...$ & $...$ & $...$ & A & (Assoc. 2) \\
HEC3 & $00\,36\,31.7$ & $+44\,44\,16.5$ & $49.9$ & $342.4$ &  $0.81$ & $0.75$ & $-5.36$ & H14 & $17.6$ & H14 & $1.00$ & H14 & $...$ & $...$ & $...$ & N & $...$ \\
H11 & $00\,37\,28.0$ & $+44\,11\,26.5$ & $42.1$ & $342.1$ &  $0.35$ & $0.52$ & $-7.88$ & H14 & $3.0$ & H14 & $0.93$ & H14 & $-213 \pm 7$ & $94 \pm 8$ & V14 & N & $...$ \\
\hline
\label{t:fulldata}
\end{tabular}
\end{table}
\end{landscape}

\addtocounter{table}{-1}

\begin{landscape}
\begin{table}
\centering
\caption{-- {\it continued.}}
\begin{tabular}{lccccccccccccccccc}
\hline \hline
Name & \multicolumn{2}{c}{Coordinates (J2000.0)} & $R_{\rm proj}$ & PA & $\zeta_{\rm MP}$ & $\zeta_{\rm MR}$ & $M_V$ & Source & $r_{\rm h}$ & Source & $(V-I)_0$ & Source & $V_{\rm helio}$ & $V_{\rm M31}$ & Source & Class & Note \\
 & RA & Dec & (kpc) & (deg) & & & (mag) & & (pc) & & (mag) & & (km$\,$s$^{-1}$) & (km$\,$s$^{-1}$) & & & \\
\hline
H12 & $00\,38\,03.9$ & $+37\,44\,00.2$ & $49.9$ & $194.7$ & $0.30$ & $0.45$ & $-8.19$ & H14 & $2.9$ & H14 & $0.88$ & H14 & $-396 \pm 10$ & $-98 \pm 10$ & V14 & N & $...$ \\
PAndAS-30 & $00\,38\,29.0$ & $+37\,58\,39.2$ & $46.4$ & $194.3$ &  $0.38$ & $0.44$ & $-5.42$ & H14 & $10.9$ & H14 & $0.96$ & H14 & $...$ & $...$ & $...$ & N & $...$ \\
HEC6 & $00\,38\,35.5$ & $+44\,16\,51.4$ & $42.5$ & $346.2$ &  $0.58$ & $0.50$ & $-5.92$ & H14 & $26.7$ & H14 & $0.94$ & H14 & $-132 \pm 12$ & $174 \pm 12$ & V14 & N & $...$ \\
PAndAS-31 & $00\,39\,59.8$ & $+43\,03\,19.7$ & $25.4$ & $343.8$ &  $0.48$ & $0.67$ & $-4.41$ & H14 & $18.5$ & H14 & $0.99$ & H14 & $...$ & $...$ & $...$ & N & $...$ \\
H15 & $00\,40\,13.2$ & $+35\,52\,36.7$ & $74.0$ & $185.4$ &  $0.24$ & $0.27$ & $-6.60$ & H14 & $10.3$ & H14 & $0.83$ & H14 & $-367 \pm 10$ & $-74 \pm 10$ & V14 & A & (SW Cloud) \\
PAndAS-33 & $00\,40\,57.4$ & $+38\,38\,10.2$ & $36.3$ & $187.6$ &  $0.39$ & $0.35$ & $-5.39$ & H14 & $35.8$ & H14 & $0.88$ & H14 & $...$ & $...$ & $...$ & N & $...$ \\
H17 & $00\,42\,23.7$ & $+37\,14\,34.7$ & $55.0$ & $181.0$ &  $0.07$ & $0.08$ & $-7.23$ & H14 & $2.4$ & H14 & $0.96$ & H14 & $-246 \pm 16$ & $49 \pm 16$ & V14 & A & ($\zeta_{\rm MP}$) \\
HEC7 & $00\,42\,55.1$ & $+43\,57\,27.3$ & $36.7$ & $0.7$ &  $0.11$ & $0.31$ & $-6.57$ & H14 & $24.9$ & T12 & $0.99$ & H14 & $...$ & $...$ & $...$ & A & ($\zeta_{\rm MP}$) \\
H18 & $00\,43\,36.1$ & $+44\,58\,59.3$ & $50.8$ & $2.4$ &  $0.59$ & $0.58$ & $-8.09$ & H14 & $3.0$ & H14 & $0.90$ & H14 & $-206 \pm 21$ & $99 \pm 21$ & V14 & N & $...$ \\
H19 & $00\,44\,14.9$ & $+38\,25\,42.2$ & $39.0$ & $174.1$ &  $0.19$ & $0.11$ & $-7.29$ & H14 & $3.8$ & H14 & $0.92$ & H14 & $-272 \pm 18$ & $24 \pm 18$ & V14 & A & (GSS) \\
PAndAS-36 & $00\,44\,45.6$ & $+43\,26\,34.8$ & $30.1$ & $9.6$ &  $0.25$ & $0.31$ & $-7.30$ & H14 & $4.5$ & H14 & $0.94$ & H14 & $-399 \pm 7$ & $-96 \pm 8$ & V14 & A & ($\zeta_{\rm MP}$) \\
G339 & $00\,47\,50.2$ & $+43\,09\,16.5$ & $28.8$ & $26.2$ &  $0.07$ & $0.02$ & $-7.58$   & H14 & $3.9$ & B07 & $0.94$ & H14 & $-97 \pm 6$     & $204 \pm 7$   & V14 & A & ($\zeta_{\rm MP}$) \\
PAndAS-37 & $00\,48\,26.5$ & $+37\,55\,42.1$ & $48.1$ & $161.3$ &  $0.00$ & $0.01$ & $-7.35$ & H14 & $3.4$ & H14 & $1.13$ & H14 & $-404 \pm 15$ & $-111 \pm 15$ & V14 & S & GSS \\
H22 & $00\,49\,44.7$ & $+38\,18\,37.8$ & $44.4$ & $155.0$ &  $0.06$ & $0.05$ & $-7.65$ & H14 & $3.1$ & H14 & $0.93$ & H14 & $-311 \pm 6$ & $-18 \pm 7$ & V14 & A & (GSS) \\
PAndAS-38 & $00\,49\,45.7$ & $+47\,54\,33.1$ & $92.3$ & $10.1$ &  $0.59$ & $0.46$ & $-4.50$ & H14 & $24.4$ & H14 & $0.83$ & H14 & $...$ & $...$ & $...$ & N & $...$ \\
G353 & $00\,50\,18.2$ & $+42\,35\,44.2$ & $26.4$ & $46.1$ &  $0.19$ & $0.15$ & $-7.60$   & H14 & $3.7$ & B07 & $0.87$ & H14 & $-295 \pm 26$ & $4 \pm 26$ & G06 & A & ($\zeta_{\rm MP}$) \\
PAndAS-39 & $00\,50\,36.2$ & $+42\,31\,49.3$ & $26.4$ & $48.6$ &  $0.17$ & $0.16$ & $-6.19$ & H14 & $13.0$ & H14 & $1.04$ & H14 & $...$ & $...$ & $...$ & A & ($\zeta_{\rm MP}$) \\
MGC1 & $00\,50\,42.5$ & $+32\,54\,58.7$ & $116.2$ & $168.6$ &  $0.74$ & $0.57$ & $-9.20$ & M10 & $6.5$ & M10 & $0.89$ & M10 & $-355 \pm 2$ & $-73 \pm 4$   & AB09 & N & $...$ \\
PAndAS-40 & $00\,50\,43.8$ & $+40\,03\,30.2$ & $26.5$ & $128.0$ &  $0.86$ & $0.91$ & $-5.13$ & H14 & $10.0$ & H14 & $0.95$ & H14 & $...$ & $...$ & $...$ & N & $...$ \\
EXT8 & $00\,53\,14.5$ & $+41\,33\,24.5$ & $27.2$ & $80.8$ &  $0.71$ & $0.50$ & $-9.28$    & H14 & $3.2$ & H14 & $0.79$ & H14 & $-194 \pm 6$  & $102 \pm 7$   & V14 & N & $...$ \\
PAndAS-41 & $00\,53\,39.6$ & $+42\,35\,15.0$ & $33.1$ & $56.1$ &  $0.32$ & $0.32$ & $-7.07$ & dTZ15 & $6.2$ & dTZ15 & $0.80$ & dTZ15 & $-94 \pm 8$ & $203 \pm 9$ & V14 & S & Stream C/D \\
H23 & $00\,54\,25.0$ & $+39\,42\,55.7$ & $37.0$ & $124.0$ &  $0.99$ & $0.76$ & $-8.09$ & H14 & $2.8$ & T12 & $1.01$ & H14 & $-373 \pm 1$ & $-80 \pm 4$ & S15 & A & (Stream D) \\
HEC10 & $00\,54\,36.5$ & $+44\,58\,44.6$ & $58.7$ & $29.3$ &  $0.92$ & $0.54$ & $-6.14$ & H14 & $22.5$ & H14 & $0.97$ & H14 & $-98 \pm 5$ & $202 \pm 6$ & V14 & N & $...$ \\
HEC11 & $00\,55\,17.4$ & $+38\,51\,02.0$ & $46.6$ & $134.2$ &  $0.40$ & $0.58$ & $-6.65$ & H14 & $14.6$ & H14 & $0.90$ & H14 & $-215 \pm 5$ & $76 \pm 6$ & V14 & A & (Stream D) \\
H24 & $00\,55\,43.9$ & $+42\,46\,15.9$ & $38.8$ & $57.0$ &  $0.28$ & $0.29$ & $-7.10$ & H14 & $8.9$ & T12 & $0.94$ & H14 & $-121 \pm 15$ & $176 \pm 15$ & V14 & S & Stream C/D \\
PAndAS-42 & $00\,56\,38.0$ & $+39\,40\,25.9$ & $42.2$ & $120.0$ &  $0.40$ & $0.58$ & $-6.59^*$ & H14 & $15.4^*$ & H14 & $1.05^*$ & H14 & $-176 \pm 4$ & $116 \pm 5$ & V14 & A & (Stream D) \\
PAndAS-43 & $00\,56\,38.8$ & $+42\,27\,17.8$ & $38.9$ & $64.2$ &  $0.14$ & $0.34$ & $-5.27$ & H14 & $4.8$ & H14 & $0.97$ & H14 & $-135 \pm 6$ & $161 \pm 7$ & V14 & S & Stream C/D \\
PAndAS-44 & $00\,57\,55.9$ & $+41\,42\,57.0$ & $39.4$ & $79.8$ &  $0.79$ & $0.49$ & $-7.72$ & H14 & $2.2$ & H14 & $0.82$ & H14 & $-349 \pm 11$ & $-55 \pm 11$ & V14 & S & Stream C/D \\
HEC12 & $00\,58\,15.4$ & $+38\,03\,01.3$ & $60.0$ & $135.9$ &  $0.05$ & $0.19$ & $-6.16$ & H14 & $33.2$ & T12 & $0.99$ & H14 & $-288 \pm 2$ & $0 \pm 4$ & V14 & S & Stream Cp \\
HEC13 & $00\,58\,17.1$ & $+37\,13\,49.8$ & $68.8$ & $142.1$ &  $0.04$ & $0.09$ & $-5.54$ & H14 & $20.7$ & H14 & $0.85$ & H14 & $-366 \pm 5$ & $-79 \pm 6$ & V14 & S & Stream Cr \\
PAndAS-45 & $00\,58\,38.0$ & $+41\,57\,11.5$ & $41.7$ & $75.7$ &  $0.34$ & $0.40$ & $-4.06$ & H14 & $8.7$ & H14 & $0.96$ & H14 & $-135 \pm 16$ & $159 \pm 16$ & V14 & S & Stream C/D \\
PAndAS-46 & $00\,58\,56.4$ & $+42\,27\,38.3$ & $44.3$ & $67.1$ &  $0.26$ & $0.43$ & $-8.67$ & H14 & $3.2$ & H14 & $0.85$ & H14 & $-132 \pm 16$ & $163 \pm 16$ & V14 & S & Stream C/D \\
PAndAS-47 & $00\,59\,04.8$ & $+42\,22\,35.1$ & $44.3$ & $68.7$ &  $0.21$ & $0.43$ & $-5.66$ & H14 & $2.8$ & H14 & $1.14$ & H14 & $-359 \pm 16$ & $-64 \pm 16$ & V14 & S & Stream C/D \\
H26 & $00\,59\,27.5$ & $+37\,41\,30.9$ & $65.8$ & $136.6$ &  $0.06$ & $0.10$ & $-7.40$ & H14 & $4.4$ & H14 & $0.90$ & H14 & $-411 \pm 7$ & $-124 \pm 8$ & V14 & A & (Stream C) \\
PAndAS-48 & $00\,59\,28.3$ & $+31\,29\,10.6$ & $141.3$ & $159.7$ & $0.09$ & $0.15$ & $-4.80$ & M13 & $26.0$ & M13 & $0.81$ & H14 & $-250 \pm 5$ & $24 \pm 6$ & V14 & A & ($\zeta_{\rm MP}$) \\
H25 & $00\,59\,34.6$ & $+44\,05\,38.9$ & $57.2$ & $46.2$ &  $0.70$ & $0.43$ & $-7.93$ & H14 & $4.6$ & H14 & $0.94$ & H14 & $-204 \pm 14$ & $93 \pm 14$ & V14 & N & $...$ \\
B517 & $00\,59\,59.9$ & $+41\,54\,06.8$ & $44.9$ & $77.5$ &  $0.15$ & $0.23$ & $-8.17$   & P10 & $...$   & $...$ & $0.93$ & P10 & $-277 \pm 13$ & $16 \pm 13$   & V14 & S & Stream C/D \\
PAndAS-49 & $01\,00\,50.1$ & $+42\,18\,13.3$ & $48.2$ & $71.5$ &  $0.39$ & $0.33$ & $-4.81$ & H14 & $16.4$ & H14 & $1.07$ & H14 & $-240 \pm 7$ & $54 \pm 8$ & V14 & A & (Stream C/D) \\
PAndAS-50 & $01\,01\,50.7$ & $+48\,18\,19.2$ & $106.7$ & $24.1$ &  $0.03$ & $0.01$ & $-6.38$ & H14 & $17.1$ & H14 & $1.10$ & H14 & $-323 \pm 7$ & $-23 \pm 8$ & V14 & A & ($\zeta_{\rm MP}$) \\
PAndAS-51 & $01\,02\,06.6$ & $+42\,48\,06.6$ & $53.4$ & $65.3$ &  $0.90$ & $0.36$ & $...$ & $...$ & $...$ & $...$ & $...$ & $...$ & $-226 \pm 5$ & $68 \pm 6$ & V14 & N & $...$ \\
H27 & $01\,07\,26.3$ & $+35\,46\,48.4$ & $99.9$ & $136.7$ &  $0.34$ & $0.12$ & $-8.39$ & H14 & $4.0$ & T12 & $0.86$ & H14 & $-291 \pm 2$ & $-12 \pm 4$ & AB09 & N & $...$ \\
PAndAS-52 & $01\,12\,47.0$ & $+42\,25\,24.9$ & $78.1$ & $75.9$ &  $0.49$ & $0.40$ & $-7.58$ & H14 & $5.3$ & H14 & $0.96$ & H14 & $-297 \pm 9$ & $-9 \pm 9$ & V14 & N & $...$ \\
PAndAS-53 & $01\,17\,58.4$ & $+39\,14\,53.2$ & $95.9$ & $103.9$ &  $0.45$ & $0.49$ & $-9.09$ & H14 & $3.1$ & H14 & $0.85$ & H14 & $-271 \pm 1$ & $8 \pm 3$ & S15 & N & $...$ \\
PAndAS-54 & $01\,18\,00.1$ & $+39\,16\,59.9$ & $95.8$ & $103.6$ &  $0.49$ & $0.42$ & $-8.58^*$ & H14 & $3.9^*$ & H14 & $0.84^*$ & H14 & $-345 \pm 1$ & $-66 \pm 3$ & S15 & N & $...$ \\
\hline
\end{tabular}
\end{table}
\end{landscape}

\addtocounter{table}{-1}

\begin{landscape}
\begin{table}
\centering
\caption{-- {\it continued.}}
\begin{minipage}{235mm}
\begin{tabular}{lccccccccccccccccc}
\hline \hline
Name & \multicolumn{2}{c}{Coordinates (J2000.0)} & $R_{\rm proj}$ & PA & $\zeta_{\rm MP}$ & $\zeta_{\rm MR}$ & $M_V$ & Source & $r_{\rm h}$ & Source & $(V-I)_0$ & Source & $V_{\rm helio}$ & $V_{\rm M31}$ & Source & Class & Note \\
 & RA & Dec & (kpc) & (deg) & & & (mag) & & (pc) & & (mag) & & (km$\,$s$^{-1}$) & (km$\,$s$^{-1}$) & & & \\
\hline
PAndAS-56 & $01\,23\,03.5$ & $+41\,55\,11.0$ & $103.3$ & $81.7$ & $0.03$ & $0.71$ & $-7.63$ & H14 & $3.5$ & H14 & $0.89$ & H14 & $-241 \pm 2$ & $40 \pm 4$ & S15 & S & \hspace{3mm}E Cloud\hspace{3mm} \\
PAndAS-57 & $01\,27\,47.5$ & $+40\,40\,47.2$ & $116.4$ & $90.3$ & $0.03$ & $0.54$ & $-5.70$ & H14 & $10.3$ & H14 & $0.91$ & H14 & $-186 \pm 6$ & $90 \pm 7$ & V14 & S & E Cloud \\
dTZZ-21 & $01\,28\,49.2$ & $+47\,04\,22.0$ & $137.8$ & $50.1$ & $0.59$ & $0.89$ & $-7.25$ & dTZ14 & $4.9$ & dTZ14 & $0.93$ & dTZ14 & \hspace{3mm}$...$\hspace{3mm} & \hspace{3mm}$...$\hspace{3mm} & $...$ & N & $...$ \\
PAndAS-58 & $01\,29\,02.2$ & $+40\,47\,08.7$ & $119.4$ & $89.4$ & $0.06$ & $0.46$ & $-6.17$ & H14 & $9.3$ & H14 & $1.04$ & H14 & $-167 \pm 10$ & $109 \pm 10$ & V14 & S & E Cloud \\
\hline
\end{tabular}
\medskip
\vspace{-1mm}
\\
Notes: (i) The calculated galactocentric radii $R_{\rm proj}$ and position angles (PA) assume that the M31 centre has coordinates $00\,42\,44.3$ $+41\,16\,09.4$ and sits at a distance modulus $(m-M)_0 = 24.46$.\\
(ii) The listed cluster luminosities $M_V$ and sizes ($r_{\rm h}$) also assume $(m-M)_0 = 24.46$, except for clusters MGC1 \citep{mackey:10a} and PA-48 \citep{mackey:13b}.\\
(iii) Photometric measurements marked with an asterisk ($^*$) have poor quality flags in \citet{huxor:14}.\\
(iv) Classification flags are: `N' = ``non-substructure''; `A' = ``ambiguous''; and `S' = ``substructure''.\\
(v) Objects that are classified as members of the ``substructure'' class have their underlying halo structure or kinematic group listed under ``Notes''.\\
(vi) Objects that are classified as members of the ``ambiguous'' class have the likely halo structure or kinematic group listed in parenthesis; ($\zeta_{\rm MP}$) indicates a local stellar density in the top quartile.\\
(vii) List of sources: AB09 = \citet{alvesbrito:09}; B07 = \citet{barmby:07}; dTZ14 = \citet{dtz:14}; dTZ15 = \citet{dtz:15}; G06 = \citet{galleti:06}; H14 = \citet{huxor:14}; M10 = \citet{mackey:10a}; M13 = \citet{mackey:13b}; M18 = this work; P10 = \citet{peacock:10}; S15 = \citet{sakari:15}; T12 = \citet{tanvir:12}; V14 = \citet{veljanoski:14}.
\end{minipage}
\end{table}
\end{landscape}


\end{document}